\documentclass[ 11pt,english,letterpaper, leqno]{article}
\usepackage[T1]{fontenc}
\usepackage[latin9]{inputenc}
\usepackage{geometry}
\usepackage{verbatim}
\usepackage{ragged2e}
\usepackage{textcomp}

\usepackage{subfig}
\usepackage{url}
\usepackage{amsmath}
\usepackage{amsthm}
\usepackage{amssymb}
\usepackage{wasysym}
\usepackage{graphicx}
\usepackage{setspace}
\usepackage[authoryear]{natbib}
\usepackage{xcolor}
\usepackage[hidelinks]{hyperref}

\newtheorem{theorem}{Theorem}[section]

\newtheorem{remark}[theorem]{Remark}

\definecolor{bronze}{rgb}{0.8, 0.5, 0.2}
\definecolor{applegreen}{rgb}{0.55, 0.71, 0.0}
\definecolor{armygreen}{rgb}{0.29, 0.33, 0.13}
\definecolor{amber}{rgb}{1.0, 0.75, 0.0}
\definecolor{britishracinggreen}{rgb}{0.0, 0.26, 0.15}
\definecolor{brown(traditional)}{rgb}{0.59, 0.29, 0.0}
\definecolor{cadmiumorange}{rgb}{0.93, 0.53, 0.18}

\makeatletter

\providecommand{\tabularnewline}{\\}

\theoremstyle{plain}
\newtheorem{thm}{\protect\theoremname}

\renewcommand{\hat}{\widehat}

\makeatother

\usepackage{babel}
\providecommand{\theoremname}{Theorem}

\title{Boosting: Why You Can Use the HP Filter}
\date{}
\author{Peter C. B. Phillips$^{a,b,c,d}$ and
Zhentao Shi$^{e,}$\footnote{This paper is an updated and revised version of an earlier working paper entitled `Boosting the Hodrick Prescott Filter' \citep{phillips2019boosting}. 
	PCBP and ZS contributed equally to this paper.
	We thank the Editor, two referees, Michael Zheng Song and Brendan Beare for helpful comments and suggestions. 
	We thank Yang Chen for excellent research assistance. Phillips acknowledges research support from the Kelly Foundation at the University of Auckland, the NSF under Grant No. SES 18-50860, and an LKC Fellowship at Singapore Management University. Shi acknowledges support from the Hong Kong Research Grants Council Early Career Scheme
	No.\,24614817.
	Please address correspondence to: Peter C.B. Phillips,
	Cowles Foundation for Research in Economics, Yale University, Box 208281, New Haven, CT 06520-8281, U.S.A.
	E-mail: \texttt{peter.phillips@yale.edu}.
}
\\
\\
	$^{a}$Yale University, USA,$^{b} $University of Auckland, New Zealand,\\
	$^c$University of Southampton, UK, $^d$Singapore Management University, Singapore\\
	$^e$The Chinese University of Hong Kong, Hong Kong SAR, China
}

\begin{document}

\maketitle

\thispagestyle{empty}

\medskip

\normalsize

\bigskip





\bigskip

{\small
\textbf{Abstract:}
We propose a procedure of iterating the HP filter to produce a smarter smoothing device, called the boosted HP (bHP) filter, based on $L_{2}$-boosting in machine learning. Limit theory shows that the bHP filter asymptotically recovers trend mechanisms that involve integrated processes, deterministic drifts, and structural breaks, covering the most common trends that appear in current modeling methodology. A stopping criterion automates the algorithm, giving a data-determined method for data-rich environments. The methodology is illustrated in simulations and with three real data examples that highlight the differences between simple HP filtering, the bHP filter, and an alternative autoregressive approach.
}

\newpage{}
\normalsize
\singlespacing
\setcounter{page}{1}

\begin{quote}
 \emph{The principle adopted here in the construction of a trend for a time series consists in minimizing a linear combination of two sums of squares, of which one refers to the second differences of the trend values, the other to the deviations of the observations from the trend values ... this procedure seems a particularly natural one when dealing with economic time series. The resulting family of trends may be described as quasi-linear trends.} \citet{leser1961simple}
\end{quote}

\begin{quote}
	\emph{Our statistical approach does not utilize standard time series analysis. The maintained hypothesis based on growth theory considerations is that the growth component of aggregate economic times series varies smoothly over time.} \citet{hodrick1997postwar}
\end{quote}


\section{Introduction}

Two prominent features of macroeconomic data are trending long-run growth in aggregate economic activity and a cyclical component that represents fluctuations in this activity over shorter periods known as business cycles. Modern macroeconomic theory of the business cycle, evident in the vast literature on RBC and DSGE modeling, seeks to explain the cyclical movement and co-movement of macroeconomic variables about long-run trends. Both aspects of economic activity are important in economic analysis and policy making. Trends are an intrinsic element in determining long term economic prospects and the overall health of an economy. Cyclical behavior is especially important to policy makers, who are interested in understanding past contractions with a view to foreseeing the onset of future recessions and minimizing their impact on employment and business activity.  

To analyze business cycles in observed data it is necessary to isolate the cyclical component from the trend. Rigorous study requires clarity concerning the trend mechanism and various definitions have been used in past work to distinguish slow moving and cyclical mechanisms in the data.\footnote{Readers are referred to
\cite{phillips1998new,phillips2003laws,phillips2005automated,phillips2010mysteries}, \cite{phillips2004local}, \cite{white2011consideration}, and \cite{muller2018long} for general discussion and limitations of trend formulations commonly used in empirical work.} Decomposition into trend and cycle is commonly achieved by regression or filtering. The latter is primarily motivated by the prior view that a trend is distinguished as a smoothly varying component in relation to the observed data, a concept reflected in the header quotation from \citet{hodrick1997postwar} (hereafter HP), leading to the so-called HP filter, to the use of spectral methods \citep{hannan1963, christiano2003band}, and to the use of orthonormal polynomial regression \citep{phillips1998new,phillips2005automated, phillips2014optimal}. The HP filter belongs to the statistical approach that was introduced by \citet*{whittaker1922new} and \citet*{whittaker1960calculus}, who provided a probabilistic framework of penalized maximum likelihood estimation to deliver a quantitative measure of trend (or graduation in their terminology). The explicit form of the smoothness measure penalty involving squared second differences in HP (1997) was used in earlier work by \citet{leser1961simple}, who emphasized its relevance to trend determination with economic data on the grounds of its `quasi-linear' trend-producing properties, as indicated in the primary header quotation.  

The HP method is now widely used in applied macroeconomic work by economists in central banks, international economic agencies, industry, and government. Its use in academic work is less extensive, partly because it has been subject over many years to considerable criticism and analyses that have revealed a myriad of its limitations for empirical studies in economics, an early example being \citet{cogley1995effects}. For recent discussion of the merits and demerits of the filter, see \citet{phillips2015business}  (PJ, henceforth)  and \citet{hamilton2017you} and the many references cited therein. 

Given time series data $\left(x_{t}:t=1,\ldots,n\right)$, the HP method decomposes the series into
two additive components --- a trend component $\left(f_{t}\right)$,
and a residual or cyclical component $\left(c_{t}\right)$, estimated as 
\begin{equation} \label{tt0}
\left(\widehat{f}_{t}^{\mathrm{HP}}\right)=\arg\min_{\left(f_{t}\right)}\left\{ \sum_{t=1}^{n}\left(x_{t}-f_{t}\right)^{2}+\lambda\sum_{t=2}^{n}\left(\Delta ^ 2 f_{t}\right)^{2}\right\}, 
\text{ and}
\quad \left(\widehat{c}_{t}^{\mathrm{HP}}\right)=\left(x_t-\widehat{f}_{t}^{\mathrm{HP}}\right)
\end{equation}
where $\Delta f_{t}=f_{t}-f_{t-1}$, $\Delta^2 f_{t}= \Delta f_{t}- \Delta f_{t-1} = f_{t}- 2 f_{t-1} + f_{t-2} $,  and $\lambda\geq0$
is a tuning parameter that controls the extent of the penalty. As is apparent from this criterion, the method is nonparametric and the choice of $\lambda$ inevitably plays a major role in determining the shapes of the fitted trend and cycle functions. If $\lambda$ is selected too large, the fitted trend becomes nearly linear, as implied by Leser's (1961) characterization, and a linear trend is indeed the solution when $\lambda \to \infty$. 
In consequence, if the trend is nonlinear and $\lambda$ is too large, the HP fitted trend leaves a residual trend that inevitably contaminates the cyclical component. If $\lambda$ is selected too small, the fitted trend becomes highly flexible, so that it closely tracks the data and thereby embodies elements of short-term fluctuations. In the absence of an underlying model that defines and enables direct quantification of trend and cycle, some degree of cycle distortion with the HP filter is inevitable. Of course, precisely the same criticism applies to other smoothing techniques, as well as regression methods of trend extraction when the trend model is misspecified, as is nearly always the case in practical work.  

In the application of modern nonparametric methods, serious efforts are normally made to choose tuning parameters based on some well-defined optimization criterion with data-determined versions of these criteria that can be implemented in empirical work. This approach is greatly facilitated by the use of an underlying model representing the generating mechanism. By contrast, empirical practice with the HP filter almost universally relies on standard settings for the tuning parameter that have been suggested largely by experimentation with macroeconomic data and heuristic reasoning about the form of economic cycles and trends. For quarterly data the standard choice is $\lambda=1600$, as recommended by HP (1997) based on their experimentation with US data. This value has served as a gold standard in converting the tuning parameter to other sampling frequencies such as annual or monthly data \citep{ravn2002adjusting}. Importantly, these HP filter smoothing parameter settings are normally employed irrespective of the sample size of the data in contrast to standard nonparametric methods.

HP filtering retains an agnostic position with regard to the exact functional form of the trend and cycle. This approach has the advantage of generality, but the central weakness is that its properties with respect to consistent trend estimation are poorly understood, especially with regard to the relevant choice of the tuning parameter. Recent research by PJ has shown that the performance of the filter as a potentially consistent estimator of trends of certain types can be assessed directly in relation to the choice of the tuning parameter, just as in standard nonparametric analysis as the sample size $n \to \infty$. That work revealed the constraints that are needed on $\lambda$ in order to achieve consistent trend estimation for polynomial trends and stochastic process trends. The paper also analyzed the effect of the filter on trends with structural breaks. In doing so, PJ showed how the standard HP filter settings may not be adequate in removing trends --- particularly stochastic trends --- in economic data, a conclusion that reverses earlier thinking \citep{king1993low} to the effect that the HP filter removes up to four unit roots in the original data. PJ further showed that the common choice of the tuning parameter $\lambda=1600$ for quarterly data is typically too large to completely
remove stochastic process trends given the length of the macroeconomic time series usually encountered in practice. It is likely therefore that remnants of stochastic trends linger in the fitted cycles of much applied research, a conclusion supported by the recent examples exhibited in \citet{hamilton2017you} and the earlier work by \citet{cogley1995effects}.

The present paper proposes an easy-to-implement modification that is designed to make the HP filter
more effective in trend fitting and trend elimination, and is formulated with data-determined smoothing choices to aid practical application. The idea is simple. Since the cyclical component may often retain trend elements, we feed the data into the filter again to clean the leftover elements. The notion of refitting the residual in statistical applications goes back to \citet{tukey1977exploratory}
under the name \emph{twicing}, where procedures were employed twice to assist in data-cleaning exercises. 
The notion of twicing can be continued into ``$n$--ing'', as discussed in \citet{buja1989linear}. This approach motivates the present proposal of repeated application of the HP filter. Since the solution of the HP filter can be
explicitly expressed as a linear operation, which will be made clear in
Section \ref{sec:Boosted-HP-Filter}, repeated HP fitting is closely
related to the $L_{2}$-boosting \citep{buhlmann2003boosting} procedure which is now commonly used in
machine learning. 

The existing statistical theory on boosting is developed mostly
for environments with independent identically distributed (iid) data. We are unaware of any earlier work concerned with the use of $L_{2}$-boosting on stochastically trending data or nonstationary time series. 
A primary contribution of the present paper is to apply and analyze the idea of repeated fitting algorithms in machine learning to trend detection and nonstationary time series environments. 
In particular, we establish asymptotic theory to justify the use of the boosted version of the HP filter, extending earlier work by PJ on the asymptotics of the HP filter. We find that if the number of iterations
slowly diverges with the sample size, the boosted HP filter (bHP filter, hereafter) can recover the underlying trend irrespective of whether the time series contains a stochastic process trend, a deterministic polynomial drift, or a polynomial drift with a structural break. An interesting implication of the asymptotic theory in the latter case is that the bHP filter effectively delivers a consistent estimator of the break point. This result extends to the case of multiple structural breaks, so the boosted filter provides a new device for consistently estimating multiple break points, delivered automatically without additional methods of detection.    

Use of machine learning methods in econometrics has grown quickly in recent years with a particular focus on applied microeconomics where large cross sectional and wide panel datasets are available. The low-frequency nature of most macroeconomic time series, on the other hand, means that the volume of today's macroeconomic databases is much smaller by comparison and is, of course, completely dwarfed by those generated from Internet communication.\footnote{
For instance, the size of the latest version of FRED-MD (as of August
2020) is 600kb for monthly data or 170kb for quarterly data. 
FRED-MD is the Monthly Databases for
Macroeconomic Research (\url{https://research.stlouisfed.org/econ/mccracken/fred-databases/}).
See \citet{mccracken2016fred} for details.} Nonetheless, the phenomena that macroeconomists study are no less complex
than those that confront scientists working with `big data' and the scope for machine learning methods is large. For example,
in empirical analysis using country-level macro indicators, macroeconomists
handle data collected from a large number of heterogeneous economies in highly varying stages of economic development. In such cases, where there are many possible determining factors and diverse time series trajectories, automated econometric procedures \citep{phillips2005automated} can be of tremendous appeal in practical work, as has been argued recently by many authors since the development of high dimensional regression methods such as Lasso \citep{tibshirani1996regression}. 

This line of thinking partly motivates the present econometric implementation of machine learning methods in which we propose two data-driven stopping rules to terminate iterations of the bHP filter. One rule ceases the iteration
according to the outcome of a unit root test, and is appropriate when stationary time series is considered a prerequisite for further investigation, such as business cycle analysis. The second rule relies on a new version of the Bayesian information criterion (BIC) that is developed for the present bHP filter framework to take into account sample fit and effective degrees of freedom after each iteration. The latter approach accords with the now common practice of using BIC-type information criteria as stopping rules in econometric work. 

An important additional advantage to boosting is that it provides robustness to the original setting of the tuning parameter $\lambda$ in the HP filter prior to the boosting iteration. In particular, simulation experiments (reported in Appendix \ref{simu_lambda_robust}) show  that the boosted filter is robust to the HP tuning parameter setting over a wide region amounting to a factor of $4$ on either side of the conventional $\lambda=1600$ used in empirical work with quarterly data. This means that practitioners using the boosted HP filter can employ HP algorithms with standard parameter settings in conjunction with the boosting iteration and its automated termination criteria with some confidence concerning the robustness of the results. In effect, this all but eliminates dependence of the bHP on the original HP tuning parameter and provides empirical investigators some protection against critiques that findings are dependent on an arbitrary choice of tuning parameter. This advantage is  particularly useful in cross country panel comparisons of cycles and trends where multiple time series of different lengths are commonly studied in applications.    

We conduct three real data applications of the HP and bHP filters to study the impact of our iterative fitting algorithm. The first application revisits Okun's law, which posits an empirical association
of comovement between real GDP and unemployment, in an international cross country setting. \citet{ball2017okun}
extend the scope of \citet{okun1962potential}'s original focus
on the United States to 20 OECD countries, applying the standard HP filter to each time series to obtain deviations from long run levels. Accordingly, when the standard filter fails to remove the time trend, the resulting Okun law regression can suffer spurious regression effects. The bHP filter helps to mitigate the effects of potential contamination from long run influences.  

A second example conducts a cross country comparison of business cycles. In an extensive application of HP methods to remove trend, \citet{aguiar2007emerging} suggest that cyclical components are more persistent and volatile in emerging economies than in developed countries. In revisiting this application using the bHP filter, we find that the time series collected from the emerging economies are
much shorter than those from the developed ones, and so the use of the standard $\lambda=1600$ tuning parameter uniformly across all countries tends to over-penalize the shorter series. Repeated fitting helps to regularize the unbalanced panel and robustify the finding by \citet{aguiar2007emerging} of the empirical distinction in cyclical behavior between emerging and developed economies. 

The third application examines US industrial production over the last century from 1919 to 2018. Like many other macroeconomic time series this series displays strong trend characteristics with some major fluctuations over subperiods that include two world wars and the great depression during the early part of the period, and the financial crisis and great recession over the latter period. As such, the series presents challenges in trend determination that include the complex issue of whether such subperiods are better interpreted as part of the trend or part of the evolving cyclical processes of modern industrialized economies. In this application, we provide a detailed comparison of the HP and bHP filters with the alternative autoregressive modeling approach recently advocated by \citet{hamilton2017you}. 

We close this introduction with a brief discussion of related literature on filtering and boosting. First, there are now many competing methods of data filtering to remove trend such as the band pass filter methods of \citet{baxter1999measuring}, \citet{christiano2003band}, and \citet{corbae2006extracting}. Most of these share many common characteristics with the HP filter. Nonetheless, the HP filter remains the most popular\footnote{As of August, 2020, the article by \cite{hodrick1997postwar} had over 9,500 listed citations and the article by \cite{baxter1999measuring} nearly 4,000 citations in Google Scholar.} in practical work and serves as a benchmark for other agnostic methods of trend extraction. In addition, there has been renewed recent interest in the theoretical properties
of the HP filter, useful algebraic representations, and computational algorithms. \citet{phillips2010two} and PJ provide exact matrix and operator representations and new asymptotics. \citet{cornea2017explicit} gives an explicit algebraic formula 
for the HP filter in finite samples. \citet{de2016econometrics} and \citet{sakarya2017property} provide further finite sample results, including another representation of the HP filter as a symmetric weighted average
plus some adjustments. \citet{hamilton2017you} provides a cautionary note concerning the limitations
of the HP filter approach, reinforcing earlier warnings and suggesting the alternative of scalar autoregression.

Second, machine learning methods have been employed to generate
new statistical procedures specifically tailored for economic applications in recent work by  
\citet{belloni2012sparse}, \citet{belloni2014inference}, 
\citet{chernozhukov2015post},
\citet{fan2015power},
\citet{hirano2017forecasting}
and \citet{caner2018asymptotically}, to name a few. 
Boosting is one of the most successful machine learning methods. Originally
proposed for classification problems \citep{freund1995desicion},
boosting has given rise to many useful variants (e.g. \citet{hastie2009bible}).
\citet{buhlmann2003boosting} extended the idea of refitting to linear regression with the $L_{2}$ norm for the residuals, opening up a wide range of potential applications. In high dimensional regression, component-wise boosting is also related
to forward stage selection and the greedy algorithm \citep{buehlmann2006boosting}.
The idea of refitting (specifically twicing) was introduced by \cite{tukey1977exploratory} and appeared in econometrics in \citet{newey2004twicing} and most recently \citet{lee2018lasso}. 
In high-dimensional regression, \citet{bai2009boosting} employed boosting in macroeconomic forecasting. \citet{shi2016econometric}
used boosting to select relevant moments in structural models defined by many moment conditions. \citet{ng2014viewpoint}
and \citet{luo2017_2} applied boosting to recession forecasting and
other economic examples. 

The rest of the paper is organized as follows.
Section \ref{sec:Boosted-HP-Filter} introduces the iterative algorithm for boosting the HP
filter and develops asymptotic theory that characterizes the behavior of the boosted filter, giving conditions for consistent estimation of stochastic process and deterministic polynomial trends with possible structural breaks.
Stopping rules are provided to automate the procedure for practical work.
Simulations are conducted in Section \ref{sec:simulations} to reveal the effect of boosting along with
the stopping rules.
Section \ref{sec:emp-application} reports three empirical applications of the bHP methodology. 
Section \ref{sec:conclusion} concludes with a summary of arguments in support of the bHP filter as a trend determination device in empirical research and a response based on our present findings to the recent critique of the HP filter by \citet{hamilton2017you}. Proofs are given in Appendix A and additional graphical demonstrations in Appendix B.

\section{The Boosted HP Filter} \label{sec:Boosted-HP-Filter}

\subsection{The Boosting Algorithm}

The optimization problem (\ref{tt0}) leading to the HP filter and related criteria for general filters of this type have closed-form algebraic solutions in convenient matrix form.\footnote{See \cite{phillips2010two}, PJ, \cite{de2016econometrics}, and \cite{cornea2017explicit} for recent work on exact matrix forms and other exact representations of the HP and related filters.} In the HP case, if $D'$ is the rectangular $\left(n-2\right)\times n$ matrix
with second differencing vector $d=(1,-2,1)^{\prime}$ along
the leading tri-diagonals and  $I_{n}$ is the $n\times n$ identity
matrix, the explicit form of the trend solution is
\begin{equation} \label{tt1}
\widehat{f}^{\mathrm{HP}}=Sx,
\end{equation}
where $S=\left(I_n+\lambda DD'\right)^{-1}$ is a deterministic operator and $x=(x_1,...,x_n)'$ is the sample data.
The smoothed component $\widehat{f}^{\mathrm{HP}}$ is interpreted as the estimated 
trend and 
\begin{equation} \label{ttt}
\widehat{c}^{\mathrm{HP}}=x-\widehat{f}^{\mathrm{HP}}=\left(I_{n}-S\right)x
\end{equation}
as the estimated cyclical or stationary component. 

The behavior and asymptotic properties of the estimated trend $\widehat{f}^{\mathrm{HP}}$ crucially depend 
on the choice of the tuning parameter and the underlying generating mechanism of $x_t$. For macroeconomic data the mechanism may reasonably be expected to involve a stochastic trend, possibly accompanied by some deterministic drift component that may be well modeled by a low order polynomial or a similar deterministic function subject to breaks. In the prototypical case of a unit root process, $x_t$ satisfies under quite general conditions the functional law \citep{phillips1992asymptotics}
\begin{equation} \label{tt4}
n^{-1/2}  x_{\left\lfloor nr\right\rfloor }  \rightsquigarrow B(r)
\end{equation}
where $\lfloor \cdot \rfloor $ is the floor function, $B$ is Brownian motion with variance $\omega^2$ given by the long run variance of $\Delta x_t$, and $\rightsquigarrow$ signifies weak convergence, here on the Skorohod space $D[0,1]$. Asymptotics analogous to \eqref{tt4} with related Gaussian limit processes such as linear diffusions hold for other stochastic trends after appropriate normalization and the methods we discuss are applicable in such cases.  

The problem of consistent HP filter estimation of the trend then amounts to whether as $n \to \infty$ we have 
$ n^{-1/2} {\widehat{f}_{\left\lfloor nr\right\rfloor }^{\mathrm{HP}}} \rightsquigarrow B(r)$, 
in which case the filter asymptotically captures the underlying stochastic process trend in $x_t$. PJ show that this reproduction of the asymptotic form of the trend holds only under special restrictions on the smoothing parameter $\lambda$ that ensure it does not diverge too quickly. In particular, if  $\lambda = O(n^4)$ or greater as $n \to \infty$, then the HP filter trend 
$ \widehat{f}_{\left\lfloor nr\right\rfloor }^{\mathrm{HP}}$
 is inconsistent. In such cases, 
$ n^{-1/2}  {\widehat{f}_{\left\lfloor nr\right\rfloor }^{\mathrm{HP}}} \rightsquigarrow f^{\mathrm{HP}}\left(r\right)$ 
where $f^{\mathrm{HP}}\left(r\right)$ is a smooth stochastic process different from $B(r)$, which implies that the cyclical component $\widehat{c}^{\mathrm{HP}}=x-\widehat{f}^{\mathrm{HP}}$ inevitably inherits elements of the stochastic trend even in the limit. Similar issues arise in the case of time series with stochastic trends coupled with deterministic drift or deterministic drift with breaks (see PJ for details). In all these cases, the HP filter fails to recover the underlying trend in $x_t$ asymptotically. The limit theory therefore confirms much informal commentary in the literature concerning the shortcomings of the HP filter as a suitable trend determination mechanism for economic data.     

We propose an easy remedy to establish consistent estimation of stochastic process and deterministic trends in the data. If the cyclical component $\widehat{c}_{t}^{\mathrm{HP}}$ still exhibits
trending behavior after HP filtering, we continue to apply the HP filter to
$\widehat{c}^{\mathrm{HP}}$ to remove the leftover trend residual. After a second fitting, the cyclical component can be written as
\[
\widehat{c}^{\left(2\right)}=\left(I_{n}-S\right)\widehat{c}^{\mathrm{HP}}=\left(I_{n}-S\right)^{2}x,
\]
where the superscript ``$\left(2\right)$'' indicates that the HP
filter is fitted twice. The corresponding trend component becomes
\[
\widehat{f}^{\left(2\right)}=x-\widehat{c}^{\left(2\right)}=\left(I_{n}-\left(I_{n}-S\right)^{2}\right)x.
\]
If $\widehat{c}^{\left(2\right)}$ continues to exhibit trend behavior, the filtering process may be continued for a third or further time. 
After $m$ repeated applications of the filter, the cyclical and trend component are 
\begin{eqnarray} \label{pp0}
	\widehat{c}^{\left(m\right)} & = & \left(I_{n}-S\right)\widehat{c}^{\left(m-1\right)}=\left(I_{n}-S\right)^{m}x\\
	\widehat{f}^{\left(m\right)} & = & x-\widehat{c}^{\left(m\right)}=B_{m}x, \label{pp1}
\end{eqnarray}
where $B_{m}=I_{n}-\left(I_{n}-S\right)^{m}.$ We call
this iterated process the \emph{boosted HP filter} or bHP in view of its similarity to $L_{2}$-boosting in terms of  
numerical implementation.

Boosting is so-called because it has the capacity to enhance the flexibility of what is called in the machine learning literature a \emph{weak base learner} as the starting point. In machine learning language, this process is known as a mechanism for achieving the `strength of weak learnability' \citep{schapire1990strength}. 
The intuition behind the asymptotic validity of the bHP filter is the observation that 
the HP filter, with a conventional choice of the tuning parameter $\lambda$,
serves as a weak base learner.\footnote{In quarterly economic time series for example, accumulating evidence has shown that 
the setting $\lambda = 1600$ is often too large given the length of the time series 
typically encountered in empirical macroeconomics \citep{schlicht2005estimating,phillips2015business, hamilton2017you}.}
In consequence, the crude HP filter is too weak by itself to fully capture the underlying trend, particularly when the trend involves a stochastic process. 
Initiating from the conventional HP filter, 
we iterate the procedure to strengthen this filter as a weak base learner. 
As discussed in the following section, under certain conditions on the number of iterations $m$ 
 the boosted version is able to recover the trend as $n\to \infty$, 
even if the base learner itself is too weak for consistent estimation.

\subsection{Asymptotic Theory}

The criterion underlying the HP filter is agnostic about the data generation
process. The key element in controlling the capacity of the filter to capture underlying trend behavior of various forms lies in the choice of the smoothing parameter $\lambda$ and controls that are implemented on its asymptotic behavior in relation to sample size. The latter is particularly important and is presently almost universally neglected in empirical work. As we now discuss, suitable controls may be implemented on the boosted HP filter to ensure that underlying trend behavior is captured consistently.  

It will be convenient to start with the case where the time series $x_t$ has a stochastic trend and satifies the functional law (\ref{tt4}). In this case, Theorem 3 of PJ shows that if $\lambda=\mu n^{4}$ for some fixed constant $\mu>0$
independent of the sample size $n$, then the limiting form of the HP filter is given by a smooth random function whose series representation is
\begin{align} 
	n^{-1/2}  \widehat{f}_{\left\lfloor nr\right\rfloor }^{\mathrm{HP}} 
	\rightsquigarrow f^{\mathrm{HP}}\left(r\right)=\sum_{k=1}^{\infty}\frac{\lambda_{k}^{2}}{\mu+\lambda_{k}^{2}}\sqrt{\lambda_{k}}\varphi_{k}\left(r\right)\xi_{k}\label{eq:PJ_sum}, 
\end{align}
where $\xi_{k}\sim\mathrm{iid\ }N\left(0,\omega^{2}\right)$, $\varphi_{k}\left(r\right)=\sqrt{2}\sin\left(r/\sqrt{\lambda_{k}}\right)$
and $\lambda_{k}=\left[\left(k- 0.5 \right)\pi\right]^{-2}$.
This asymptotic form of the HP filtered data is deduced by analyzing the asymptotic impact of the HP operator on the Karhunen-Lo\`{e}ve (KL) representation\footnote{Readers may refer to \citet*{phillips1998new} for further details of KL representations and their relevance in the asymptotic analysis of nonstationary time series.} of the Brownian motion limit function given in (\ref{tt4}), viz,
\begin{align} 
B(r)=\sum_{k=1}^{\infty}\sqrt{\lambda_{k}}\varphi_{k}\left(r\right)\xi_{k}\label{tt7} 
\end{align}
 The representations (\ref{eq:PJ_sum}) and (\ref{tt7}) are orthonormal series in the trigonometric polynomials $\varphi_{k}\left(r\right)$ as well as the random coefficients $\xi_k$ and the series converge almost surely and uniformly for $r \in [0,1]$. But whereas Brownian motion is everywhere non-differentiable, the asymptotic form of the HP filter given in (\ref{eq:PJ_sum}) is differentiable to the fourth order and converges almost surely and uniformly for $r \in [0,1]$. As discussed in PJ, for typical time series of quarterly macroeconomic data, the limit form in (\ref{eq:PJ_sum}) produces a smoothed version of the time series that closely matches output from an HP filter with $\lambda=1600$ when the constant $\mu$ is set so that $\mu =1600/n^{4}$ to ensure comparability of the tuning parameter with the standard setting that is used in practical work with quarterly data. Thus, (\ref{eq:PJ_sum}) may be regarded as an asymptotic approximation to the trend output from HP filtering typical quarterly macroeconomic time series. The upshot is that when the HP filter is conducted under standard settings for $\lambda$, it fails to deliver a consistent estimate of an underlying stochastic trend in the data.   

The limit formula on the right-hand side of (\ref{eq:PJ_sum}) is particularly convenient as a starting point in understanding the effects of repeated HP fitting. The following theorem confirms that, in contrast to (\ref{eq:PJ_sum}), the bHP filter $\widehat{f}_{t}^{\left(m\right)}$ captures
the stochastic trend in the data when the number of iterations $m$ in the  boosted filter is allowed to diverge and the primary tuning parameter setting is $\lambda=\mu n^{4}$. 

\begin{thm}
	\label{thm:Brownian} Suppose that $x_{t}$ satisfies the functional limit
	law (\ref{tt4}) 	and the HP filter is iterated $m$
	times according to the boosted HP algorithm with $\lambda=\mu n^{4}$ and $\mu$ fixed. If $m\to\infty$ as $n\to\infty$ then 
	\begin{equation} \label{tt6}
	 n^{-1/2} \widehat{f}_{\left\lfloor nr\right\rfloor }^{\left(m\right)} \rightsquigarrow B(r).
    \end{equation}
\end{thm}

\begin{description}
	\item{\textbf{Remarks}}
	\item{\textbf{(i)}}
	This result shows that repeated application of the HP filter to the cyclical component residual from each pass of the filter is successful in asymptotically eliminating remnants of the stochastic trend from the estimated cyclical component of the time series. The bHP filter algorithm thereby assures consistent estimation of a stochastic process trend like Brownian motion. Importantly and distinct from the asymptotic result in PJ, consistent estimation of the stochastic process trend applies for the boosted filter even with the primary filter setting retained as $\lambda=\mu n^{4}$. This property is relevant in applications because the setting $\lambda=\mu n^{4}$ closely matches practical outcomes with the tuning parameter $\lambda=1600$ common in empirical research. Moreover, as simulations reported later in the paper show, the bHP filter turns out to be robust to variations in the choice $\lambda=1600$ for typical sample sizes in applied work. This robustness means that practitioners using our methods can employ standard HP algorithms in conjunction with the boosting iteration and the automated termination criteria discussed in Section \ref{subsec:Stopping-Criterion} below with some confidence concerning the robustness of the results.
	\item{\textbf{(ii)}}
	The heuristic explanation of (\ref{tt6}) is as follows. Both series representations (\ref{eq:PJ_sum}) and (\ref{tt7}) converge almost surely and uniformly in $r$ and therefore admit further linear operations associated with the boosted filter. Successive operations of the filter then proceed to remove the remaining stochastic trend components from the cycle, leading to consistent estimation of the stochastic trend. The proof of the theorem makes use of the operator 
	$G_{\lambda}= 1/[ \lambda L^{-2} (1-L) ^{4}+1]$,
	which is the asymptotic form (apart from end corrections) of the HP operator on the time series. As shown in PJ, the operator $G_\lambda$ may be interpreted as a pseudo-integral operator,\footnote{\label{pseudo-operator}Pseudo-differential and pseudo-integral operators extend the conventional concept of differential and integral operators. For example, if $D=d/dx$ is the usual differential operator and $D^{-1}$ the integral operator, these definitions may be extended to include operators such as $D^{-z}$ for complex $z$ with $\Re(z)>0$ using the integral representation $D^{-z}f(x) =\Gamma(z)^{-1}\int_0^{\infty}e^{-Dt}f(x)t^{z-1}dt=\Gamma(z)^{-1}\int_0^{\infty}f(x-t)t^{z-1}dt$. Thus, when $f(x)=e^{ax}$ for some constant $a$, direct calculation yields the pseudo-derivative $D^{-z}e^{ax}=a^{-z}e^{ax}$. Much more general functional operators such as $g(D)$ may be represented under certain conditions in integral form using related Fourier integral transforms, see \citet{treves1980introduction} and \citet{ross2006fractional}. Similarly, matrix functional operator representation are possible \citep{phillips1987fractional}. These methods are used by PJ in the derivation of \eqref{eq:PJ_sum}.}
	which facilitates the analysis of its asymptotic properties. The corresponding operator that delivers the cyclical component is $1-G_\lambda$ and $m$ successive operations in the boosted HP filter then lead to the operator $(1-G_\lambda)^m$. The asymptotic result is obtained by using the approximation 
	$\left(1-G_\lambda \right)\varphi_{k}\left( t/n\right)\approx
	\mu / (\mu+\lambda_{k}^{2}) \cdot \varphi_{k}\left(t/n\right),$
	which holds with a well-controlled approximation error. Repeated fitting leads to
\begin{align}\label{pp5} 
\left(1-G_\lambda \right)^{m}\varphi_{k}\left(\frac{t}{n}\right)\approx\left(\frac{\mu}{\mu+\lambda_{k}^{2}}\right)^{m}\varphi_{k}\left(\frac{t}{n}\right) \to 0, \quad \text{as } m \to \infty.
\end{align}
	Then, $\left[1-\left(1-G_\lambda \right)^{m}\right]
\varphi_{k}\left( \left\lfloor nr\right\rfloor / n \right) \approx \varphi_{k}\left(r\right)$, {as } $m,n \to \infty$.
Pursuing this line of argument, the proof of Theorem \ref{thm:Brownian} establishes (\ref{tt6}) rigorously by
verifying that the approximation errors accumulated throughout the series summation are asymptotically negligible.
	\item {\textbf{(iii)}} 	When it is viewed as a special case of linear penalized spline smoothing, the HP filter
places knots on the $n-2$ observed time points $x_t,\  t=2, \ldots, n-1$ omitting the first and last observations 
\citep[Equation(2.2)]{paige2010hodrick}. The nonstationary nature of the knots in trending time series cases requires new technical tools in analyzing the asymptotic behavior of the boosting procedure. In this respect, the present results go beyond the scope of existing work, such as \citet{buhlmann2003boosting}'s Section 3.2 which is concerned with boosting nonparametric mean models based on penalized spline smoothing 
with fixed or iid knots.

\end{description}	

We next proceed to consider the effect of boosting the HP filter when the time series involves a deterministic
trend. PJ have shown that the HP filter itself asymptotically preserves a polynomial
trend up to the 3rd order. In consequence, the bHP filter also asymptotically
maintains the presence of a polynomial trend up to the 3rd order. The following result further shows that the bHP filter consistently estimates any higher order polynomial trends that may accompany a stochastic process trend in the data, as well as the limiting stochastic trend itself, thereby capturing the full limiting trend process.  

\bigskip
\begin{thm} \label{thm:drift}
	Let $x_{t}=\alpha_{n}+\beta_{n,1}t+\cdots+\beta_{n,J}t^{J}+x_{t}^{0}$
	where $x_{t}^{0}$ follows the functional limit law (\ref{tt4}) 
	and the coefficients in the polynomial $\alpha_{n}/\sqrt{n}\to\alpha$,
	$n^{j-1/2}\beta_{n,j}\to\beta_{j}$ for $j=1,\ldots,J$. Suppose the HP filter is iterated $m$ times with $\lambda=\mu n^{4}$ and $\mu$ fixed. If $m,n\to\infty$, then 
\begin{equation} \label{tt20}
	 n^{-1/2}   x_{\left\lfloor nr\right\rfloor }, 
	 n^{-1/2} \widehat{f}_{\left\lfloor nr\right\rfloor }^{\left(m\right)} 
	 \rightsquigarrow\alpha+\beta_{1}r+\cdots+\beta_{J}r^{J}+B\left(r\right).
\end{equation}
\end{thm}
\bigskip

The polynomial component of $x_t$ in Theorem \ref{thm:drift} is specified with sample size dependent coefficients, assuring that the standardized time series $  n^{-1/2}  x_{\left\lfloor nr\right\rfloor }  $ satisfies the functional law (\ref{tt20}), giving a limit stochastic process trend with polynomial drift of degree $J$. The result extends Theorem 4 of PJ by showing that boosting the HP filter with primary tuning parameter setting $\lambda=\mu n^{4}$ asymptotically preserves a polynomial trend of any finite order as $m \to \infty$. The implication is that when using the conventional $\lambda=1600$ setting of the smoothing parameter for quarterly macroeconomic time series, the boosting algorithm ensures that the filter delivers a consistent estimator of the limiting form of the trend in a time series that has a stochastic process trend with a finite order polynomial drift. 

A closely related result might be expected in the case of boosting the HP filter applied to a time series that has a stochastic process limiting trend function accompanied by a deterministic drift that is piecewise continuous with a finite number of break points. Theorem 5 of PJ shows by using the Fourier series representation of the drift function that the asymptotic effect of the HP filter is to smooth a piecewise continuous limit drift function into a smooth curve in which the breaks in the deterministic trend are represented by smooth transitions over adjacent neighborhoods. In view of Theorem \ref{thm:drift}, we might expect that boosting the HP filter would enable the filter under some conditions to capture the continuous parts of a polynomial trend as $m,n \to \infty$. The following result provides asymptotic theory and conditions for the case of a time series with a stochastic trend and time polynomial drift with a single break point. 

To fix ideas, suppose $g_{n}\left( t\right) $ is a trend break polynomial with a single
break point at $\tau _{0}=\lfloor nr_{0}\rfloor $ with $r_0 \in (0,1)$ that takes the form 
\begin{equation*}
g_{n}\left( t\right) =\left\{ 
\begin{array}{cc}
\alpha _{n}^{0}+\beta _{n,1}^{0}t+...+\beta _{n,J}^{0}t^{J} & t<\tau
_{0}=\lfloor nr_{0}\rfloor  \\ 
\alpha _{n}^{1}+\beta _{n,1}^{1}t+...+\beta _{n,J}^{1}t^{J} & t\geq \tau
_{0}=\lfloor nr_{0}\rfloor 
\end{array}%
\right. ,
\end{equation*}%
with $\alpha _{n}^{\delta } / \sqrt{n} \rightarrow \alpha ^{\delta }$
and $  \{ n^{j- 0.5 }\beta _{n,j}^{\delta }\rightarrow \beta
_{j}^{\delta }:j=1,..,J \} $ for $\delta =0,1.$ The limiting form of
this polynomial break function is 
\begin{equation*}
n^{-1/2}g_{n}\left( \lfloor nr\rfloor \right) \rightarrow g\left( r\right)
=\left\{ 
\begin{array}{cc}
\alpha ^{0}+\beta _{1}^{0}r+...+\beta _{J}^{0}r^{J} & r<r_{0} \\ 
\alpha ^{1}+\beta _{1}^{1}r+...+\beta _{J}^{1}r^{J} & r\geq r_{0}%
\end{array}%
\right. ,
\end{equation*}%
giving a piecewise continuous polynomial function with a single break at $%
r=r_{0} \in (0,1).$ If the generating mechanism of the observed data $x_{t}$ is $%
x_{t}=g_{n}\left( t\right) +x_{t}^{0}$, where $x_{t}^{0}$ is a stochastic
trend that satisfies the functional limit law (\ref{tt4}), then the normalized process
 $ n^{-1/2} x_{t=\lfloor nr\rfloor } $ has the following limit 
\begin{equation}
n^{-1/2}  x_{t=\lfloor nr\rfloor } \rightsquigarrow g\left( r\right) +B\left(
r\right) =:B_{g}(r)  \label{tt30}
\end{equation}%
as $n\rightarrow \infty $.
PJ explore the asymptotic form of the HP filter applied to such a time
series $x_{t},$ showing that when $\lambda =\mu n^{4}$ the limiting form of
the normalized HP filtered time series is a smooth function $B_{g}^{\mathrm{HP}}(r):=g^{\mathrm{HP}}\left( r\right)
+B^{\mathrm{HP}}(r)$ where $g^{\mathrm{HP}}\left( r\right) $ is a continuous function
approximation to $g\left( r\right) $ that smooths over the break point of $%
g\left( r\right) $ at $r_{0}$ and $B^{\mathrm{HP}}(r)$ is a smooth functional
approximation to the Brownian motion $B\left( r\right) $ of the same form as
(\ref{eq:PJ_sum}). 

The following result shows that the bHP filter can consistently
estimate the limit function $B_{g}(r)$ for $r\not=r_{0.}$, thereby capturing the polynomial drift function at all points except the break point, as well as the stochastic trend process. 

\begin{thm} \label{thm:break}
Let $x_{t}=g_{n}\left( t\right) +x_{t}^{0}$ where $x_{t}$ satisfies $%
n^{-1/2}x_{t=\lfloor nr\rfloor }\rightsquigarrow g\left( r\right) +B\left(
r\right) .$ Suppose the HP filter is iterated $m$ times with $\lambda =\mu
n^{4}$ and $\mu $ fixed. If $ 1/m + m/n  \rightarrow 0$ as $n \to \infty$, then 
\begin{equation}
 n^{-1/2}   \widehat{f}_{\left\lfloor nr\right\rfloor }^{\left( m\right) }  
 \rightsquigarrow g^{ \mathrm{bHP} }\left( r\right) +B\left( r\right) :=\left\{ 
\begin{array}{cc}
g\left( r\right) +B\left( r\right),  & r \neq  r_{0} \\ 
0.5 \left\{ g\left( r_{0}^-\right) +g\left( r_{0}^+\right) \right\}
+B\left( r_{0}\right),  & r=r_{0}%
\end{array}%
\right. .  \label{tt78}
\end{equation}%
for each $r\in \left[ 0,1\right]$ and $r_0 \in (0,1).$
\end{thm} 

Compared with Theorem \ref{thm:drift}, this result imposes the additional condition that 
$ m / n \rightarrow 0$ as $m,n \to
\infty$. The extra condition is useful in the proof in deriving the limit behavior of the boosted filter around the break point
$r=r_0$. When $r \approx r_0$, the HP filter and boosted filter both smooth the time series trajectory of $x_t$ using
observations on either side of the break point. Like the Fourier series approximation of a piecewise
continuous function (and the Gibbs phenomenon), the HP filter and the boosted filter do not converge to the true value, $B_g(r_0)$, of the limit function at the break point
$r=r_0$. PJ show that when $\lambda=\mu n^4$ the HP filter converges to a smoothed version of the limit process $B_g(r)$ for all $r
\in (0,1)$. The above result shows that for the same tuning parameter setting of $\lambda$ the boosted filter provides a substantial
enhancement by consistently estimating $B_g(r)$ for all $r \ne r_0$ in the limit as $m,n \to \infty$ when $
m/n \to 0$. An implication of this result is that the bHP filter provides a consistent estimate of the break point $r_0$ in an
arbitrary polynomial trend. By consistently estimating $B_g(r)=g(r)+B(r)$ for all $r \ne r_0$ the boosted filter effectively reveals
the break point $r_0$ by virtue of the fact that the deterministic limit function $g(r)$ has a finite right limit to the value
$g(r_0^+)=g(r_0)$ on the right and has a finite left limit to a value $g(r_{0}^{-}) \ne g(r_0)$. The asymptotic form of
$ n^{-1/2}  \widehat{f}_{\left\lfloor nr\right\rfloor }^{\left( m\right)} $ 
captures this same behavior, showing that at the break point $r=r_0$ the limit of the bHP filter is the simple average of the
left and right limits, viz., $  0.5 \left\{ g\left( r_{0}^-\right) +g\left( r_{0}^+\right) \right\}$, thereby mimicking the
behavior of the Fourier series representation of the function $g(r)$ at the break point $r_0$. 

Theorem \ref{thm:break} is proved for a polynomial of arbitrary finite order with a single break point. The proof of this theorem reveals that under the same conditions the result may be extended to any  piecewise continuous polynomial function with a finite number of break points. In effect, the extended result shows that the bHP filter can consistently estimate a stochastic trend together with a breaking polynomial drift that has multiple structural breaks. Consistency applies for all points in the domain with the exception of the break points themselves. But in the same manner as Theorem \ref{thm:break} the fact that consistency holds almost everywhere with exceptions at the break points ensures that the bHP filter delivers consistent estimates of the break points themselves. These results hold in the presence of discrete break points. The present asymptotic development does not provide for local break point departures with breaks that decay with the sample size. The analysis of the asymptotic properties of the bHP filter in such cases is left as a topic of future research.

\subsection{Numerical Illustrations with the Boosted HP Filter}\label{subsec:num-illu}

It is common for modern machine learning methods --- such as boosting, random forest
and artificial neural network modeling --- to have 
multiple tuning parameters. The bHP filter has been formulated with two tuning parameters, one primary ($\lambda$) and one secondary ($m$). The near-universal choice for the primary smoothing parameter is $\lambda=1600$ for quarterly data and Theorems \ref{thm:Brownian} and \ref{thm:drift} show that with this choice, assuming that $\lambda=\mu n^4=1600$, the boosted filter can successfully consistently estimate and extract both stochastic and deterministic trends. As will be shown later in Section \ref{subsec:Stopping-Criterion}, the bHP filter is robust to considerable variation in the initial setting $\lambda=\mu n^4=1600$. There is therefore good reason to continue using the standard value $\lambda=1600$ with quarterly data and to focus attention on a suitable choice for the secondary parameter $m$. This approach matches \citet{buhlmann2003boosting}'s recommendation of using a relatively large primary parameter for smoothing and designating the boosting parameter as the sole tuning parameter. This regularization method of terminating the algorithm after several iterations is called \emph{early stopping} in statistical learning theory.

As the results of the last section show, the asymptotic effect of increasing $m$ is similar to reducing the
value of $\lambda$ in the simple HP filter, as both approaches can lead to consistent estimation of stochastic trends. In particular, PJ's results imply that consistent trend estimation is restored if smaller values of $\lambda$ are used so that $\lambda/n^{4}$ shrinks to $0$ fast enough as $n\to\infty$. But implementing such a scheme would require a grid system $\left(\lambda^{\left(1\right)},\lambda^{\left(2\right)},\ldots,\lambda^{\left(m\right)}\right)$
to be specified and the performance of the HP filter over these choices to be monitored and evaluated by some other criterion. Such a regularization scheme itself is an iterative procedure that is conceptually no more appealing than early stopping and is difficult to implement absent suitable criteria for the selection process and supporting asymptotic theory. Instead, use of HP filter settings such as $\lambda=1600$ for initialization and iteration in the bHP algorithm enable practitioners to employ standard software, and the resulting robustness of the bHP filter to the initial setting of $\lambda$ 
(See Appendix \ref{simu_lambda_robust}) gives reassurance that the procedure is not reliant on arbitrary tuning parameter choices.   

Macroeconomic time series are now available internationally in great abundance
and computation is therefore a relevant consideration in all such big data applications when machine learning methods are employed \citep{aruoba2010globalization,
aruoba2010real}. A key computational advantage in the use of a bHP filter
and early stopping procedure is its lower computational complexity and higher numerical stability in comparison to choosing $\lambda$ values on a grid system for the simple HP filter \citep{raskutti2014early, reiss2017optimal}. Early stopping
computes only once the inverse matrix operator $S=S\left(\lambda\right)=\left(I_{n}+\lambda DD'\right)^{-1}$
for a given $\lambda$. Using the same matrix $S(\lambda)$ stored in computer memory, successive iterations in the bHP filter involve simple matrix-scalar multiplication 
 $(I_n - S(\lambda)  ) \hat{c}^{(m-1)}$, which amounts to $2n^2 - n$ linear operations. In contrast,
searching for a suitable $\lambda$ on a grid system involves inverting an $n\times n$ matrix to obtain $S\left(\lambda\right)$ or carrying out QR decomposition\footnote{
	Instead of directly computing the matrix inverse 
	it is common to use a QR decomposition of the $(2n-2)\times n$ matrix 
	$\left[I_{n}\ \brokenvert\ \lambda D \right]'$ to reduce computation cost.
	However, even for such a QR decomposition the leading term of the 
	number of linear operations is $ (10/3) n^3$ by Householder transformation.  }
   for every value of $\lambda$ on the grid, compared to which the computational cost of
 matrix-scalar multiplication is negligible. 

\begin{figure}[htpb]
	\centering
		\includegraphics[scale=.45]{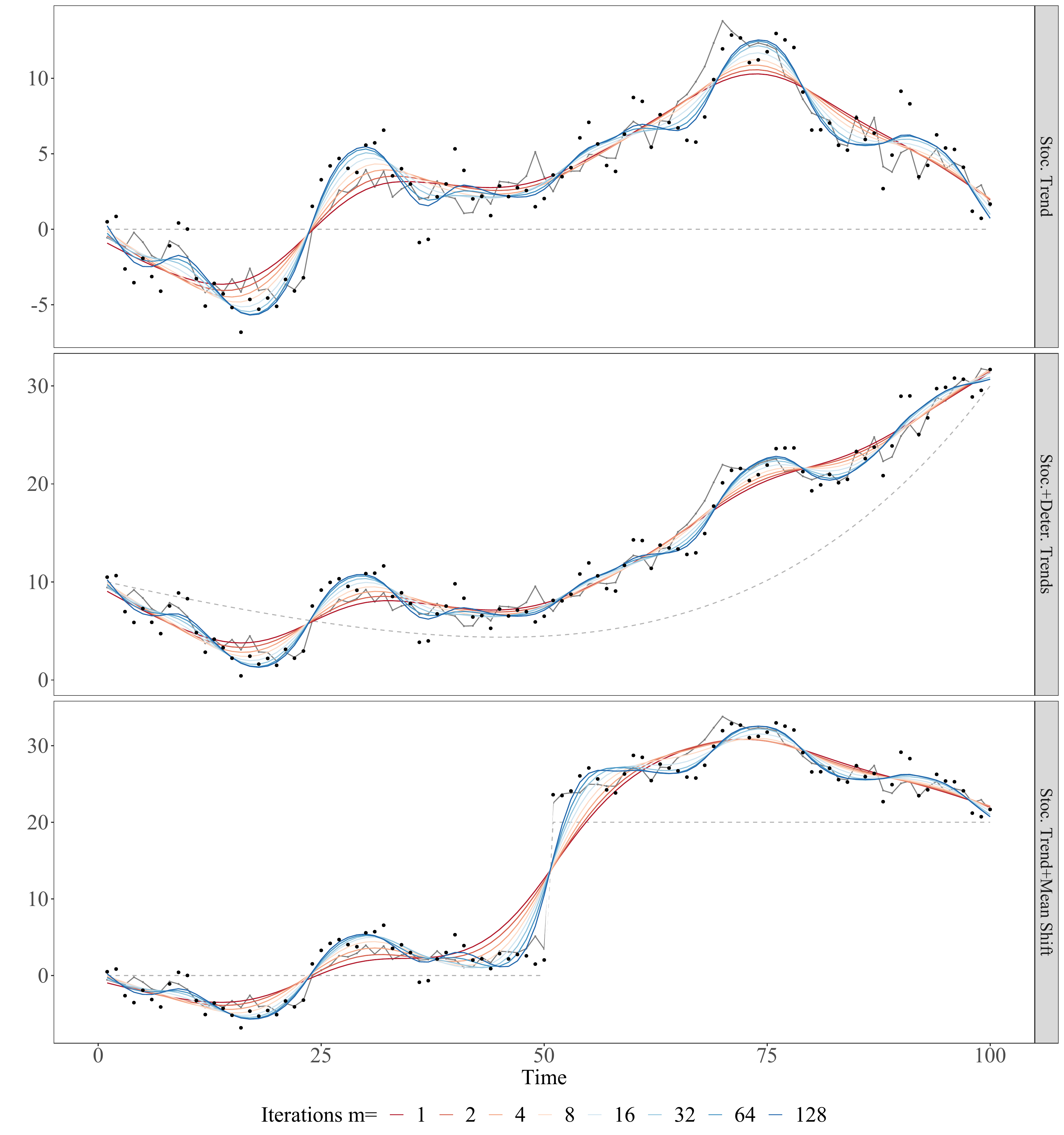}
	\caption{In each panel, observations $x_t$ are shown in black dots, the underlying trend process $\tilde{x}_t $ by the solid grey lines, the deterministic trend $g_n(t)$ by the dashed grey lines, and the fitted trend lines obtained by the HP filter (red line) and bHP filters (colored shades of orange progressing to shades of blue).
	A sequential decomposition of the middle panel of the figure into the component graphics is shown in Figure \ref{fig:fig1-decomp}, highlighting the trend capture performance of the HP and bHP filters in comparison to that of an AR(4) autoregression.}\label{fig1}
\end{figure}

We discuss various stopping rules for determining the number $m$ of boosting iterations in Section \ref{subsec:Stopping-Criterion} below. Before doing so, we conduct a numerical exercise to observe the effects of repeated
fitting in three prototypical cases involving a stochastic trend, and a stochastic trend with a drift and a mean break. 
Let $u_{t}^{(z)}\sim \mathrm{iid\ } N(0,1)$ and 
$u_{t}^{(e)}\sim \mathrm{iid\ } N(0,1) $ be independent innovation sequences, and $g_n(t)$ be a deterministic sequence.
Define
\begin{eqnarray} \label{eq:simulate_x0}
z_t &= & z_{t-1} + u_{t}^{(z)}, \nonumber \\
e_t & = &  0.5e_{t-1} + u_{t}^{(e)} + u_{t-1}^{(e)}, \nonumber \\
\tilde{x}_t & = & g_n(t) + z_t, \nonumber \\
x_t & = & \tilde{x}_t + e_t,
\end{eqnarray}
where $(z_t)$ is a random walk, $(e_t)$ is an ARMA(1,1) stationary process,
$(\tilde{x}_t)$ is a trend consisting of a non-random drift component $g_n (t)$ and the stochastic trend component $z_t$, and $(x_t)$ is the observed time series, which allows for stationary deviations or measurement error in observations of $\tilde{x}_t$.

Given the same realized stochastic trend (measured with error) $x_t^0 = z_{t} + e_{t} $ of length $n=100$, we generate the observations  $x_t = g_n (t) + x_t^0  $ shown in the panels 
of Figure \ref{fig1} by varying the deterministic component as follows to accommodate two prototypical trends:\footnote{The third prototypical trend is a simple stochastic trend with no deterministic component and results for this case are given in Table \ref{tab:fig1_MSE} below.}
a 4th order polynomial trend
$ g_n(t) = 10^{-3} \cdot (n-t)^2 + 3 \times 10^{-7} \cdot t^4 $ in the 
upper panel, and a mean shift $g_n(t) = 20 \cdot 1\{t \ge 0.5n +1 \} $
in the lower panel. 
The observations $x_t$ are represented by 
the black scattered dots, the trend $\tilde{x}_t$ by the solid grey line, and the deterministic trend $g_n(t)$
by the dashed grey line.    
The HP filter with $\lambda=1600$ is used to extract the trend and is shown as the red curve  ($m=1$) in the figure. The other curves are the fitted trends obtained by iterating
the HP filter $m$ times (the $m$ values are given in the figure legend) according to the boosted filter.

The upper panel reveals that the repeated fitting tracks the random wandering behavior of the random walk uniformly better than the HP filter.
The bHP filter gives a superior fit to the true trend (represented by the solid grey line in Figure \ref{fig1}) with a smaller $L_2$ distance to the grey line than the HP filter ($m=1$) for all values of  $2 \le m \le 128$.  
The curves are insensitive to the number of iterations once $m$ becomes large. This phenomenon is known as boosting's `resistance to overfitting'.\footnote{Further evidence of resistance to overfitting is given later by Figure \ref{fig:bias-variance} in the simulation study.}
It is corroborated analytically in the proof of Theorem \ref{thm:Brownian} where it is shown that as $m$ becomes large for given $n$ the boosted filter stabilizes and approximates a finite number of terms in the orthonormal series representation of the limiting trend process. This finite term orthonormal representation is a smooth approximation to the true limit process, explaining the smooth form of the boosted HP filter. Further analysis of this example by  decomposition of the component graphics is given in Figure \ref{fig:fig1-decomp}, which includes a comparison of the trend capture performance of the bHP filter with that of an AR(4) autoregression

In the lower panel, it is evident from the plots that boosting the filter goes a long way towards enhancing performance in the region of the structural break by eliminating a substantial amount of the transition smoothing in the HP filter around the break point. For large $m \approx n$, the boosted filter trajectory is strongly suggestive of a structural break around observation $t=50$ with a mid-point estimate of the value at the break point, corroborating the implications of Theorem  \ref{thm:break}.

\subsection{Stopping Criterion\label{subsec:Stopping-Criterion}}

The residual component after trend extraction by smoothing methods such as the HP filter has long been a building
block for applied macroeconomists in studying business cycles and the interactions
between macroeconomic aggregates and indicators. By definition, the cyclical component
is a time series that exhibits no long run trending behavior, so that its spectrum has no unit root or deterministic trend asymptote at the zero frequency. In practice this criterion can be implemented by the elimination of all low frequency elements, an approach that band-pass filter methods use directly in filtering the data \citep{baxter1999measuring, christiano2003band,corbae2006extracting}. 


A natural and somewhat analogous approach in the present context is to refilter the data until there is no evidence of a non-stationary zero frequency asymptote. This can be conveniently achieved by monitoring the outcome of unit root tests on the residual series $\widehat{c}_{t}^{\left(m\right)}$. Standard procedures for unit root testing such as the augmented Dickey-Fuller (ADF) or Phillips-Perron \citep{phillips1988testing} tests can be used and the boosting iterations can be continued until the test statistic is smaller than a specified $p$-value, such as 0.05 or 0.01. Such test-based stopping criteria are easy to implement and are well-tailored to existing applied macroeconomic practice, echoing \citet{kozbur}'s test-based stopping criterion for forward selection, and \citet{diebold2000unit}'s test-based
forecasting approach. Relatedly, HP (1997) use unit root tests to assist in determining an appropriate setting for the primary smoothing parameter $\lambda$.  
In our simulations and empirical examples, we will use the ADF test conducted with significance level 0.05 to illustrate implementation of this approach. The boosted HP filter that results from this ADF test-based selection will be denoted bHP-ADF.

Information criteria offer an alternative approach to a stopping criterion. These criteria are routinely employed in statistics to achieve bias-variance trade-offs and to prevent overfitting in modeling and forecasting. We therefore consider the following BIC 
for the selection of the stopping time for $m$
\begin{equation} \label{BBIC}
IC\left(m\right)=\frac{\widehat{c}^{\left(m\right)\prime}\widehat{c}^{\left(m\right)}}{\widehat{c}^{\mathrm{HP}\prime}\widehat{c}^{\mathrm{HP}}}
+\log\left(n\right) \frac{ \mathrm{tr}\left(B_{m}\right)
}{\mathrm{tr}\left(I_{n}-S\right)}.
\end{equation}

Similar to BIC, this criterion penalizes fit by adding $\log(n)$ times a term that quantifies the relative weight of the $m$ additional iterations that are involved in the boosted filter. The first term of (\ref{BBIC}) measures the residual sum of squares fit of the boosted HP filter, $\widehat{c}^{\left(m\right)\prime}\widehat{c}^{\left(m\right)}$,
relative to the HP filter itself, $\widehat{c}^{\mathrm{HP}\prime}\widehat{c}^{\mathrm{HP}}$. The penalty term involves the usual $\log(n)$ scale factor multiplied by a ratio that measures the effective degrees of freedom of the boosted filter after $m$ iterations to the effective degrees of freedom of the HP filter. To interpret this ratio, it is useful to think of the linear operator $\left(I_{n}-S\right)$ that produces the residual cyclical component $\widehat{c}^{\mathrm{HP}}=\left(I_{n}-S\right)x$ of the HP filter as analogous to a linear regression projector or hat matrix, so that $\mathrm{tr}\left(I_{n}-S\right)$ is analogous to the degrees of freedom in the sample after projection. The operator $B_{m}=I_n-(I-S(\lambda))^m$ is the $m$-fold operator corresponding to the boosted filter and the quantity $\mathrm{tr}(B_m)$ may therefore be interpreted in a similar way as the effective degrees of freedom after successive fitting by the boosted HP. This interpretation corresponds to usage in the machine learning literature  \citep{tutz2006generalized}. It is convenient from now on to refer to the criterion $IC\left(m\right)$ simply as BIC and to the bHP filter that results from this selection rule as bHP-BIC.
.

\begin{figure}[htbp]
	\centering
	\includegraphics[width=12cm,height=6cm]{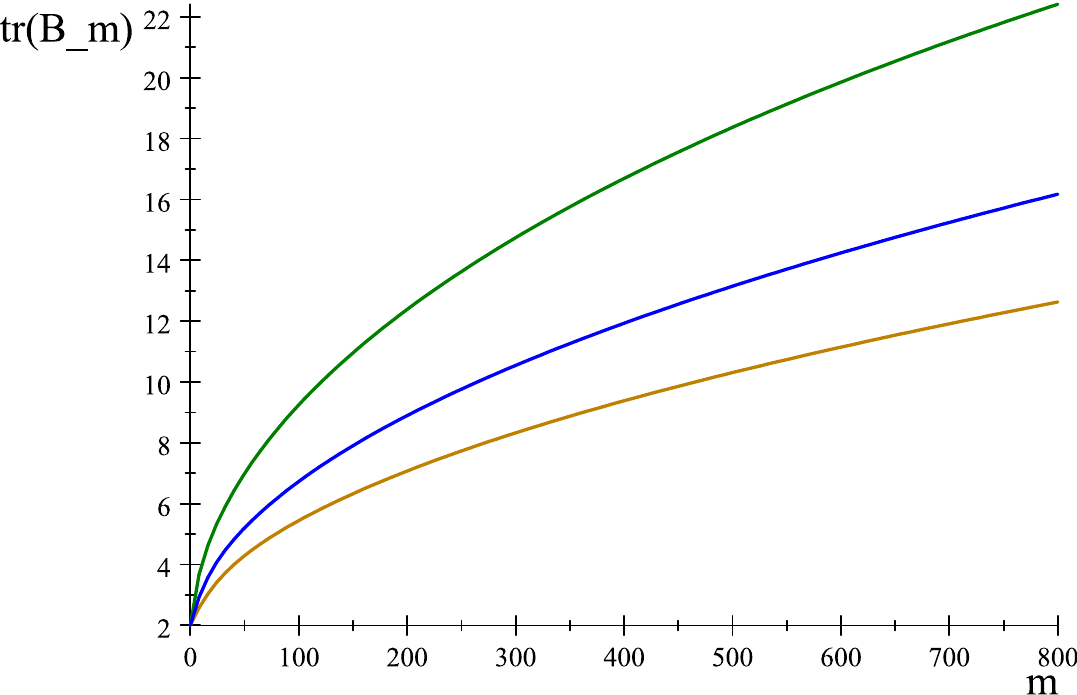}
	\caption{\label{fig:Bm_grf} Plots of $\mathrm{tr}(B_m)=n-\sum_{k=1}^{n-2}
	 (\lambda \delta_{k}^2)^{m} \big/ (1+\lambda \delta_{k}^2)^m 
	 \left\{1+o\left(1\right)\right\}$ for $n=50,75,100$ (green, blue, sienna).}
\end{figure}

It is shown in the Appendix that $\mathrm{tr}(B_m)$ can be asymptotically approximated by the following simple analytic expression as $n \to \infty$ 
	\begin{equation} \label{tt26}
\mathrm{tr}(B_m) =\mathrm{tr}(I_n-(I_n-S(\lambda))^m)=n-\sum_{k=1}^{n-2}\frac{(\lambda \delta_{k}^2)^{m}}{(1+\lambda \delta_{k}^2)^m}\left\{1+o\left(1\right)\right\}, \quad \delta_{k}^2=4\left(1-\cos{\frac{k\pi}{n-1}}\right)^2.
	\end{equation}
Figure \ref{fig:Bm_grf} graphs $\mathrm{tr}(B_m)$ against this approximation as a function of $m$, showing how the penalty term coefficient $\mathrm{tr}(B_m)$ increases monotonically and nonlinearly with $m$ for any given value of the sample size $n$. Differentiating (\ref{tt26}) with respect to $m$ gives
\begin{equation*}
\frac{\partial \mathrm{tr} \left( B_{m} \right) }{\partial m}=\log \left( 1+%
\frac{1}{\lambda \delta _{k}^{2}}\right) \sum_{k=1}^{n-2}\left( \frac{1}{%
	1+1/\left( \lambda \delta _{k}^{2}\right) }\right) ^{m} \left\{1+o\left(1\right)\right\} >0,
\end{equation*}%
so that $\mathrm{tr}(B_{m} ) $ is increasing in $m$ with
decreasing derivative as $m$ increases, as is evident in Figure \ref{fig:Bm_grf}.
 Moreover, as is clear from formula (\ref{tt26}) and the graph, the penalty coefficient $\mathrm{tr}(B_m) \to 2$ as $\lambda \delta_{k}^2 \to \infty$, which happens for all $k=\lfloor nr\rfloor $ with $r\in (0,1]$ and $m$ fixed when $\lambda = \mu n^4$. Thus, the impact of the penalty on the choice of $m$ is attenuated as $n \to \infty$.   

With the implementation of one of these stopping rules, the boosted HP fitting algorithm is automated and data-determined, making it ready for practical use like other non-parametric procedures with data-determined bandwidth selectors. 
The following sections assess the performance of these stopping rules in 
simulated experiments and provide three real data applications.

\begin{table}[htbp]
\caption{ MSE of the Estimated Trend in Figure \ref{fig1} } \label{tab:fig1_MSE}
	\begin{center}
		\begin{tabular}{lcccc}
		\hline 
		 & HP & ADF & BIC & AR(4)\tabularnewline
		\hline 
		Stoc. Trend & 2.274 & 1.561 & 1.336 & 3.915\tabularnewline
		Stoc. + Deter. Trends & 2.344 & 1.585 & 1.336 & 4.046\tabularnewline
		Stoc. Trend + Mean Shift & 8.600 & 6.495 & 4.212 & 8.687\tabularnewline
		\hline 
		\end{tabular}
	\par\end{center}
\small{Note: Use of the fitted AR(4) autoregression follows \citet{hamilton2017you}'s recommended approach, 
namely $\hat{f}_t = \hat{\beta}_0 + \sum_{k=1}^4 \hat{\beta}_k x_{t-k} $, with coefficients $(\hat{\beta}_k)_{k=0}^4$ 
obtained from an AR(4) regression with fitted intercept.
}
\end{table}

As a numerical illustration for the data displayed in Figure \ref{fig1}, 
the deviation of the estimated trend $\hat{f}_t$ from the underlying trend $\tilde{x}_t$ is measured in terms of the mean squared error (MSE) calculated as 
$\mathbb{M}_n  = (n-8)^{-1} \sum_{t=5}^{n-4} (\hat{f}_t - \tilde{x}_t)^2 $. The end points are trimmed in $\mathbb{M}_n$ to accommodate start-up in the AR(4) process and end points in the HP filter. Calculating the MSE using $\mathbb{M}_n$ without trimming did not materially affect the results.
In this particular experiment, both ADF and BIC significantly reduce the MSE of the HP filter 
while AR(4) evidently does not fit the underlying trend well. For further analysis and more detailed graphical displays see Figure 
\ref{fig:fig1-decomp} in Appendix \ref{subsec:fig1-decomp}.

\section{Simulations}\label{sec:simulations}

We conduct simulation exercises with eight data generating processes to observe 
the finite sample performance of the bHP filter 
in practice when the trend process involves both stochastic and deterministic elements.
Similar to the models used in Section \ref{subsec:num-illu}, in DGPs 1 and 2 below  we add a stationary component to the deterministic trends. With the
addition of this component to the data iterating an excessive number of times in the boosting process in finite samples can potentially lead to overfitting. 
The experimental design therefore reveals the bias-variance tradeoff that occurs in such cases and 
the effectiveness of the two stopping criteria in preventing saturation fitting in practical applications of the boosted filter.

DGPs 3-8 focus on fitting a trend
under alternative plausible generating mechanisms that include a pure random walk, a structural break, a sinusoidal trend, 
a cosine cycle, and various combinations of these components. The experiments also provide performance comparisons of the boosted filter approach to trend extraction with the autoregressive model estimation approach advocated in \citet{hamilton2017you}.

\subsection{The Bias-Variance Tradeoff in Boosting}\label{subsec:bias-variance}

According to Theorem \ref{thm:drift}, the 
boosted HP filter can asymptotically remove any finite-order polynomial drift,
whereas the HP filter can only handle a polynomial drift up to the 3rd order. 
Higher order time polynomials are known to be useful in modeling
the nonlinear growth of both macroeconomic and microeconomic time series and as sieve approximations to more general nonlinear trend functions (\citet{baek2015testing}; \citet{cho2018sequentially}). We are therefore interested in the capability of the boosting mechanism to enable the HP filter to capture these general deterministic trend elements in addition to stochastic trends. 

The following two experimental designs involve finite degree polynomial drift functions, $g_n (t)$, to accompany the stochastic trend generating mechanism as in (\ref{eq:simulate_x0}). The specification illustrates the potential gains that can be obtained in trend determination by boosting the HP filter even in the presence of simple deterministic drifts.

\begin{figure}[htbp] 
	\centering
	\subfloat[DGP 1: stochastic trend with 3rd-order polynomial drift]{\begin{centering}
			\includegraphics[scale=0.4]{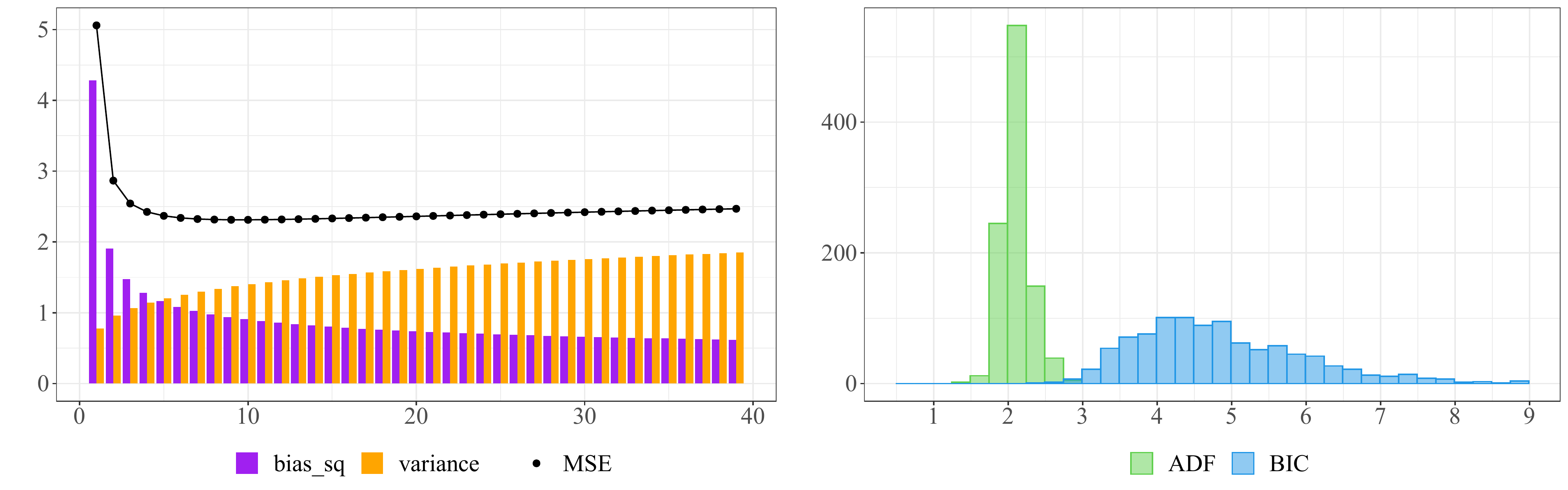}
			\par\end{centering}
		
	}\bigskip{} \bigskip{}
	\subfloat[DGP 2: stochastic trend with 4th-order polynomial drift]{\begin{centering}
			\includegraphics[scale=0.4]{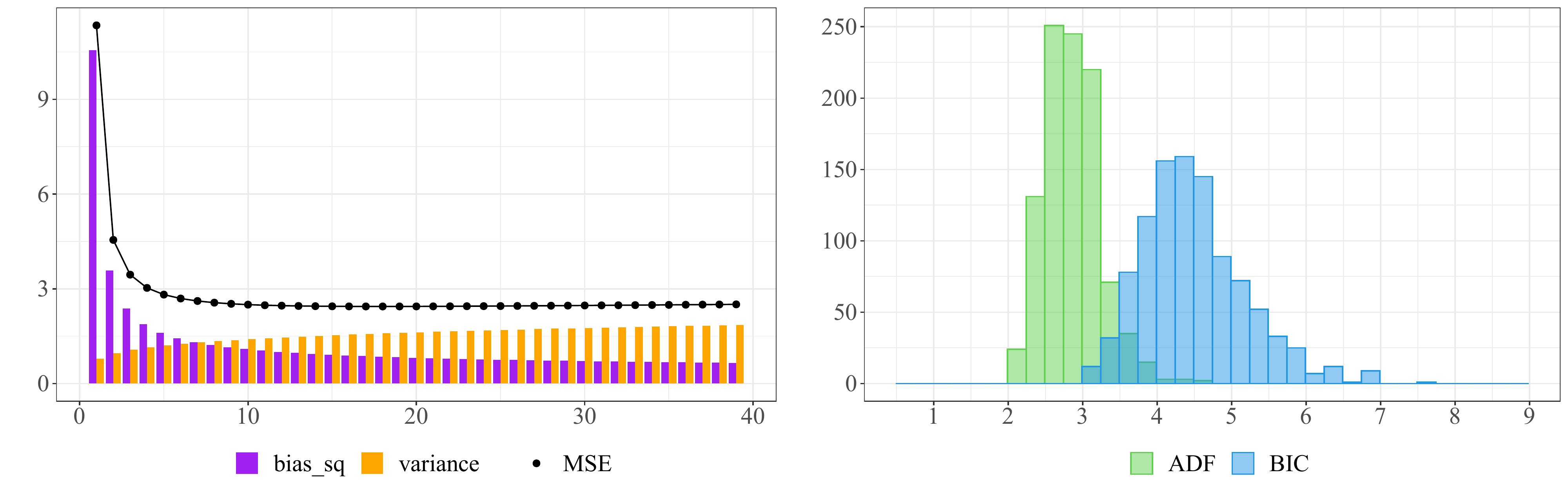}
			\par\end{centering}		
	}	
	\caption{Bias-variance tradeoffs in trend estimation (left panel) and distributions of stopping times calculated by the ADF (green) and BIC (blue) criteria (right panel) for the HP ($m=1$) and boosted HP filters ($m>1$). The number of iterations ($m$) is shown on the horizontal axis in each figure.}	
	\label{fig:bias-variance}
\end{figure}

\begin{description}
	\item[DGP 1] Set the sample size $n=100$ (25 years in quarterly data),
	and the deterministic trend $ g_n (t)= 0.0005 t^3 $. 
	Step 1: Generate a stochastic trend plus drift process $ \tilde{x}_t = g_n(t) + z_t$ as  defined in (\ref{eq:simulate_x0}).
	 Step 2: Given the realized trend 
	$ \tilde{x}_t $ in Step 1, simulate the stationary random component $e_t$,
	also defined in (\ref{eq:simulate_x0}), to produce the measured observation $x_t = \tilde{x}_t + e_t$.
	Step 3: Repeat Step 2 for 50 times (calling this the \emph{inner loop}) in order to compute the bias and variance of the filters given the trend $\tilde{x}_t$. Step 4: Repeat Steps 1-3 for 1000 replications (we call this the \emph{outer loop}) to average the bias and variance over the various realizations of the trend process.  
	
	\item[DGP 2] This experimental design is identical to DGP 1 except for the fact that  
	the deterministic trend component is generated by a 4th order polynomial $   g_n (t) = 5 \times  10^{-6} \cdot t^4$ rather than a 3rd order polynomial.
\end{description}

Given a realized $\tilde{x}_t$,  in the inner loop of 50 replications  we compute for  
fixed $t$ and $m$  the empirical versions of the bias $\mathbb{B}^{(m)}  =  E [ \hat{f}_{t}^{(m)}  ] - \tilde{x}_t $  and the variance  $\mathrm{var}  [ \hat{f}_{t}^{(m)} ] $.
Then over the realized trend trajectory $ \tilde{x} = (\tilde{x}_t)_{t=1}^n $  we calculate the squared-bias  
$\mathbb{Q} ^{(m)} =  n^{-1} \sum_{t=1}^{n} \left[ \mathbb{B}^{(m)} \right] ^2 $ and the variance
$ \mathbb{V}^{(m)} =  n^{-1}  \sum_{t=1}^{n} \mathrm{var}  [ \hat{f}_{t}^{(m)} ] $.
Finally,  in the outer loop for each $m$ we average over the 1000 replications of
$\mathbb{Q} ^{(m)} $ and $\mathbb{V} ^{(m)} $. The 
squared bias and the variance are displayed 
in the left subgraph of Figure \ref{fig:bias-variance} for each $m=1,\ldots,40$.
The black dotted line above the bars sums the underlying two bars and gives the mean squared error (MSE).

In both DGPs, similar patterns of bias-variance tradeoff are evident.
Initiating the iteration process from the HP filter ($m=1$), we observe 
a sizable drop in the squared bias and MSE in the first few iterations of boosting.
The squared bias continues to decrease as the iterations proceed, whereas variance slowly increases. 
After it reaches a minimum, the MSE remains insensitive as a rather flat curve as 
$m$ continues to grow, which reflects the boosting saturation that occurs in finite samples. 

To evaluate the effect of the data-driven stopping criteria,  
we save the number of iterations 
in each instance and take the sample average in the inner loop. 
The outer loops provide 1000 such average stopping times and histograms of these average stopping times are shown in the 
right subgraph of Figure \ref{fig:bias-variance}. 
In DGP 1 the 3rd-order polynomial trend can be asymptotically 
removed by fitting the HP filter only once.
Setting the test size to be 0.05, we find that only 25.9 percent of the average ADF stopping times are smaller than two, indicating that  
some remnants of the stochastic trend appear in the residual cyclical component with nontrivial probability. 
The BIC criterion requires at least two iterations in all replications and often three or four fittings. The effect of these fittings is evident in the large reductions in the squared bias as observed in the left panel.

The stopping time data is more intriguing in DGP 2 where we replace the cubic trend of DGP 1 by a 4th-order time polynomial. 
According to the limit theory, without the use of boosting the HP filter cannot asymptotically
remove such a higher order polynomial trend.
This asymptotic theory is clearly supported in the finite 
sample computations. The bottom-right subgraph of Figure \ref{fig:bias-variance} shows that the average ADF stopping criterion 
is at least two and the BIC criterion requires at least three fittings and often as many as four or five.

\begin{figure}[htbp]
	\centering
	\includegraphics[width=12cm,height=6cm]{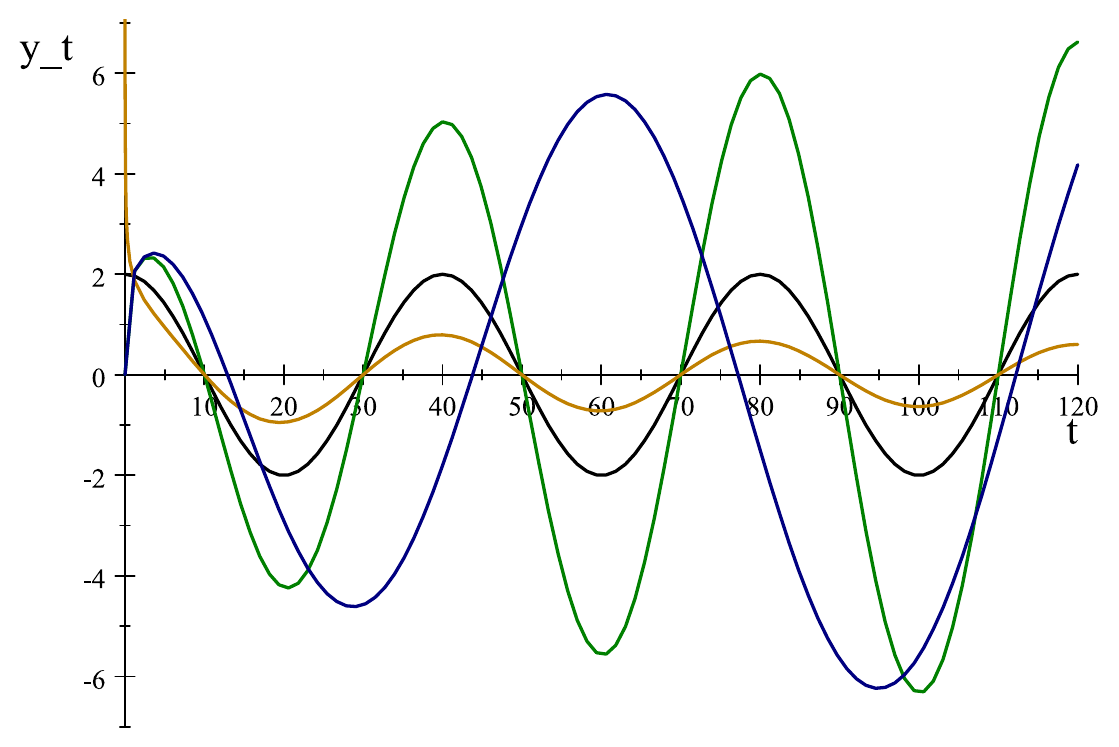}
	\caption{\label{fig:Sinusoidal_grf} Plots of various sinusoidal trend functions $y_t$: trigonometric $2 \cos(0.05 \pi t)$ (black); trending trigonometric $2t^{0.25} \cos(0.05 \pi t)$ (green); evaporating trigonometric $2t^{-0.25} \cos(0.05 \pi t)$ (sienna); and evolving duration trigonometric $2t^{0.25}\cos(0.05 \pi t^{0.90})$ (blue).}		
\end{figure}

\subsection{Goodness of Trend Determination}\label{subsec:goodness}

In the previous subsection, DGPs 1 and 2 were designed as mechanisms to produce a polynomial trend plus a stochastic trend. 
Whatever the precise nature of the trend, conditional on its realized form computations of the bias and variance of the bHP filter reveal the tradeoff that occurs in these measures of fit as the number $m$ of iterations in the boosted filter rises. As the results with DGPs 1 and 2 show, bias typically falls quickly as $m$ begins to rise, demonstrating immediate gains from boosting. But with increasing $m$ bias reductions diminish and variance rises to a point where mean squared error stabilizes. Thus, in finite samples there are limits to what can be accomplished by boosting just as in any nonparametric procedure.  

Many empirical studies model time series data in terms of integrated or near-integrated processes augmented with various complementary mechanisms such as polynomial drifts, similar drifts with breaks, sinusoidal trends, or trends induced by time varying coefficients, all of which are intended to improve harmony with the observed data but with no certainty concerning the true specification of its generating mechanism. This section considers the performance of the bHP filter in such cases and compares the performance of the bHP filter with  \citet{hamilton2017you}'s alternative recommendation of the use of autoregressive (AR) modeling with a small number of lags, typically an AR(4) which is expected to be well suited to quarterly data applications. 

The following six models are used to illustrate the performance characteristics of these approaches.  The pure random walk case is used as a baseline in DGP 3 and DGPs 4-6 couple this integrated process with various other complementary trend specifications that progressively enhance the complexity of the generating mechanism. DGPs 7-8 employ a specific fixed-period cyclical component. The notation follows the framework of (\ref{eq:simulate_x0}) and sample size is set to $n=100$.

\begin{description}

	\item[DGP 3]   The observed time series is $x_t^{(3)} = z_t$,  a random walk with independent Gaussian increments.

	\item[DGP 4]  Real economic activity may involve long duration cycles that are time-dependent and evolve in a non-replicative manner, for example with varying magnitudes or cycle lengths. We use a deterministic sinusoidal trend of the form
	$g_n(t) = 5 t^{1/5} \cos( 0.05\pi t^{0.9} ) $
	to embody this type of complexity. Figure \ref{fig:Sinusoidal_grf} graphs the form of various expanding and decaying sinusoidal trends of this type. The observed time series is expressed in the form $x_t^{(4)} = g_n(t) + x_t^{(3)}$. 
	
	\item[DGP 5] This model serves as a simple prototype of GDP takeoff that can be used to represent a successful emerging economy growth trajectory. The model has a structural break in the middle of the sample and takes the form  
	 $$x_t^{(5)} = u^{(z)}_{t} \cdot \mathbf{1}\{t < 0.5n \} + 
		\big( t-0.5n + \sum_{s=0.5n}^t u^{(z)}_s  \big )\cdot \mathbf{1}\{t \geq 0.5n \} . $$
		The first half of the sample is a stationary sequence and the second half is an integrated process with a linear upward drift.  
	
	\item[DGP 6] This model is formed from the composition of the deterministic sinusoidal trend 	$g_n(t) = 5 t^{1/5} \cos( 0.05\pi t^{0.9} ) $ of DGP 4 with the structural break model in DGP 5 leading to the time series $x_t^{(6)} = g_n(t) + x_t^{(5)}$.


	\item[DGP 7] This model is the same as DGP 4 except that the deterministic component $g_n(t)$ is replaced by the periodic function $\check{g}_n(t) = \cos\left(\pi t/2\right)$ which repeats every 4 periods. 

	\item[DGP 8] In DGP 6 $g_n(t)$ is replaced by $ \check{g}_n(t) \cos\left(\pi t/2\right)$.

\end{description}

The goal in the simulation exercise is to determine the trend from data generated by these different mechanisms using the HP filter, the bHP filter, and the AR(4) regression technique of \citet{hamilton2017you}. In each replication the observed time series is filtered or regressed to obtain the corresponding fitted trend estimate $\hat{f}_t$. Deviation from the underlying trend $\tilde{x}_t$ is measured in terms of the MSE calculated 
using $\mathbb{M}_n$ as that in Table \ref{tab:fig1_MSE}.
The trend processes
$\tilde{x}_t$ are produced from the generating processes prescribed above so that
$\tilde{x}_t^{(3)} = x_t^{(3)}$, 
$\tilde{x}_t ^{(4)} = g_n(t) + \tilde{x}^{(3)}$, 
$\tilde{x}_t^{(5)} = ( t-0.5n + \sum_{s=0.5n}^t u^{(z)}_s  )\cdot \mathbf{1}\{t \geq 0.5n \}$, 
and  
$\tilde{x}_t ^{(6)} = g_n(t) + \tilde{x}^{(5)}$ for DGPs 4--6, respectively. In DGPs 7 and 8, the deterministic function $ \check{g}_n(t) = \cos\left(\pi t/2\right)$ is periodic and does not exhibit trending behavior when $n \gg 4$. We therefore take this function as a component of the cycle. So, the trends in DGP 7 and DGP 8 are simply 
$\tilde{x}_t^{(7)} = \tilde{x}_t^{(3)} $ and $\tilde{x}_t ^{(8)} = \tilde{x}^{(5)}$.

\medskip
\begin{table}[htbp] 
	\caption{MSE of Trend Estimation and Number of Iterations} \label{tab:MSE}

\centering{}%
\begin{tabular}{c|cccc}
	\hline 
	DGP & HP & ADF  & BIC & AR(4)\tabularnewline
	& \multicolumn{4}{c}{ MSE }\tabularnewline
	3 & 1.5982 & 1.5033 & 0.8540 & 0.9295\tabularnewline
	4 & 2.6204 & 1.4697 & 0.9943 & 1.1536\tabularnewline
	5 & 1.0719 & 0.9001 & 0.5787 & 1.0091\tabularnewline
	6 & 1.8795 & 0.8913 & 0.6329 & 1.2881\tabularnewline
	7 & 1.5983 & 1.5704	& 0.9845 & 1.4159\tabularnewline
	8 & 1.0721 & 0.8799	& 0.6569 & 1.4270\tabularnewline
	\hline 
	& \multicolumn{4}{c}{Average number of iterations}\tabularnewline
	3 & 1 & 1.23 & 9.48 & N.A. \tabularnewline
	4 & 1 & 2.10 & 5.73 & N.A. \tabularnewline
	5 & 1 & 1.54 & 5.33 & N.A. \tabularnewline
	6 & 1 & 2.32 & 4.91 & N.A. \tabularnewline
	7 & 1 & 1.42 & 5.43 & N.A. \tabularnewline
	8 & 1 & 3.14 & 3.41 & N.A. \tabularnewline
	\hline 
\end{tabular}

\end{table}

Table \ref{tab:MSE} reports the empirical average of $\mathbb{M}_n$ and the observed number of iterations in the bHP filter over 5000 replications. 
In DGP 3, where the time series is generated from a random walk, the fitted AR(4) is particularly well suited since the regression model includes the true generating mechanism. Unlike the AR(4) regression
which is based only on past information in forecasting the trend, the two-sided nature of the HP filter uses all sample information, including future observations to determine the current period trend value. There are notable differences in the results between the ADF selected and BIC selected stopping times for the iteration. These differences reveal the importance of iterating the filter. The BIC selector leads to a substantially lower MSE in trend determination from the boosted filter. The ADF selector tends to stop the iteration too early to achieve optimal improvement with an average number of iterations of 1.23, which is close to the HP filter itself (with $m=1$) and substantially lower than the average number of 9.49 iterations for the BIC selector. With the BIC selector the bHP filter provides a substantial reduction in MSE over the HP filter. The bHP-BIC filter also produces a smaller MSE to the underlying trend than the AR(4) regression, an interesting result given that the AR(4) regression model encompasses the simple random walk model DGP 3 and the bHP has none of these explicit features. 

The HP filter methods are all nonparametric in nature and, as the asymptotic theory suggests,  when the tuning parameters are chosen appropriately these methods can adapt to complex trend processes and generating mechanisms. The simulation evidence supports this theory. In particular, once a slowly moving smooth deterministic trend is added to the random walk in DGP 4, the differences in performance are magnified and the MSE of the AR(4) regression deteriorates more than the bHP-BIC filter. Interestingly, the presence of a deterministic trend triggers more iterations in the bHP-ADF filter and it reduces the MSE to 1.47 from the value 1.50 in DGP 3. 

Since the first half of the DGP 5 sample is a white noise for which the constant level trend function is easy to predict in a nonparametric method, the filter methods each obtain a smaller MSE than their counterparts in DGP 3. However, as a global parametric method, the AR(4) regression is inevitably misspecified when this structural break from an I(0) to an I(1) process is present in the observed series. In this case, the MSEs of the bHP-ADF and bHP-BIC filters are both substantially smaller than that of the AR(4). DGP 6 raises the level of trend complexity further by including an evolving sinusoidal trend. For this DGP, the boosted filter again provides much better trend determination. In fact, bHP-BIC has MSE less than half that of the AR(4). Comparison of the results for DGP 4 and DGP 6 
shows that bHP provides a very effective tool that adapts well to increasing complexity in the underlying trend mechanism. The HP filter, on the other hand, has MSE that is almost three times the size of that of the bHP-BIC filter. 

The results for DGPs 7 and 8 in Table 2 reveal that the MSE of trend estimation by bHP-BIC are substantially smaller than those of the AR(4) with a difference greater than $0.43$ in both cases. These results show that the bHP estimated trend correctly excludes the cosine function. The $0.43$ difference between the MSEs of bHP-BIC and AR(4) for DGP 7 is roughly the naive `sample variance' $s_w^2= n^{-1}  \sum_{t=1}^nw_t^2-(n^{-1} \sum_{t=1}^nw_t)^2 =0.50$ of the periodic function \{$w_t:=\cos\left(\pi t/2\right)$, $t=1,\ldots,n=100$\}. In effect, bHP-BIC treats the cosine function correctly as a cyclical component, which it is with sample size $n=100 \gg 4$, whereas AR(4) regression does not make this distinction leading to the larger MSEs in trend determination. For DGP 7, the average number of iterations for bHP-ADF is 1.42 giving MSE results close to those of the HP filter which shows that bHP-BIC, with an average of 5.43 iterations in this case, is more effective in refining HP. 

\section{Empirical Examples}\label{sec:emp-application}

We illustrate the effects of using the bHP filter in three real data examples.
The first example revisits empirical support for Okun's law across 20 OECD countries. The second explores business cycle behavior in a panel of 78 heterogeneous time series covering emerging and developed markets with various degree of persistence and volatility. 
The third studies the behavior of the filters in trend determination using US industrial production data over the past century. In this last application we use the HP and bHP filters as well as the AR(4) parametric approach. 

\subsection{Okun's Law}

\begin{figure}[htbp]
	\centering
	\includegraphics[width=14cm,height=19cm]{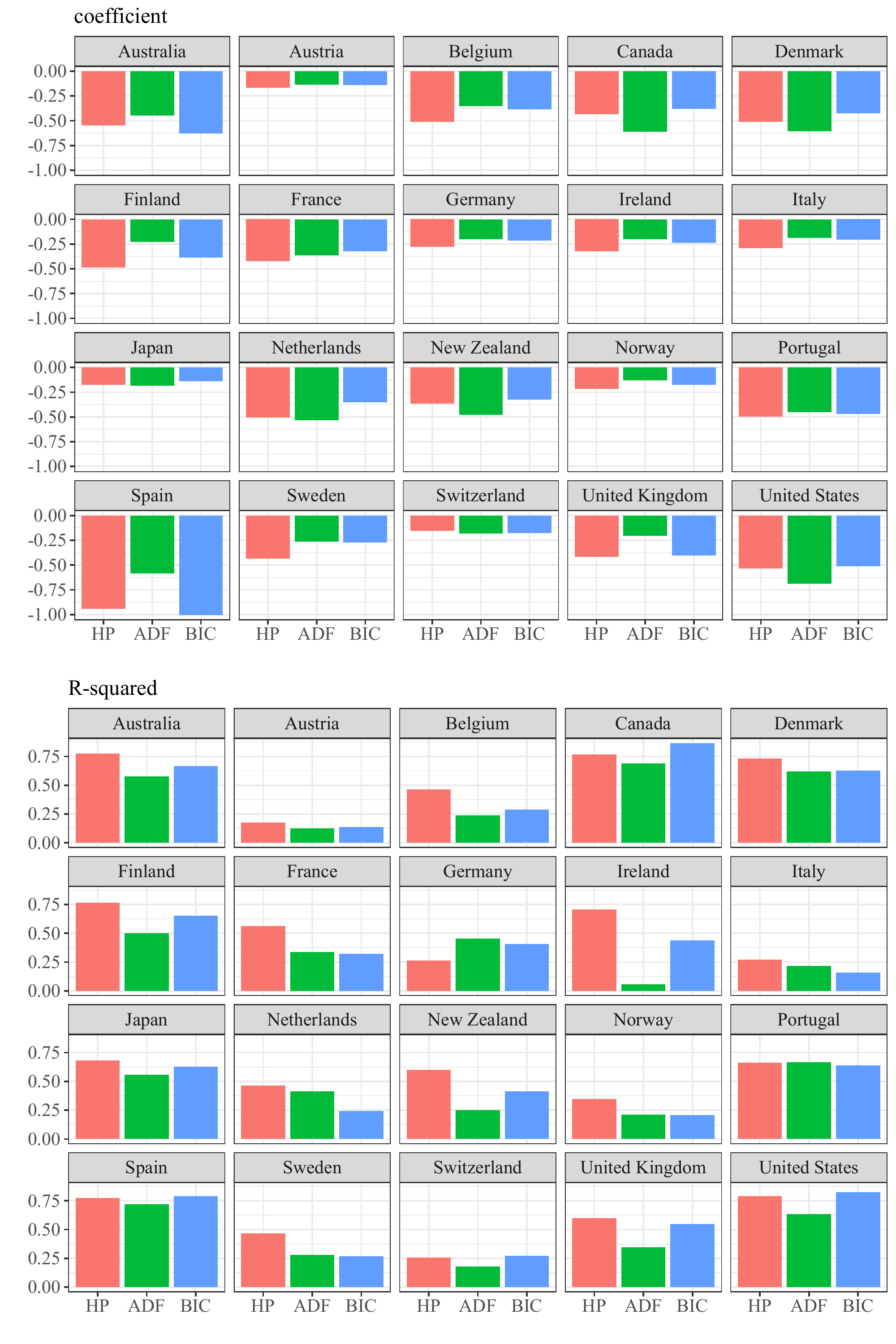}
	
	\caption{\label{fig:Okun_coef} Fitted OLS coefficients and $R^{2}$ statistics for equation (\ref{eq:Okun}) with data obtained by simple HP	and bHP filtering using ADF and BIC tuning parameter selection. Annual data over 1980 to 2016.}
\end{figure}

Okun's law \citep{okun1962potential} posits an empirical association between output and the unemployment
rate that has received wide attention among practicing economists and policy makers as well as academic economists and authors of undergraduate economics texts. For the United States, Okun's law is stated as relating a 1 percent increase in GDP (relative to potential GDP) to a 0.5 percent reduction in the unemployment
rate (relative to the natural rate of unemployment). Following the original formulation by Okun, \citet{ball2017okun} specify the empirical model in terms of the following empirical regression equation
\begin{equation}
U_{t}\text{\textminus}U_{t}^{*}=\beta\left(Y_{t}-Y_{t}^{*}\right)+\varepsilon_{t},\label{eq:Okun}
\end{equation}
where $U_{t}$ is the unemployment rate, $Y_{t}$ is the logarithm
of GDP, and $U_{t}^{*}$ and $Y_{t}^{*}$ are the natural rate of
unemployment and the potential GDP. The sign and the
magnitude of $\beta$ signify the direction and strength of the relationship.
In view of its potential policy implications, Okun's law has been extensively tested
over time and cross countries. Most recently, \citet{ball2017okun}
testify to its robustness in 20 advanced economies. These authors, as many others, estimate the
long-run levels of $U_{t}^{*}$ and $Y_{t}^{*}$ by means of the HP filter
under the primary parameter setting $\lambda=100$ for annual data. Equation (\ref{eq:Okun}) is therefore
a simple regression between two cyclical components produced by the HP filter. 

The primary motivation of using trend extraction techniques prior to the regression (\ref{eq:Okun}) is to focus on cyclical variates. A secondary motivation is to eliminate the possibility of spurious regression in the variables, which would distort inference \citep{granger1974spurious, phillips1986understanding} unless there is strong justification for residual stationarity and a cointegrating relationship between the variables. Use of the ADF test in the implementation of the boosted filter assists in addressing both these issues and rationalizing the regression. 

\begin{figure}[htbp]
\centering
		\includegraphics[scale=.9]{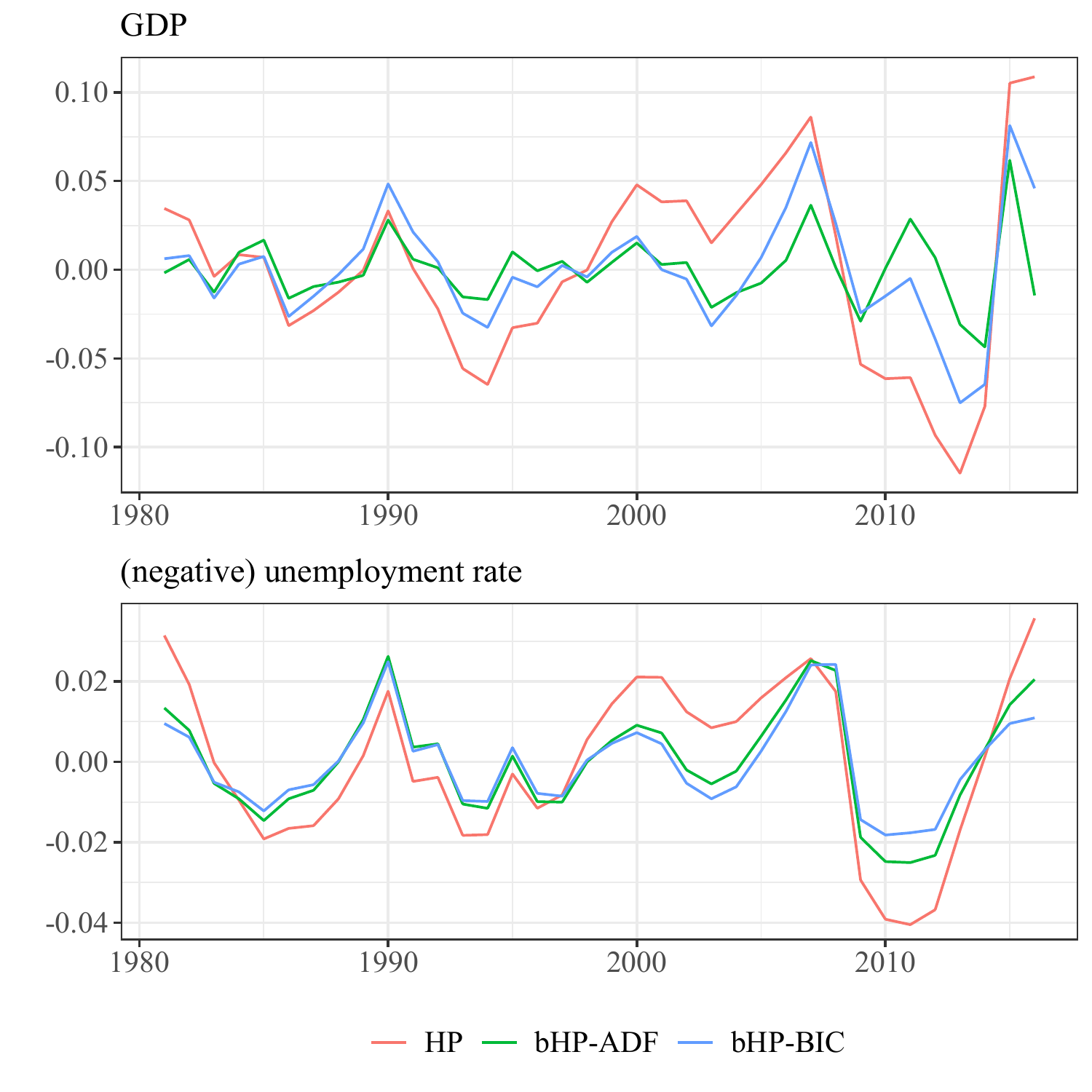}
	\caption{\label{fig:ireland} Cyclical components of GDP and the unemployment rate in Ireland. The (negative) unemployment rate is shown in the second panel. Annual data over 1980 to 2016.}
\end{figure}

We collect annual GDP data from the OECD (\texttt{OECD.stat}) and annual unemployment
rates from the World Bank. We follow \citet{ball2017okun} in studying
the same 20 economies over the period 1980 to 2016. The dataset is mostly balanced, except for a few countries with 1 or 2 missing values at the beginning of the time period. We maintain the primary parameter setting of 
$\lambda=100$ for the simple HP filter, and we apply the bHP filter based on the same tuning parameter $\lambda$. For each
country, Figure \ref{fig:Okun_coef} reports the OLS coefficient estimate
of $\beta$ in the upper panel, and the regression $R^2$ in the lower panel.
The regressions are conducted with cyclical components extracted
by the simple HP filter (shown by red bars), the boosted HP filter with iterations stopped
by (i) ADF test outcomes at the 5 percent level (shown by green bars), and (ii) use of the information criterion (\ref{BBIC}) (shown by blue bars). 

For most countries, the fitted coefficients and $R^2$ are similar across
the filtering methods. For example, in United States the coefficient
is approximately -0.5, and $R^2$ is around 0.8. These figures accord with established results for the USA and the recent findings of \citet{ball2017okun}. In particular, the results from using the boosted filter tend to confirm the conclusion of the latter authors that `Okun's law is a strong relationship in most countries'. 

One country where there is a surprisingly
large contrast among the methods is Ireland, where the equation $R^2$ is $0.71$ after simple HP filtering but only 0.10 after bHP-ADF filtering and 0.43 after bHP-BIC filtering. To
explore these differences, we display the relevant data for Ireland in Figure
\ref{fig:ireland}. The upper panel graphs the estimated cyclical components
of GDP obtained by HP, bHP-ADF and bHP-BIC. The red line produced by the HP filter shows a long upward trend from the mid 1990s to 2007, followed by a sustained slump until 2013. These trends are evident in the data from inspection and it is apparent that the HP filter fails to remove them in estimating the cycle. In fitting the boosted filter using the ADF procedure to select the boosting tuning parameter 19 iterations of the filter were needed, the largest number of iterations among all the 40 series in this experiment. The associated cyclical component is represented by the green
line. This GDP cyclical component fluctuates around the mean in a smaller range, shows no evidence of a residual trend, and it appears much more stable than the cycle determined by the HP filter. The shape of the cycle obtained by using BIC selection is very similar after $m=5$ iterations. 

The lower panel displays the three fitted curves of the (negative) unemployment rate for Ireland. The negative rate is used in the figure to better visualize the association with the GDP fitted cycles shown in the upper panel. For the unemployment rate series, the boosted filter is stopped by ADF after 2 iterations and by BIC after 5 iterations. In both cases, the use of repeated
filtering clearly mitigates residual trend behavior in the unemployment rate in comparison with the HP filter. The mitigation is more evident in the case of bHP-BIC filtering where the fitted cycle in the unemployment rate appears even more regular than after bHP-ADF filtering, especially during the aftermath of the 2007-2008 financial crisis. These adjustments in the fitted cycle from bHP filtering appear not to reduce the evident association with the GDP cycle and the fitted regression coefficients after bHP filtering indeed have similar values, as shown in the upper panel of Figure \ref{fig:Okun_coef}.

In sum, this application continues to support the robustness of Okun's law across developed
countries, thereby reinforcing the conclusion of \citet{ball2017okun}. But the results also expose the insufficiency of the standard HP filter to remove stochastic trend components in the case of Ireland. Repeated fitting in this case helps to isolate the cyclical component in each time series.

\subsection{International Business Cycles: Emergent and Developed Economies}

The HP filter was originally motivated in HP (1997) through its usefulness in the empirical study of business cycles in the USA. In an influential paper with a similar thematic concerning international evidence of business cycles, \citet{aguiar2007emerging} find that emerging
markets (represented by 13 economies) in general are more persistent
in the cyclical components of the three series they consider (GDP, consumption and investment)
than those of the developed markets (represented by another group
of 13 countries). In summarizing their study they declared that `[for emerging markets] the cycle
is the trend.'

We revisit this conclusion using the methods of the present paper to analyze the same data that the authors provide online.\footnote{Downloadable at\texttt{
	\url{https://scholar.harvard.edu/gopinath/pages/data-and-codes}}.}
Within each country, the three time series have the same length but across countries the length of the time series varies considerably. For example, the median length
is 52 quarters for the emerging economies, with Argentina the shortest
(1993Q1\textendash 2002Q4, 40 quarters), whereas the median is 94
quarters for the developed countries, with Australia, Finland, Netherlands
and Norway the longest (1979Q3\textemdash 2003Q2, 95 quarters). The
authors established their empirical results after HP-filtering all
78 time series with the standard setting $\lambda=1600$. As discussed earlier in the paper, the analysis in 
PJ shows that the implied penalty from using this standard setting is heavier for shorter time series, making stochastic trend identification difficult in international comparisons with series of differing lengths. As our asymptotic theory shows, the bHP filter provides a mechanism for adapting the standard setting to account for shorter and longer sample sizes. We employ the iterated procedure to the logarithm
of GDP, consumption and investment to study whether the cyclical patterns noted by \citet{aguiar2007emerging} in the
two groups of countries remain distinguishable.\footnote{\citet{aguiar2007emerging} report the moments after processing the cyclical components in a macroeconomic structural model. We directly work
	with the raw data here to keep the comparison as simple and straightforward
	as possible.} 

\begin{table}
	\caption{\label{tab:app_busi_cycle} Number of iterations and some moments
		(median in each group)}
	\medskip
	\centering{}%
	\begin{tabular}{lrrrrrr}
		\hline 
		& \multicolumn{2}{c}{HP} & \multicolumn{2}{c}{bHP-ADF} & \multicolumn{2}{c}{bHP-BIC}\tabularnewline
		& emerging & developed & emerging & developed & emerging & developed\tabularnewline
		\hline 
		& \multicolumn{6}{c}{Number of iterations}\tabularnewline
		GDP (Y) & 1 & 1 & 2 & 3 & 10 & 7\tabularnewline
		Consumption (C) & 1 & 1 & 2 & 2 & 10 & 7\tabularnewline
		Investment (I) & 1 & 1 & 4 & 2 & 12 & 6\tabularnewline
		\hline 
		& \multicolumn{6}{c}{Variance and correlation coefficient}\tabularnewline
		$\sigma\left(Y_{t}\right)$ & 0.0251 & 0.0134 & 0.0228 & 0.0094 & 0.0173 & 0.0076\tabularnewline
		$\sigma\left(C_{t}\right)$ & 0.0339 & 0.0127 & 0.0325 & 0.0093 & 0.0233 & 0.0070\tabularnewline
		$\sigma\left(I_{t}\right)$ & 0.0960 & 0.0413 & 0.0786 & 0.0332 & 0.0607 & 0.0268\tabularnewline
		$\rho\left(C_{t},Y_{t}\right)$ & 0.7592 & 0.7234 & 0.6284 & 0.4772 & 0.6610 & 0.4370\tabularnewline
		$\rho\left(I_{t},Y_{t}\right)$ & 0.8327 & 0.7024 & 0.7527 & 0.5435 & 0.7177 & 0.5135\tabularnewline
		$\rho\left(Y_{t},Y_{t-1}\right)$ & 0.7608 & 0.7528 & 0.6201 & 0.5624 & 0.5444 & 0.4475\tabularnewline
		\hline 
	\end{tabular}
\end{table}

\begin{figure}[htbp]
	\centering
	
	\includegraphics[scale=0.55]{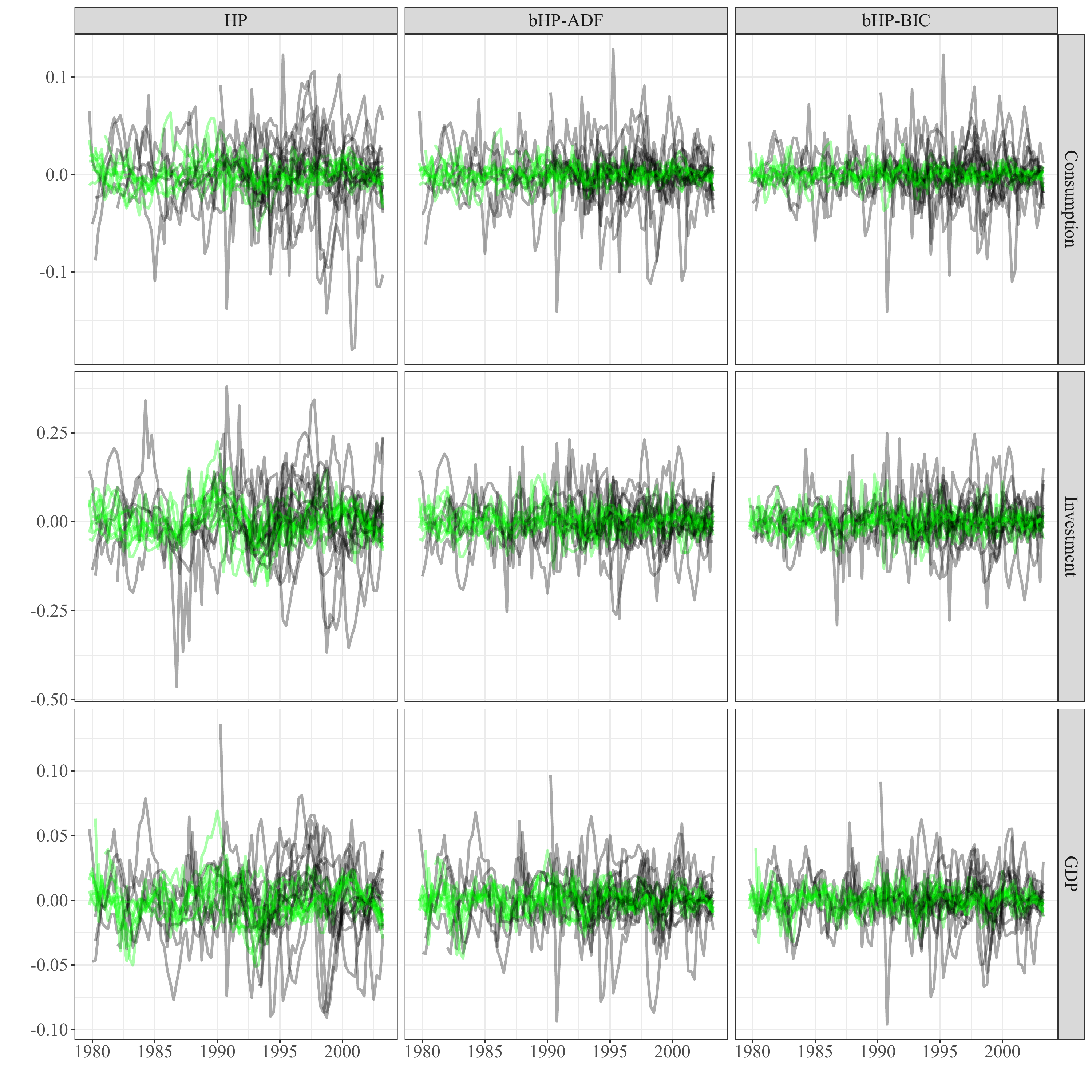}
	
	\caption{\label{fig:busi-cyc} Cyclical components of GDP, Consumption, and Investment obtained by the HP, bHP-ADF and bHP-BIC filters, the latter with data-determined stopping. Developed nations are displayed in green and emerging nations are shown in black.}
\end{figure}

For each of the 78 time series, we apply the HP filter and automated bHP filters, all with the same
$\lambda=1600$ setting, to extract trend and save the cyclical component. In general, the emerging
economies need more iterations than the developed countries to isolate trend, manifesting the differences in persistence. Table \ref{tab:app_busi_cycle} displays within each group of 13 countries
the medians of the number of iterations,
standard deviations, and correlation coefficients. The standard deviations typically
become smaller as boosting progresses, while the relative magnitude
between the emerging and developed markets remains stable. Similar relative
sizes are observed in the correlation coefficients. The repeated
filtering changes absolute values, but the relative magnitudes of
the volatility and persistence are largely maintained in the two groups
of countries.

Figure \ref{fig:busi-cyc} shows the cyclical components of each time
series, with the developed nations in green and the emerging nations in black. Despite the small number of iterations involved, the bHP-ADF filter provides noticeably greater smoothing of the time series. With  only a few more iterations taken by the bHP-BIC filter, the cyclical components appear more stable around the mean. The contrast in the volatility of the
two groups of countries is strongly manifest in the graph. Overall, this application of the boosted filter therefore  
confirms that \citet{aguiar2007emerging}'s findings are robust when
machine learning methods are used to assist in compensating for the differing lengths of the time series across countries.

\subsection{US Industrial Production Index}\label{subsec:univariate}

In this final application of our methods, we analyze a single macroeconomic time series of industrial production that has visually evident trend and (somewhat irregular) cyclical components over a long historical period. The US \emph{industrial production index} used here is an indicator of aggregate economic activity that measures real production output of manufacturing, mining, and utility industries based on hundreds of individual time series. The series is seasonally adjusted, covers the last century from 1919:Q1--2018:Q2, and comprises 398 observations. It is one of the longest US quarterly macroeconomic series available from the Federal Reserve data base.\footnote{ 
Downloadable at \url{https://fred.stlouisfed.org/series/IPB50001SQ}. 
}

\begin{figure}[htbp]
 \centering
  \subfloat[HP]{
  	\includegraphics[width=.7\linewidth]{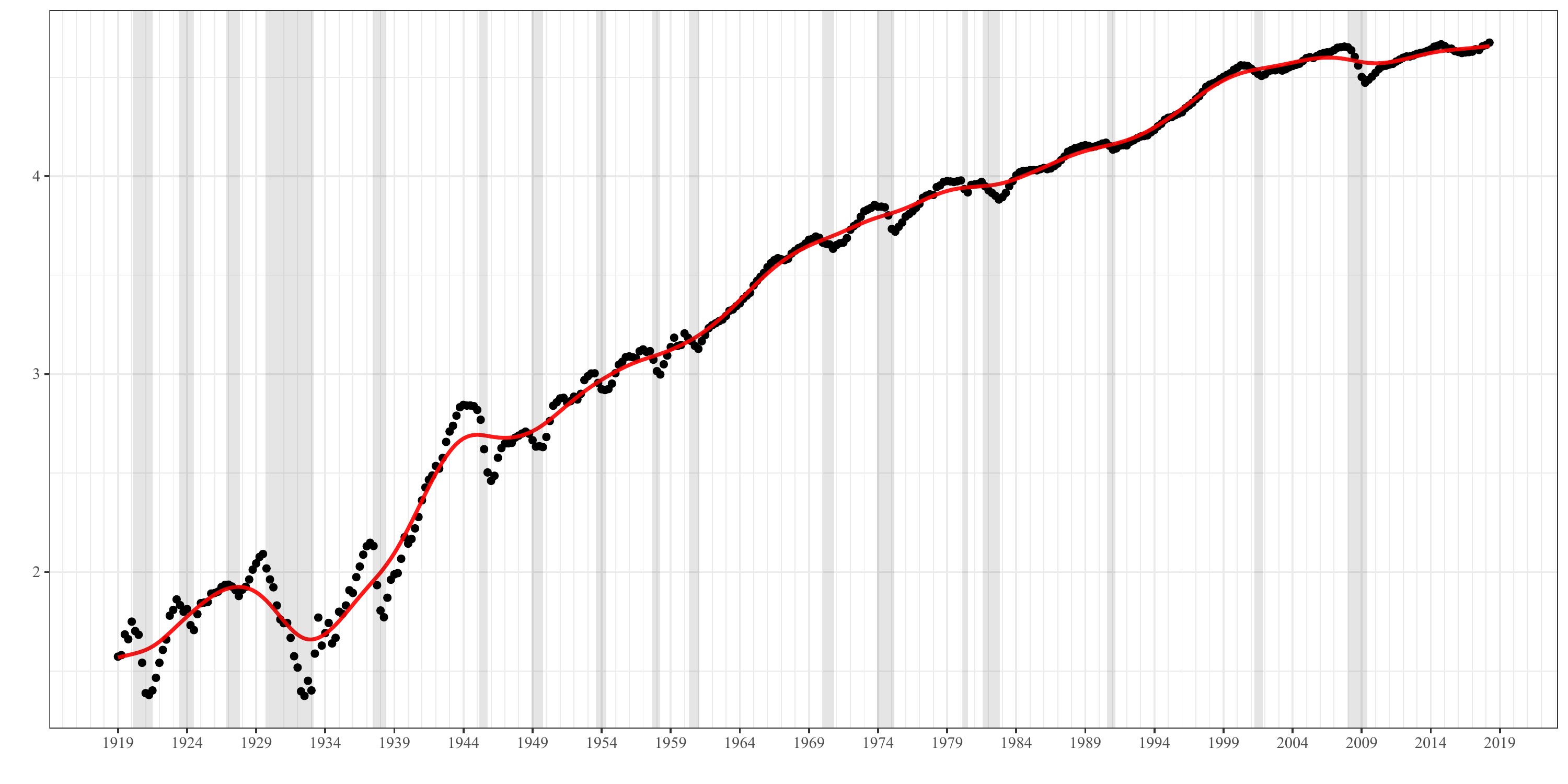}  
  }

  \subfloat[bHP-BIC]{
  	\includegraphics[width=.7\linewidth]{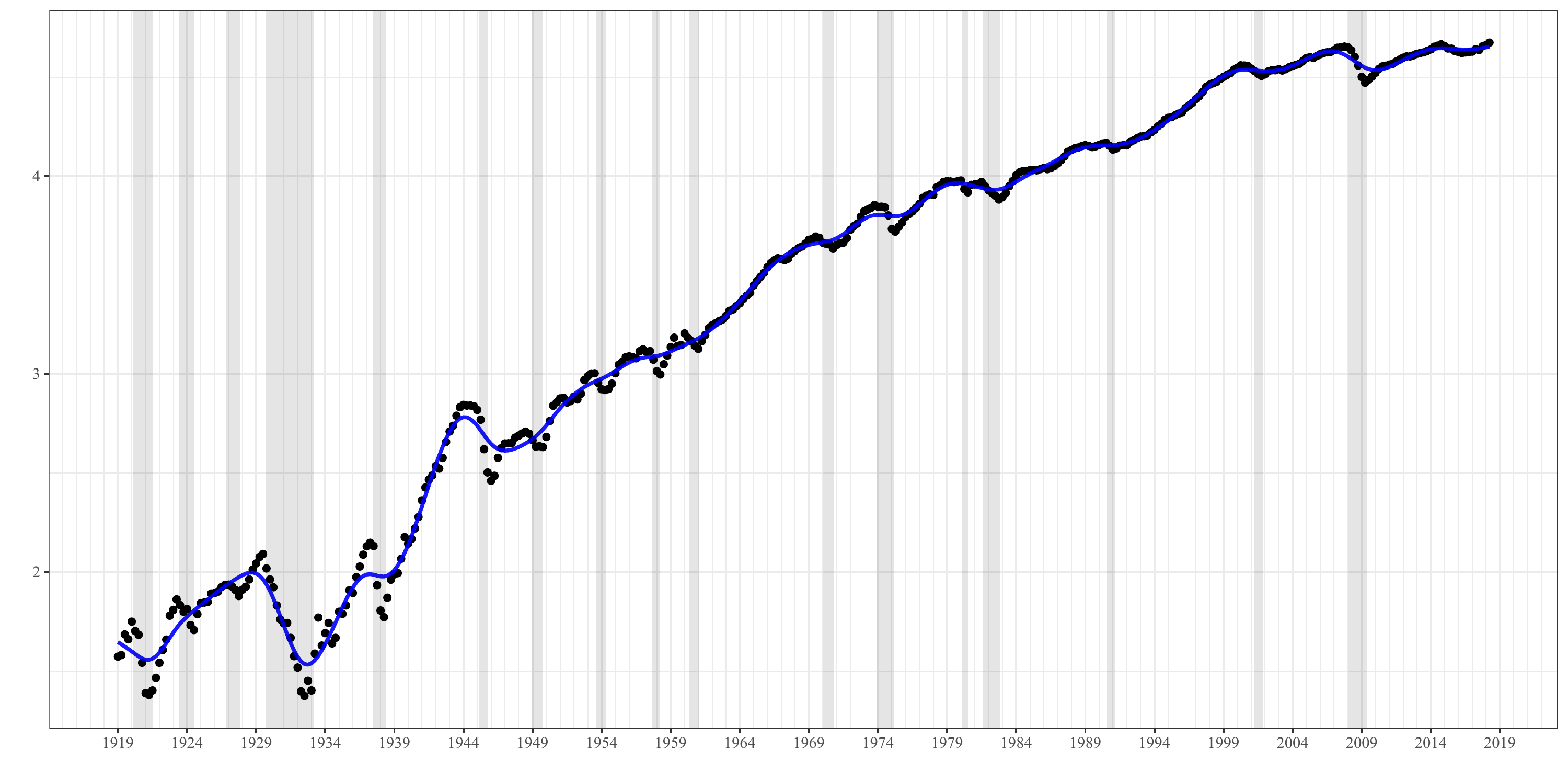}  
  }

  \subfloat[AR(4)]{
    \includegraphics[width=.7\linewidth]{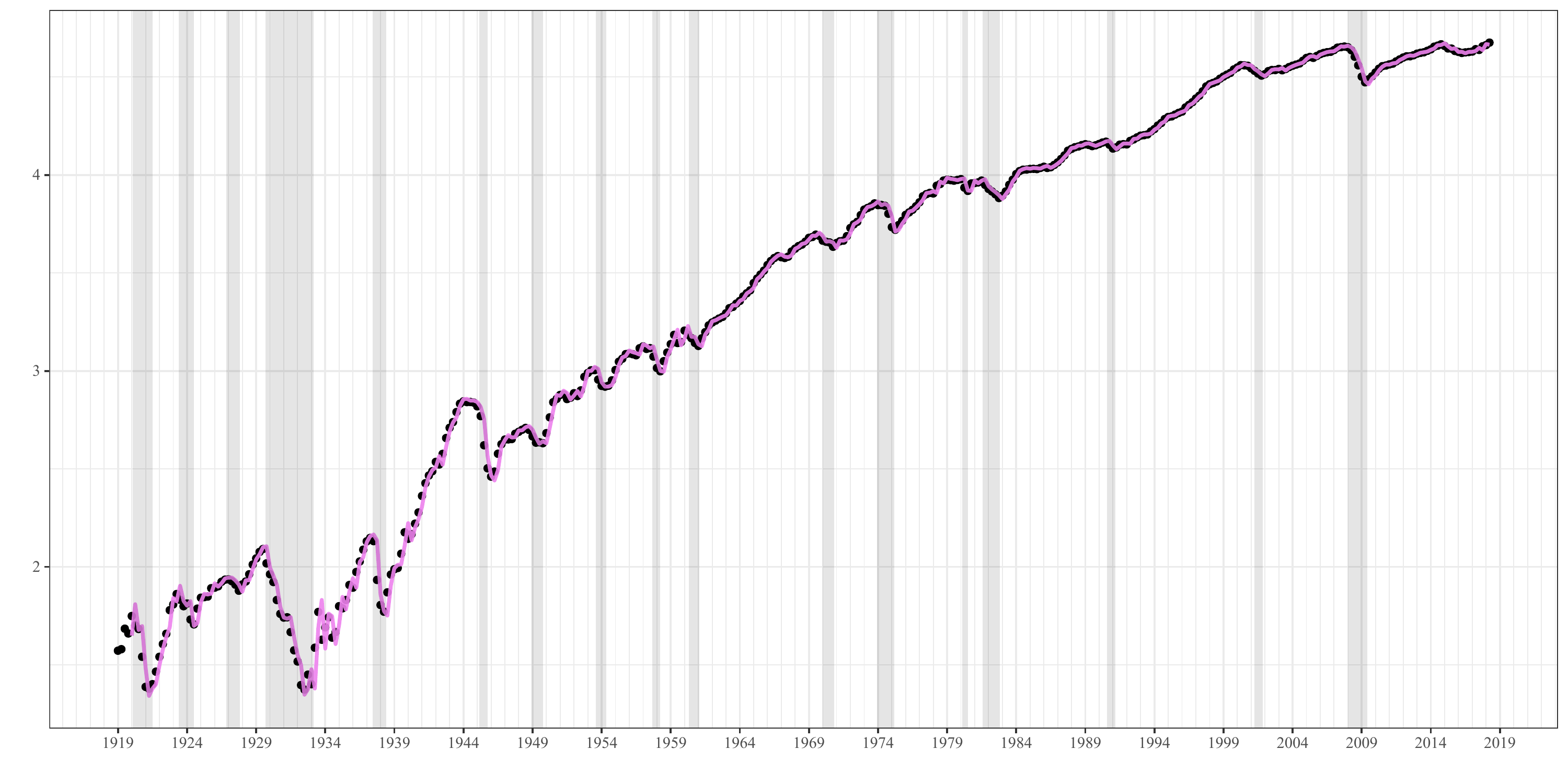} 
    }

\caption{US Quarterly Industrial Production and fitted trends over 1919-2019. The shaded periods show the recessions dated by the NBER.
}
\label{fig:IP1919}
\end{figure}

\begin{figure}[htbp]
 \centering
  \subfloat[HP]{
  	\includegraphics[width=.7\linewidth]{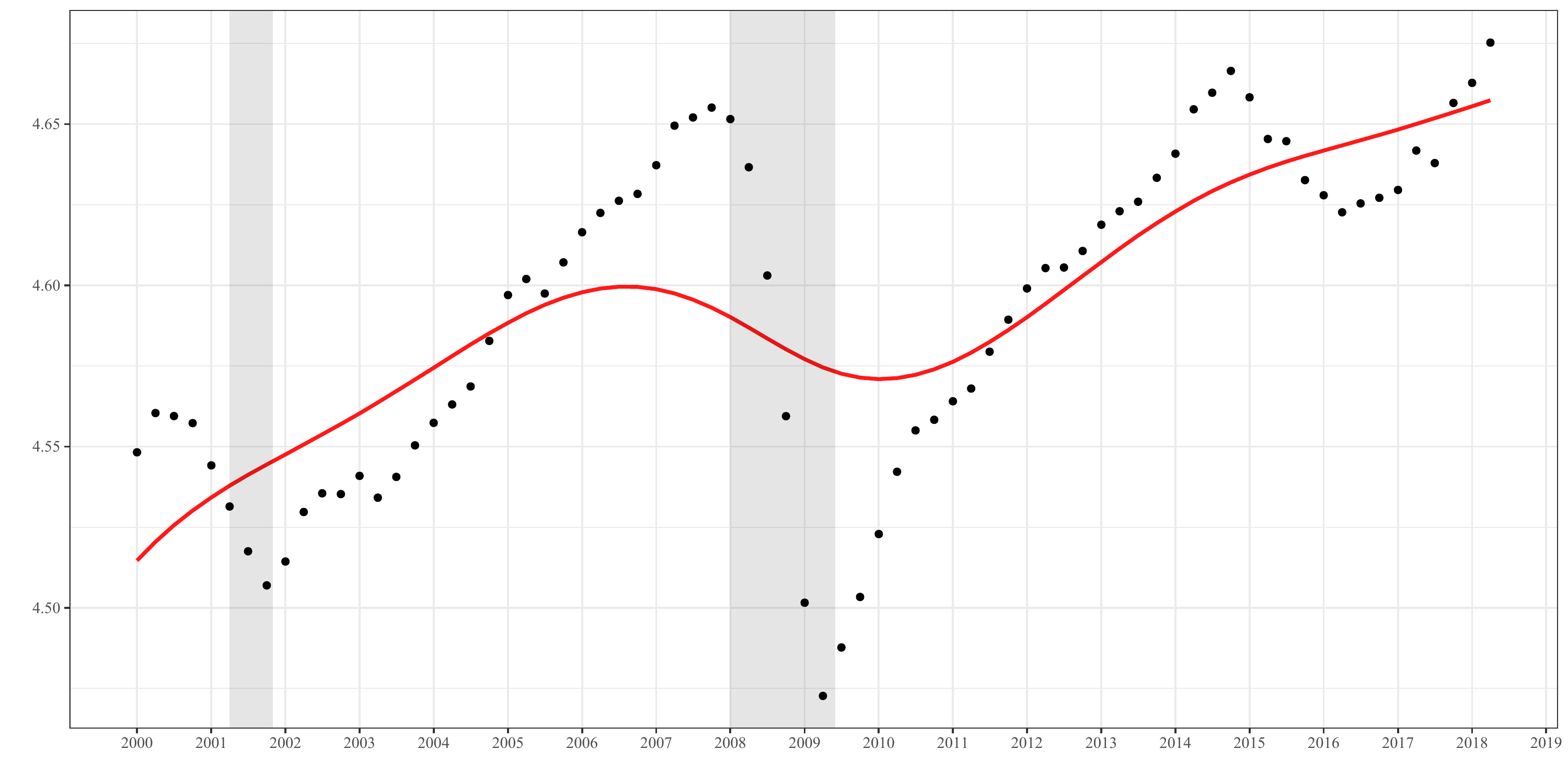}  
  }

  \subfloat[bHP-BIC]{
  	\includegraphics[width=.7\linewidth]{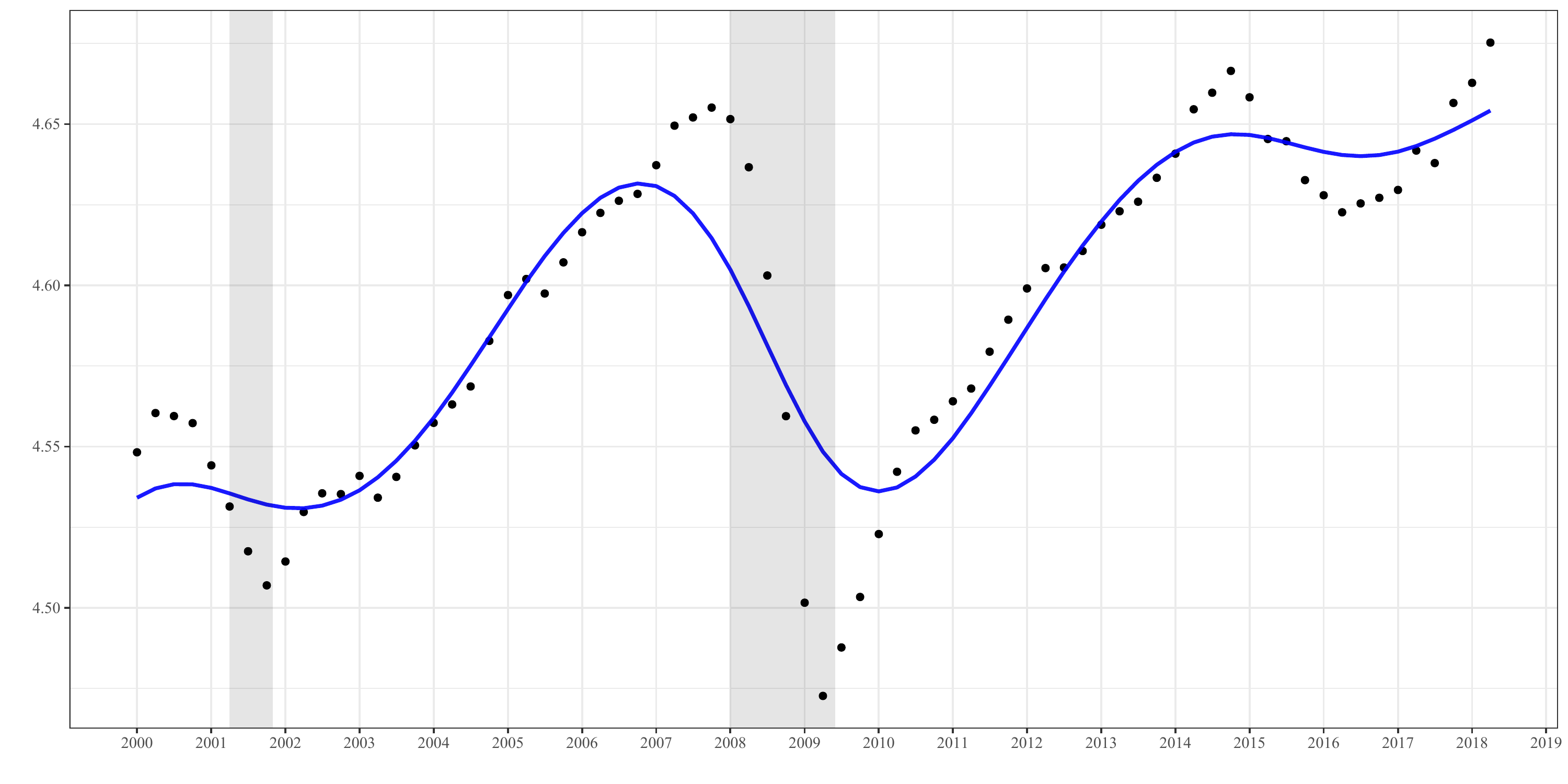}  
  }

  \subfloat[AR(4)]{
    \includegraphics[width=.7\linewidth]{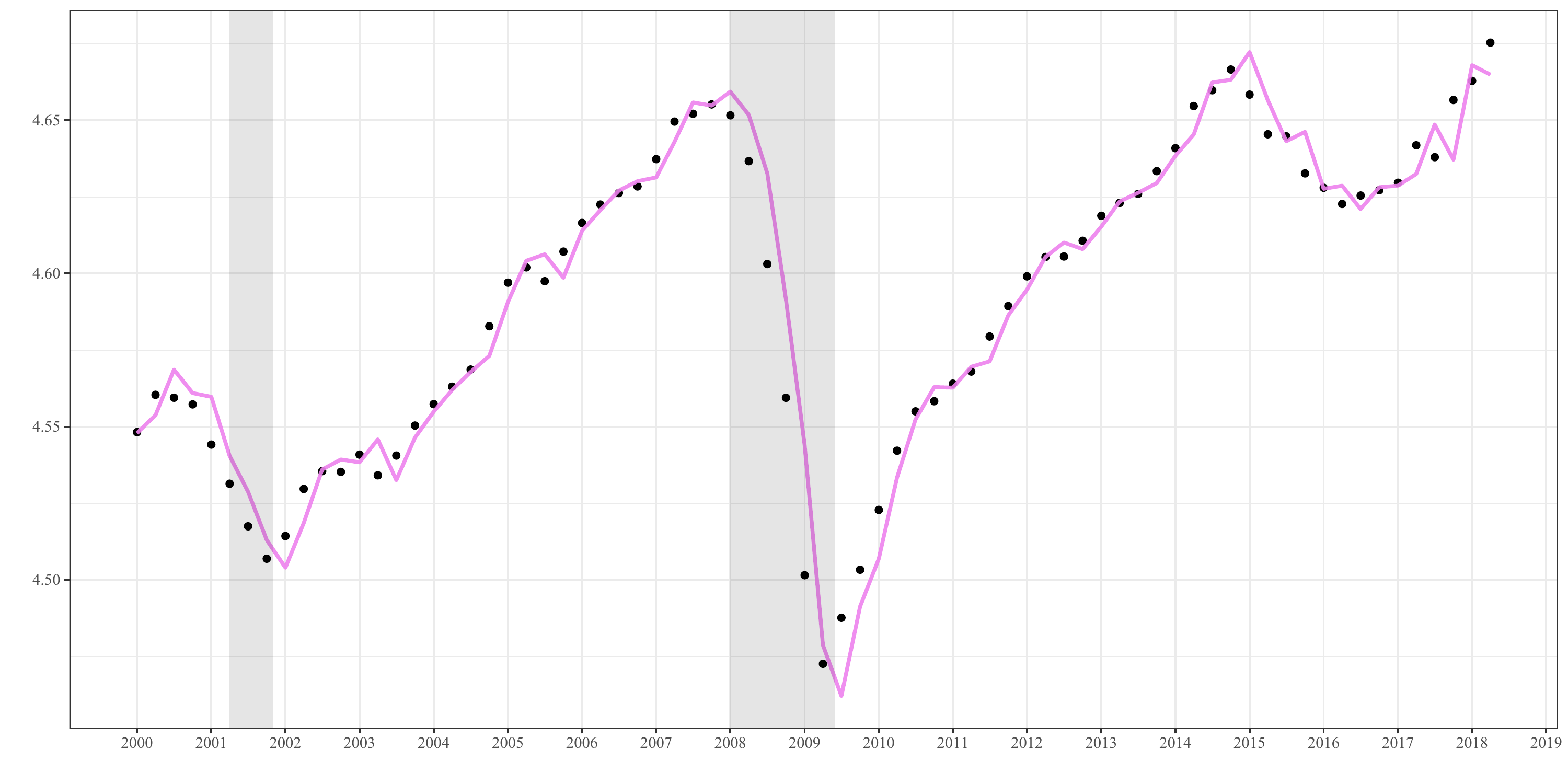} 
    }

\caption{US Quarterly Industrial Production and fitted trend lines in the 21st century, zoomed versions of Figure \ref{fig:IP1919}. }
\label{fig:IP2000}
\end{figure}

In Figure \ref{fig:IP1919}, the black dots plot 
the logarithm of the raw time series.
The shaded regions are the recessions dated by NBER, 
where both the Great Depression and the recent Great Recession are clearly visible. 
The index is very volatile before the Second World War. Following the Second World War, fluctuations around the upward path of the index moderate but occur regularly until the end of the 20th century. Figure \ref{fig:IP2000} zooms in on the more recent and more dramatic period of 21st century experience over 2000:Q1--2018:Q2.

It is common for macroeconomists, for example \citet{romer1999changes}, to study the many changing features of long time series of this type
by analyzing subperiods and comparing their defining characteristics across such periods. The HP filter approach, as well as other forms of trend extraction, seeks a unified econometric representation of the trend component of the entire series. The HP filter in Figure
\ref{fig:IP1919}(a)
is created with smoothing parameter setting $\lambda = 1600$, and this filter accordingly smooths out the peaks and valleys of the index.\footnote{bHP-ADF is stopped after one iteration 
and thereby producing the same result as the HP filter. 
It is discussed in Section \ref{subsec:additional_IP_figures}.
}
The bHP-BIC filter, shown in Figure \ref{fig:IP1919}(b),
involves 7 iterations.
Compared to the HP filter, it is more responsive to the downturn of industrial production during the episode of the financial crisis. 

Figure \ref{fig:IP2000} zooms in the period after 2000. 
The HP filter completely ignores the dot-com bubble collapse in 2001-2002 whereas the bHP-BIC filter
declines in 2001, indicating an impact of this collapse on trend and with the residual deviations (the bHP cycle) corresponding closely to the NBER dated 2001 recession shown by the shaded area of the graph. The bHP filter subsequently reflects the serious impact of the Great Recession on the upward trend path of production, matches the first part of the NBER dated 2008-2009 recession, and extends the recession period to 2010. As a measure of potential industrial production, the estimated impact on trend from the boosted HP filter is more consistent with the fundamental deterioration that many macroeconomists, such as \citet{krugman2012}, perceived to have occurred in the aftermath of the financial crisis.

Figures \ref{fig:IP1919}(c) and \ref{fig:IP2000}(c) report additional findings from the alternative AR(4) approach.
During both crashes in the first decade of the 21st century, the AR(4) fitted trend overshoots the extremes of the realized observations both before 
the burst of the financial bubble and at the end of the collapse that produced the downturn in the real economy that is reflected in the production index. This phenomenon is a typical feature of highly autoregressive (near unit root) fitting. In the present case the fitted AR(4) has long run autoregressive coefficient $0.9978$ which is virtually unity.\footnote{
	The estimated coefficients of the AR(4) model intercept and the first to fourth lags are
	$0.011,       1.421,       -0.514,        0.216$      and  $-0.1252$, 
	respectively. The long run autoregressive coefficient, based on the sum of the four autoregressive coefficients, is $0.9978$. The characteristic polynomial has roots 
	$[0.6123,0.9959, -0.0936\pm0.4433 \mathbf i]$  and the complex roots correspond to a damped (amplitude $0.4531$) cycle of $4.61$ quarters or
	$1.15$ years. The seasonally unadjusted industrial production series was also analyzed and produced very similar results to those given here, so they are not reported.}  
Autoregressive fitting of time series tends to capture the fine grain as well as the global features of a time series trajectory, thereby removing most of the variation in the time series and reducing the residual closer to a series with martingale difference characteristics
(see also Figure \ref{fig:IP_cycle1919}(c) in Appendix B). 
This approach seems too rigid in its goal of removing variation to separate slow moving trend components from irregular cyclical and stationary elements in time series. In particular, as this example demonstrates, the AR approach seems unable to effectively differentiate between trends and cycles in the economic environment, especially during episodes of extraordinary change that impact both trend and cyclical elements in economic activity. In the present case, the slow economic recovery following the crisis may be viewed as a distinguishing feature of the great recession cycle and the overall downturn that was induced may be considered as an inevitable impact on the trend \citep[c.f.,][]{krugman2012}.

\begin{table}
	\caption{\label{tab:residual_variation} Residual variance ($\times 1000$) from HP and bHP filtering and AR(4) autoregression } 
	\centering
	\begin{tabular}{cccc}
		\hline 
		residual variance  & HP & BIC & AR(4)\tabularnewline
		\hline 
		$x_{t}-\widehat{f}_{t}$ & 4.945 & 2.438 & 1.298\tabularnewline
		$x_{t+1}-\widehat{f}_{t}$ & 5.078 & 2.713 & 0.286\tabularnewline
		\hline 
	\end{tabular}
\end{table}

Our assessment of these findings of trend and cycle extraction with a century-long time series is that the bHP-BIC filter works well during periods of normal growth and mild cyclical activity and is also capable of capturing more complex heterogeneous features of the downturns in trend and slow variable recovery from the major recessions that arose in earlier and later years of this long historical period.

This application reveals some major differences between the filter and autoregressive approaches to cycle estimation. Table \ref{tab:residual_variation} shows the sample variance of the fitted cyclical components. In the first row of the table, it is clear that the autoregression has very small residual variance, amounting to approximately one half of that of the cyclical residual of the bHP-BIC filter and one quarter of that of the HP filter residual. This substantial gain in fitting the time series so well compared with the filtering methods arises from the capacity of the autoregression to capture a stochastic trend component in the time series parametrically (effectively by means of a near-unit root autoregression --- here with a long run autoregressive coefficient 0.9978 that is so close to unity) --- and to employ a parametric damped cycle of amplitude $0.4531$ with period around 1.15 years that is induced by a pair of complex autoregressive roots in the characteristic equation of the autoregression. This fitted cycle has high frequency and is much closer to representing a damped seasonal oscillation in the production series than a business cycle, even though the time series that is used here is seasonally adjusted.    

A more dramatic illustration of the forecast-oriented and fit-driven nature of the autoregression is apparent with a simple one-period phase shift. In particular, if we move the observed time series one period forward, then in the second row of the table the sample variance
of $(x_{t+1} - \hat{f}_t)$ is only about 20 percent that of the unshifted fit in the first row. Thus, the AR(4) is far better at predicting past values than future values of the trending trajectory, just as  may be expected from a unit root autoregression fit of a far more complex time series trajectory. Such a one-period-ahead forecasting mechanism is highly effective in reducing residual variation. But this success in the AR mechanism can compromise its role in capturing more slowly varying (multi-period) trend effects in a time series that involve features over many time periods. Of course, by construction, the parametric AR(4) filter is fundamentally different from the HP filter and has different traditional modeling goals, such as forecasting using past information and studying the impact of past shocks on the system variable.\footnote{In his pioneering study of business cycles in the United States between 1919-1932, 
	\citet[p.140]{tinbergen1939business} employed a fourth-order difference equation 
	to describe the `systematic cyclical forces' in the US economy, analyzing the amplitude and period of the cycle, which turned out to be 4.8 years. Tinbergen claimed that the influence of further lags was `found to be small', a conclusion that matches the recommendation of \citet{hamilton2017you} in using the AR(4) approach. In our present example, there is some evidence that extending the number of lags is beneficial in capturing in-sample fluctuation. Use of an AR(6), for example,  reduces the residual variance of 1.298 (shown in the first row of Table \ref{tab:residual_variation}) of the AR(4) to 1.087. Further, autoregressive lag determination by the standard BIC criterion clearly prefers AR(6) ($\mathrm{BIC} = -6.75$) to AR(4) ($\mathrm{BIC}=-6.58$) for the industrial production time series. The roots of the characteristic equation of the estimated AR(6) are 
	$\{0.0256, 0.9971,  -0.3934 \pm 0.6608 \mathbf i, 0.6360 \pm 0.4128 \mathbf i\}$, in which the two sets of complex roots indicate damped cycles of periods $1.51$ years and $2.73$ years.}

By comparison, phase shifting in the case of the HP and bHP-BIC filters leads naturally to deterioration in residual variation. As seen in the table, the sample variance grows in the second row of the table for both HP filters. This is explained by the fact that these filters are designed as nonparametric penalized and smoothly varying best fits to each individual observation in the time series trajectory, taking into account what is happening in neighboring observations. Such a mechanism seems more closely aligned with the general objective of historical trend determination when the concept of trend is based on the hypothesis articulated by Hodrick and Prescott (1997) in the header quotation that `the growth component of aggregate economic time series varies smoothly over time'.

\section{Conclusion}\label{sec:conclusion}

This paper explores the use of a machine learning modification of the HP filter that is designed for trend extraction in studying business cycles in macroeconomic data. The algorithm is based on the idea of repeated HP
filtering and is linked to $L_2$-boosting methods that are now commonly used in machine learning approaches to linear regression. The boosted HP filter allows empirical investigators to continue to use a primary tuning parameter setting such as the standard $\lambda=1600$ setting for quarterly data applications but alleviates the concern of using a single choice of this parameter in filtering time series of various lengths and persistence. To enhance the asymptotic performance of the HP filter, the boosted filter introduces a secondary tuning parameter that controls the degree of boosting while holding the primary parameter $\lambda$ fixed at customary levels such as 1600 for quarterly data. In practical applications, data-determined methods that rely on nonstationarity tests or information criteria may be used to select this secondary parameter in a convenient way. Asymptotic theory shows that the boosted HP filter has the capacity to consistently estimate, and thereby remove, a stochastic trend with time polynomial drift as well as a stochastic trend with deterministic polynomial drift and multiple structural breaks. These results seem relevant for many empirical applications in which standard HP tuning parameter settings are currently used.

The limit theory reveals some of the capabilities of HP filter methods as trend fitting and extraction processes for practical work that have heretofore been little understood. The methodology allows empirical researchers to rely on existing software for standard implementation of the HP filter in the boosting environment. Like ordinary least squares and vector autoregressions, empirically convenient methods such as the HP filter are unlikely ever to be put out of business, in spite of concerns that have been repeatedly raised  over many years about their usefulness and their effects on subsequent analysis. 

It is hoped that the present contribution will help empirical researchers to better understand the capabilities and limitations of the HP filter and to guide the implementation of a simple machine-learning vehicle for its improvement in applications, thereby mitigating some of the limitations of the HP filter itself. Our position is therefore more optimistic than that of \citet{hamilton2017you}. As we have shown, a key advantage of the boosted filter is that it provides a new device for consistently estimating in a nonparametric manner a wide class of trending mechanisms including processes that involve structural breaks, while remaining agnostic about the precise form of the trend nonstationarity in the data. These are central advantages of the boosted filter that justify some optimism, support continuing empirical use, and clearly distinguish the methods from competitor approaches that are typically reliant on correct model specification. 

To close the paper, we provide a summary response based on our present findings to the recent critique of the HP filter by \citet{hamilton2017you}, which takes up a long tradition of critiquing the HP filter as a tool of applied macroeconomics. Hamilton specifically argues for disuse of the HP filter on the following grounds: (i) it induces spurious cycles; (ii) it is inappropriate for a random walk; (iii) it is two-sided, giving future-informed predictions; (iv) a long autoregression, such as an AR(4) should be used instead. We consider each of these points in turn.

Point (i) repeats the central thesis of \citet{cogley1995effects}. The possible presence of spurious cycles in the residual is explained by the asymptotic theory of Theorem 3 of PJ and the smoothness of the limiting form of the filter for popular choices of the smoothing parameter. But point (i) no longer holds if the HP filter consistently estimates the trend function. In particular, Section 5 of PJ shows that much slower expansion rates than $O(n^4)$ for the smoothing parameter do lead to consistent estimates of stochastic and deterministic trend functions. Moreover, the present paper demonstrates that consistent estimation of a wide class of such trends is possible even with popular choices of the smoothing parameter by `boosting' the HP filter using machine learning techniques.
Point (ii) is invalid when the HP filter consistently estimates the limiting stochastic process corresponding to a random walk or a more general stochastic trend. Again, as shown in PJ, suitable choices of the smoothing parameter in relation to the sample size achieve consistent estimation of many limiting stochastic process trend functions; and, further, methods such as boosting can accelerate this convergence to the true function, as shown here. 
Point (iii) is correct. Like a fixed design nonparametric regression, the HP filter smooths observations on either side of the current observation. So it is true that the HP filter in its standard form is not intended as a predictive device. As Whittaker explained in his original formulation, the goal of the penalized likelihood formulation is to find the `most probable' function, which in this case is the trend function. The HP filter is a trend detection device that seeks to determine the `most probable trend' using clear probabilistic principles. Notwithstanding this primary goal of the filter, one sided filtering can be used recursively for prediction and the methods of PJ and the current paper, which show how the filter may be represented in terms of a series of smooth functions can be modified to produce predictive techniques.
Point (iv) offers an alternative. We have analyzed the performance of an autoregressive approach to trend and cycle determination in our numerical and empirical work, where the findings show a clear preference for the bHP filter over autoregression.  

We end this paper with a more general response to the proposal of using autoregressions for trend and cycle determination. Autoregressions are widely used in applied economic research and represent a valid modeling approach. But as trend elimination and cycle determination mechanisms autoregressions have limitations. These seem worthy to report in detail.
Much of the motivation for using long autoregressions stems from their capacity to capture a wide class of data generating mechanisms, motivated by inversion of the Wold representation in the stationary case and by unit root or near unit root fitting in nonstationary cases via the long run autoregressive coefficient. These valid properties coupled with convenience of implementation have sustained their extensive use in applied work over many decades. Nonetheless, autoregressions are unable to consistently estimate trends of a general form beyond simple polynomials via the inclusion of intercepts and polynomial time trends in their formulation. Further, by virtue of their potential in approximating the Wold representation of the stationary component in a time series, long autoregressions tend to produce residuals whose properties approximate martingale difference innovations, a feature that has led to their extensive use in the identification and analysis of policy shocks. Accordingly, the residuals from fitted autoregressions provide poor approximants of cyclical behavior for which temporal dependence, rather than absence of correlation, is a critical element. Finally, while autoregressions may naturally generate cycles from complex conjugate dynamic roots, such induced cycles are necessarily characterized by regularity, a fact that stands in contrast to the properties of macroeconomic data where both the period and intensity of business cycles and recessions are so noted for their irregularity that these features are embodied in the many popular descriptive terminologies that are given to them, among which we may mention the terms great depression, great moderation, great recession, short sharp recession, and long recovery. There are no doubt many others. 

In counterpoint to a long autoregression, what the HP filter does and what the boosted filter of this paper does even better is to find the most probable trend, one of sufficient generality that the residuals may take many different forms, thereby accommodating time series that can capture a potentially wide class of cyclical downturns and expansions. As the graphical demonstrations in Appendix \ref{simu_lambda_robust} show, the bHP filter is robust to the choice of the tuning parameter $\lambda$ used in fitting the initial HP filter allowing investigators to proceed with standard settings such as $\lambda=1600$ for quarterly data. With the number of iterations in the bHP data-determined by BIC or ADF testing, the bHP filter is easy to use in practical work and has consistent asymptotic properties for a wide range of trend processes. For all these reasons it is our view that the HP filter and boosted enhancements may validly be used as helpful empirical devices in the search for trend and cycle in applied econometric work.    

To implement the automated boosted HP filters in practical work, we have developed documented R, Matlab, Python and Julia functions along with test examples to assist empirical researchers. These programs may be downloaded from the following website \\  \url{https://github.com/zhentaoshi/Boosted_HP_filter}.

\appendix
 
\setcounter{footnote}{0}
\setcounter{equation}{0} 
\renewcommand{\theequation}{A\arabic{equation}} 

\section{Proofs}

\begin{proof}[\textbf{Proof of Theorem \ref{thm:Brownian}}]
	
	We apply the approach used in the proof of Theorem 3 of PJ with some new modifications. The idea is to use the KL series representations (\ref{eq:PJ_sum}) and (\ref{tt7}) and the fact that these series converge almost surely and uniformly in $r$ so that successive pseudo-integral operations may be applied to them to obtain the asymptotic form of the boosted filter. The derivations here primarily relate to the asymptotic impact of boosting.  
		
	Write $X_{n}\left( r\right)=n^{-1/2}x_{\lfloor nr\rfloor }$. By Lemma 3.1 of \citet{phillips2007unit}
  it is known that an expanded probability space can be constructed with a corresponding limiting Brownian motion for which the following uniform convergence holds almost surely  	
\begin{equation*}
\sup_{0\leq t\leq n}\left\vert X_{n} (t/n) -B (t/n) \right\vert =o_{a.s.}\left(1 \right) .
\end{equation*}%
In what follows calculations are made in this expanded space where almost sure convergence applies and in the original space the results translate as usual into weak convergence.

Since the KL series representation (\ref{tt7}) of $B\left( r\right) 
$ converges almost surely and uniformly in $r$ we may use a finite series KL
approximation $B^{K_{n}}\left( r\right) =\sum_{k=1}^{K_{n}}\sqrt{\lambda _{k}%
}\varphi _{k} (t/n) \xi _{k}$ with the property that for $%
K_{n}\rightarrow \infty $ we have $\sup_{0\leq r\leq 1}\left\vert
B^{K_{n}}\left( r\right) -B\left( r\right) \right\vert =o_{a.s.}\left(
1\right) .$ Then 
\begin{equation} \label{aa1}
\sup_{0\leq t\leq n}\left\vert
 X_{n} (t/n) -B^{K_{n}} (t/n) \right\vert =o_{a.s.}\left( 1\right) ,
\end{equation}%
if $K_{n}\rightarrow \infty $ as $n\rightarrow \infty .$ 
It follows that $%
X_{n}\left( \frac{t}{n}\right) $ is almost surely uniformly well
approximated by $B^{K_{n}} (t/n) $ for all $t\leq n$ as $%
n\rightarrow \infty .$ We therefore consider the effect of the HP trend
operator $G_{\lambda }=
1/[\lambda L^{-2}(1-L)^{4}+1] $ directly upon $%
B^{K_{n}} (t/n) =\sum_{k=1}^{K_{n}}\sqrt{\lambda _{k}}%
\varphi _{k} (t/n) \xi _{k}$ as $n\rightarrow \infty.$ %

Noting that $\sqrt{\lambda _{k}}= \left[ ( k- 0.5 ) \pi
\right]^{-1} $ and $\varphi _{k} (t/n) =\sqrt{2}\sin \left( 
 (t/n)  \lambda_{k}^{-1/2} \right),$ we have, analogous to PJ, %
\begin{eqnarray}
n\left( 1-L\right) \varphi _{k}\left( \frac{t}{n}\right) & = & \sqrt{2}n\left[
\sin \left( \frac{\frac{t}{n}}{\sqrt{\lambda _{k}}}\right) -\sin \left( 
\frac{\frac{t-1}{n}}{\sqrt{\lambda _{k}}}\right) \right]   \notag \\
&=&\sqrt{2} \cdot 2n \cdot \sin \left( \frac{1}{2n\sqrt{\lambda _{k}}}\right) \cos
\left( \frac{\frac{2t-1}{n}}{2\sqrt{\lambda _{k}}}\right)    \notag \\
&=&\frac{\sqrt{2}}{\sqrt{\lambda _{k}}} \cdot \frac{\sin \left( \frac{1}{2n%
		\sqrt{\lambda _{k}}}\right) }{\frac{1}{2n\sqrt{\lambda _{k}}}}\cos \left( 
\frac{\frac{2t-1}{n}}{2\sqrt{\lambda _{k}}}\right)   =O\left( \frac{1}{%
	\sqrt{\lambda _{k}}}\right)   \label{cc1}
\end{eqnarray}%
where the result holds uniformly in $t \le n$ and uniformly for all $k\geq 1$ since $x^{-1} \sin x  $
and $\cos \left( x\right) $ are both uniformly bounded. It follows
from (\ref{cc1}) that%
\begin{equation}
n\left( 1-L\right) \varphi _{k}\left( \frac{t}{n}\right) =\frac{\sqrt{2}}{%
	\sqrt{\lambda _{k}}}\cos \left( \frac{\frac{t}{n}}{\sqrt{\lambda _{k}}}%
\right) \left\{ 1+o\left( 1\right) \right\} ,\text{ \ as }n\rightarrow
\infty ,  \label{cc2}
\end{equation}%
uniformly in $t \le n$ and uniformly in $k\leq K_{n}$ with $n\sqrt{\lambda _{K_{n}}}\rightarrow \infty ,
$ which holds when $K_n / n \rightarrow 0.$ By repeated argument as
in (\ref{cc1}) we find that 
\begin{eqnarray}
&&L^{-2}\left[ n\left( 1-L\right) \right] ^{4}\varphi _{k}\left( \frac{t}{n}%
\right)
=\frac{\sqrt{2}}{\lambda _{k}^{2}}\sin \left( \frac{\frac{t}{n}}{\sqrt{%
		\lambda _{k}}}\right) \left\{ 1+o\left( 1\right) \right\} =\frac{\varphi
	_{k}\left( \frac{t}{n}\right) }{\lambda _{k}^{2}}\left\{ 1+o\left( 1\right)
\right\} ,  \label{cc3}
\end{eqnarray}%
again uniformly in $t \le n$ and uniformly for all $k\leq K_{n}$ whenever $ K_n / n %
\rightarrow 0.$ Similarly, as in (\ref{cc1}), we have
 $L^{-2}\left[ n\left(
1-L\right) \right] ^{4}\varphi _{k}(t/n) =O\left(
1/\lambda _{k}^{2}\right) $ for all $k\geq 1$ and $t \le n$.

Next observe that when $\lambda=\mu n^4$ for some $\mu>0$ we have  
\begin{equation*} 
\left(1-G_\lambda\right)^m=\left(\frac{\mu L^{-2}\left[(n\left(1-L\right)\right]^{4}}{\mu L^{-2}\left[(n\left(1-L\right)\right]^{4}+1}\right)^m.
\end{equation*}
Using (\ref{cc3}) and the operational calculus in PJ we find that
\begin{align} \label{cc6}
\left(1-G_\lambda\right)^m\varphi _{k}\left( \frac{t}{n}\right) &=\left(\frac{\mu L^{-2}\left[(n\left(1-L\right)\right]^{4}}{\mu L^{-2}\left[(n\left(1-L\right)\right]^{4}+1}\right)^m \varphi_{k}\left(\frac{t}{n}\right) \nonumber \\
&=\left(\frac{ \frac{\mu}{\lambda_{k}^{2}}} {1+\frac{\mu}{\lambda_{k}^{2}}}\right)^{m}\varphi_{k}\left(\frac{t}{n}\right) \left\{ 1+o\left( 1\right)\right\}
=\left(\frac{ \mu} {\mu+\lambda_{k}^{2}}\right)^{m}\varphi_{k}\left(\frac{t}{n}\right) \left\{ 1+o\left( 1\right)\right\},
\end{align}
whenever $ K_n / n \rightarrow 0$, as in (\ref{cc3}).
Then, for $\lambda=\mu n^4$ with $\mu>0$ and using (\ref{cc6}), we have for the boosted HP filter
\begin{eqnarray} 
\widehat{f}_{\frac{t}{n},K_n}^{\left(m\right)}&:= &\left[1-\left(1-G_\lambda\right)^m\right]B^{K_{n}}\left( r\right) =\sum_{k=1}^{K_{n}}\sqrt{\lambda _{k}} \nonumber 
\left[1-\left(1-G_\lambda\right)^m\right] \varphi _{k}\left( \frac{t}{n}\right) \xi _{k} \\
&= & \sum_{k=1}^{K_{n}}\sqrt{\lambda _{k}}
\left[1-\left(\frac{ \mu} {\mu+\lambda_{k}^{2}}\right)^{m}\right] \varphi _{k}\left( \frac{t}{n}\right) \xi _{k} + o_{a.s.}(1)
\end{eqnarray}
as $n \to \infty$ with $ K_{n} / n \rightarrow 0$ as in (\ref{cc6}). 
Next, observe that as $m \to \infty$

$$\sum_{k=1}^{K_{n}}\sqrt{\lambda _{k}}
\left[1-\left(\frac{ \mu} {\mu+\lambda_{k}^{2}}\right)^{m}\right] \varphi _{k}\left( \frac{t}{n}\right) \xi _{k} 
\to_{a.s.}\sum_{k=1}^{K_{n}}\sqrt{\lambda _{k}}\varphi _{k}\left( \frac{t}{n}\right) \xi _{k},$$
because, for all $k \le K_n$,
$\left( \mu / (\mu+\lambda_{k}^{2} ) \right) ^{m} 
\le  \left(   \mu / (\mu+\lambda_{K_n}^{2} ) \right) ^{m}
=  \left(  1 -  \lambda_{K_n}^2  /  (\mu + \lambda_{K_n}^2 )    \right)^{m}
 \to 0$ as $m \to \infty$ and $K_n^4 / m \to 0.$
 
We deduce that
\begin{eqnarray} 
\widehat{f}_{\frac{t}{n},K_n}^{\left(m\right)} & = & \sum_{k=1}^{K_{n}}\sqrt{\lambda _{k}}
\left[1-\left(\frac{ \mu} {\mu+\lambda_{k}^{2}}\right)^{m}\right] \varphi _{k}\left( \frac{t}{n}\right) \xi _{k} + o_{a.s.}(1)\nonumber \\
&=&\sum_{k=1}^{K_{n}}\sqrt{\lambda _{k}}\varphi _{k}\left( \frac{t}{n}\right) \xi _{k} + o_{a.s.}(1), 
\end{eqnarray}
as $m,n \to \infty$ with $ K_n / n +K_n^4 / m \rightarrow 0$. 
It follows that
$\widehat{f}_{  (t=\left\lfloor nr\right\rfloor)/n,K_n}^{\left(m\right)} \to_{a.s.} B^{K_{n}}\left(r\right)$ as $m,n \to \infty$. Moreover, since $ B^{K_{n}}\left( r\right) \to_{a.s.} B(r)$ as $K_n \to \infty$, we deduce that 
\begin{equation}
\widehat{f}_{\frac{{t=\left\lfloor nr\right\rfloor }}{n},K_n}^{\left(m\right)} \to_{a.s.} B\left(r\right), 
\end{equation}
as $m,K_n,n \to \infty$ with $  K_n / n  \to 0$, which together ensure that (\ref{cc3}) and (\ref{cc6}) hold. Then, since 
$X_{n}\left( \frac{t}{n}\right) $ is almost surely uniformly well
approximated by $B^{K_{n}} \left( t/n \right) $ for all $t\leq n$ as $n\rightarrow \infty$, it follows that
$ n^{-1/2} \widehat{f}_{t=\left\lfloor nr\right\rfloor }^{\left(m\right)}\to_{a.s.} B(r),$ as $m,n \to \infty$. 
In the original probability space, this means that  
$n^{-1/2}  \widehat{f}_{t=\left\lfloor nr\right\rfloor }^{\left(m\right)} \rightsquigarrow B(r),$ and the stated result follows. 
\end{proof}

\vspace{2mm}

\begin{remark}[Shrinkage bHP]\label{rmk:shrinkage}
\hspace{1mm} Some popular boosting
algorithms \citep[See ][Chapter 10.12.1]{hastie2009bible} further weaken the base learner by scaling with another tuning parameter $\delta \in (0,1)$. In our notation we can define a shrinkage bHP iteration for trend and cycle estimation as  
\[
\check{c}^{\left(m\right)}=\left(I_{n}-\delta S\right)\check{c}^{\left(m-1\right)},\mbox{ with initialization }\check{c}^{\left(1\right)}=\left(I_{n}-\delta S\right)x,
\]
thereby shrinking the operator $S$ towards zero by the factor $\delta$, which leads to the $m$'th iteration solution $\check{c}^{\left(m\right)}=\left(I_{n}-\delta S\right)^m x$. 
The corresponding shrinkage bHP trend operator is
$
G_{\lambda,\delta}= \delta / [\lambda L^{-2}\left(1-L\right)^{4}+1 ].
$
By operational calculus as in (\ref{cc6}) in the above proof we obtain the modified expression 
\[
\left(1-G_{\lambda,\delta}\right)^{m}\varphi_{k}\left(\frac{t}{n}\right)=\left(\frac{\mu+\left(1-\delta\right)\lambda_{k}^{2}}{\mu+\lambda_{k}^{2}}\right)^{m}\varphi_{k}\left(\frac{t}{n}\right)\left\{ 1+o\left(1\right)\right\},
\]
as $n \to \infty$. For all $k\leq K_{n}$, we have
\[
\left(\frac{\mu+\left(1-\delta\right)\lambda_{k}^{2}}{\mu+\lambda_{k}^{2}}\right)^{m}\leq\left(\frac{\mu+\left(1-\delta\right)\lambda_{K_{n}}^{2}}{\mu+\lambda_{K_{n}}^{2}}\right)^{m}=\left(1-\frac{\delta\lambda_{K_{n}}^{2}}{\mu+\lambda_{K_{n}}^{2}}\right)^{m}\to0
\]
 as $m\to\infty$ and $K_{n}^{4}/m\to0$ for any fixed $0<\delta\leq1$.
The proof of Theorem \ref{thm:Brownian} as well as the proofs of Theorems \ref{thm:drift} and \ref{thm:break} for bHP then carry over in the same way for this shrinkage bHP. So shrinkage boosting in this manner does not alter the asymptotic theory but relies on the additional tuning parameter $\delta$.
\end{remark}

\vspace{2mm}
\begin{proof}[\textbf{Proof of Theorem \ref{thm:drift}}]
	The stochastic trend component $x_t^0$ of $x_t$ is handled in the same way as the proof of Theorem \ref{thm:Brownian}.
	It remains to show that repeated applications of the HP filter in the boosting algorithm preserve the deterministic polynomial component of the $J$-th degree in $x_t$.
	Let $y_{t}=\beta_{1} (t/n) +\cdots+\beta_{J}(t/n)^{J}$ denote the polynomial trend
	so that $x_t=y_t+x_t^0$. The boosting algorithm applies the operator 
	\[
	1-G_\lambda=\frac{\lambda L^{-2}\left(1-L\right)^{4}}{\lambda L^{-2}\left(1-L\right)^{4}+1}
	\]
	repeatedly $m$ times. Taking the $j$-th term of $y_t$, we note that after $m$ iterations
\begin{equation} \label{tt28}
	\left(1-G_\lambda\right)^{m}\left(\frac{t}{n}\right)^{j}=\frac{\lambda^{m}L^{-2m}\left(1-L\right)^{4m}\left(\frac{t}{n}\right)^{j}}{\left(\lambda L^{-2}\left(1-L\right)^{4}+1\right)^{m}}=0
\end{equation}
	for all $j\leq4m-1$, since in that event the numerator component of the operator yields
	\begin{equation} \label{tt41}
	\lambda^{m}L^{-2m}\left(1-L\right)^{4m}\left(\frac{t}{n}\right)^{j}=0.
	\end{equation} 
The polynomial degree $J$ of $y_t$ is finite and so whenever the number of iterations $m$ in the boosted HP algorithm
	is sufficiently large in the sense that $ 4m-1 \geq J $, we have $\left(1-\left(1-G_{\lambda}\right)^{m}\right)y_{t} = y_{t}$. The denominator component of the operator in (\ref{tt28}) may be handled as in PJ using Fourier methods.	
	An alternative approach to handling the effect of the denominator in this case is to use an integral version of the operator (see footnote \ref{pseudo-operator}). In particular, 
	\begin{eqnarray}
	&&\left( 1-G_{\lambda }\right) ^{m}\left( \frac{t}{n}\right) ^{j}=\left( 
	\frac{\lambda L^{-2}\left( 1-L\right) ^{4}}{\lambda L^{-2}\left( 1-L\right)
		^{4}+1}\right) ^{m}\left( \frac{t}{n}\right) ^{j}=\left( \frac{L^{-2}\left(
		1-L\right) ^{4}}{L^{-2}\left( 1-L\right) ^{4}+\frac{1}{\lambda }}\right)
	^{m}\left( \frac{t}{n}\right) ^{j}  \notag \\
	&=&\frac{1}{\Gamma \left( m\right) }\int_{0}^{\infty }s^{m-1}e^{-s}e^{-s%
		\left[ L^{-2}\left( 1-L\right) ^{4}+\frac{1}{\lambda }\right]
	}ds \cdot L^{-2m}\left( 1-L\right) ^{4m}\left( \frac{t}{n}\right) ^{j}  \notag \\
	&=&\frac{1}{\Gamma \left( m\right) }\sum_{k=0}^{\infty }  
	\left\{  
	\frac{\left( -1\right) ^{k}}{k!}\int_{0}^{\infty }s^{m+k-1}e^{-s}ds\left[ L^{-2}\left(1-L\right) ^{4}+\frac{1}{\lambda }\right] ^{k} 	
	\right\}	
	 \cdot L^{-2m}\left( 1-L\right) ^{4m}\left( 
	\frac{t}{n}\right) ^{j}=0,  \notag
	\end{eqnarray}%
	just as in (\ref{tt41}) above for all $m$ such that $4m \geq J+1 \geq j+1.$ 
 	Thus, (\ref{tt28}) holds and boosting removes the polynomial component of the time series. In effect, boosting the HP algorithm raises its capacity to capture accurately a polynomial trend of any finite order as $m\to\infty$. 
\end{proof}

\vspace{2mm}
\begin{proof}[\textbf{Proof of Asymptotic Approximation (\ref{tt26})}]
We need to show that as $n\rightarrow \infty $ 
\begin{equation}
\mathrm{tr} \left(B_{m} \right)  =
\mathrm{tr}\left( I_{n}-\left( I_{n}-S\left( \lambda
\right) \right) ^{m} \right)
=n-\mathrm{tr} \left[ \sum_{k=1}^{n-2}\left( \frac{4\mu
	n^{4}\left( 1-\cos \left( \frac{k\pi }{n-1}\right) ^{2}\right) }{1+4\mu
	n^{4}\left( 1-\cos \left( \frac{k\pi }{n-1}\right) ^{2}\right) }\right)
^{m}\right]  \left\{ 1+o\left( 1\right) \right\} .  \label{tt27}
\end{equation}%
Define $d_{2}^{\prime }=(1,-2,1)$ and the $(n-2) \times 1$ unit
vectors $ e_1 =(1,0,...,0)^{\prime }$ and
$  e_n =(0,0,...,1)^{\prime }.$
We then write the $(n-2)\times n$ second differencing matrix $D^{\prime }$ as 
\[
D^{\prime }=\left[ 
\begin{array}{ccccc}
d_{2}^{\prime } & 0 & 0 & \cdots  & 0 \\ 
& d_{2}^{\prime } & 0 & \cdots  & 0 \\ 
&  & d_{2}^{\prime } & \cdots  & 0 \\ 
&  &  & \ddots  & 0 \\ 
&  &  &  & d_{2}^{\prime }%
\end{array}%
\right] =\left[ e_{1},M,e_{n}\right] ,
\]%
where $M$ is the $(n-2)\times (n-2)$ tridiagonal symmetric Toeplitz matrix  
\[
M=\left[ 
\begin{array}{cccccc}
-2 & 1 & 0 &  &  &  \\ 
1 & -2 & 1 & \ddots  &  &  \\ 
& \ddots  & \ddots  & \ddots  & \ddots  &  \\ 
&  & 1 & -2 & 1 & 0 \\ 
&  &  & 1 & -2 & 1 \\ 
&  &  &  & 1 & -2%
\end{array}%
\right] .
\]%
Use the inverse matrix formula 
\begin{eqnarray*}
	S\left( \lambda \right) &=& I_{n}-\lambda D\left( I_{n-2}+\lambda D^{\prime
	}D\right) ^{-1}D^{\prime } \\
	&=& I_{n}-\lambda  \left[ e_{1},M,e_{n}\right]^{\prime}   \left( I_{n-2}+\lambda \left( e_{1}e_{1}^{\prime
	}+M^{2}+e_{n}e_{n}^{\prime }\right) \right) ^{-1} \left[ e_{1},M,e_{n}\right] 
	\\
	&\sim &I_{n}-\lambda \left[ e_{1},M,e_{n}\right]^{\prime} 
	\left( I_{n-2}+\lambda M^{2}\right) ^{-1}\left[ e_{1},M,e_{n}\right],
\end{eqnarray*}
where the last line follows by 
\[
I_{n-2}+\lambda D^{\prime }D=I_{n-2}+\lambda \left( e_{1}e_{1}^{\prime
}+M^{2}+e_{n}e_{n}^{\prime }\right) \sim I_{n-2}+\lambda M^{2},
\]%
ignoring the end matrix corrections to the first and last diagonal elements as $%
n\rightarrow \infty .$ 
The $(n-2)\times (n-2)$ central matrix elements of $ S ( \lambda )$ are given by the symmetric matrix 
\begin{equation} 
S_{n-2}\left(
\lambda \right) =I_{n-2}-\lambda M\left( I_{n-2}+\lambda M^{2}\right) ^{-1}M. 
\label{tt80}
\end{equation} 
The eigenvalues of the symmetric tridiagonal Toeplitz matrix $M$ are well known \citep{noschese2013tridiagonal} to be given
by $\delta _{k}=-2 \left(   1- \cos \left(  k \pi /(n-1) \right)  \right)  ,$ $k=1,...,n-2,$
and so the eigenvalues of $M^{2}$ are $\left\{ \delta _{k}^{2}\right\}
_{k=1}^{n-2}.$ It follows that the eigenvalues of $S_{n-2}\left( \lambda
\right) $ are given by 
\[
\left\{ 1-\frac{\lambda \delta _{k}^{2}}{1+\lambda \delta _{k}^{2}}\right\}
_{k=1}^{n-2}=\left\{ \frac{1}{1+\lambda \delta _{k}^{2}}\right\}
_{k=1}^{n-2}.
\]%
We deduce that 
\begin{eqnarray*}
	\mathrm{tr} \left  (  B_m \right )
	&=&n-\mathrm{tr} \left( \left( I_{n}-S\left( \lambda \right) \right) ^{m}\right)
	\sim n- \mathrm{tr} \left( \left( I_{n-2}-S_{n-2}\left( \lambda \right) \right)
	^{m}\right)  \\
	&=&n-\sum_{k=1}^{n-2}\left (  1-\frac{1}{1+\lambda \delta _{k}^{2}}\right)
	^{m}=n-\sum_{k=1}^{n-2}\left (  \frac{\lambda \delta _{k}^{2}}{1+\lambda
		\delta _{k}^{2}}\right) ^{m} \\
	&=&n-\sum_{k=1}^{n-2}\left[ \frac{4\mu n^{4}\left( 1-\cos \left( \frac{k\pi 
		}{n-1}\right) ^{2}\right) }{1+4\mu n^{4}\left( 1-\cos \left( \frac{k\pi }{n-1%
		}\right) ^{2}\right) }\right] ^{m},
\end{eqnarray*}%
as required for (\ref{tt27}) when $\lambda=\mu n^4$. 	
\end{proof}

\bigskip

\begin{proof}[\textbf{Proof of Theorem \ref{thm:break}}]

We follow the line of argument used in the proof of Theorem 2. The
polynomial component $g_{n}\left( t\right) $ is handled in a similar way to
the proof of Theorem 2 but with complications arising from the break
separating the two segments $t<\tau _{0}$ and $t>\tau _{0}.$ We analyze these
two segments in turn first. Then, following those arguments, we consider limit behavior of the boosted filter at the break point $r_0$ itself. 

\noindent \textbf{(i) Lower Segment}

Over the lower segment we have $t=\lfloor nr\rfloor <\tau _{0}=\lfloor
nr_{0}\rfloor $ for $r<r_{0},$ so that for large enough $n$ we have 
\begin{equation}
t+2m=\lfloor nr\rfloor +2m<\tau _{0}=\lfloor nr_{0}\rfloor ,  \label{tt50}
\end{equation}%
which implies that $\mathbf{1} \left\{ t+2m < \tau _{0}\right\} =1$ since $%
m=o\left( n\right) .$ Repeated applications of the HP filter in the boosting
algorithm lead to%
\begin{equation}
\left( 1-G_{\lambda }\right) ^{m}\left( \frac{t}{n}\right) ^{j} \mathbf{1} 
\left\{ t<\tau _{0}\right\} =\left( \frac{L^{-2}\left( 1-L\right) ^{4}}{%
	L^{-2}\left( 1-L\right) ^{4}+1/\lambda }\right) ^{m}\left( \frac{t}{n}%
\right) ^{j} \mathbf{1} \left\{ t<\tau _{0}\right\} .  \label{tt51}
\end{equation}%
Applying the numerator operator gives%
\begin{eqnarray*}
	&&L^{-2m}\left( 1-L\right) ^{4m}\left( \frac{t}{n}\right) ^{j} \mathbf{1} 
	\left\{ t<\tau _{0}\right\} =\left( 1-L\right) ^{4m}\left( \frac{t+2m}{n}%
	\right) ^{j} \mathbf{1}  \left\{ t+2m<\tau _{0}\right\} \\
	&=&\left( 1-L\right) ^{4m}\left( \frac{t+2m}{n}\right) ^{j}=0,
\end{eqnarray*}%
because $\mathbf{1} \left\{ t+2m-k<\tau _{0}\right\} =1$ for all $k\geq 0$
in view of (\ref{tt50}) and because $\left( 1-L\right) ^{4m}t^{j}=0$ for all 
$j\leq J,$ which is finite, and $m\rightarrow \infty $ which ensures that $%
4m\geq J+1.$ Next, combining the numerator and denominator operators in (\ref%
{tt51}) we have
\begin{eqnarray}
&&\left( 1-G_{\lambda }\right) ^{m}\left( \frac{t}{n}\right) ^{j}=\left( 
\frac{L^{-2}\left( 1-L\right) ^{4}}{L^{-2}\left( 1-L\right) ^{4}+\frac{1}{%
		\lambda }}\right) ^{m}\left( \frac{t}{n}\right) ^{j}\mathbf{1} \left\{
t<\tau _{0}\right\}   \notag \\
&=&\frac{1}{\Gamma \left( m\right) }\int_{0}^{\infty }s^{m-1}e^{-s}e^{-s%
	\left[ L^{-2}\left( 1-L\right) ^{4}+\frac{1}{\lambda }\right] }ds\left(
1-L\right) ^{4m}\left( \frac{t+2m}{n}\right) ^{j}\mathbf{1} \left\{
t+2m<\tau _{0}\right\}   \notag \\
&=&\frac{1}{\Gamma \left( m\right) }\sum_{k=0}^{\infty }\frac{\left(
	-1\right) ^{k}}{k!}\int_{0}^{\infty }s^{m+k-1}e^{-s}ds\left[ L^{-2}\left(
1-L\right) ^{4}+\frac{1}{\lambda }\right] ^{k}\left( 1-L\right) ^{4m}\left( 
\frac{t+2m}{n}\right) ^{j}\mathbf{1} \left\{ t+2m<\tau _{0}\right\}  
\notag \\
&=&\sum_{k=0}^{\infty }\frac{\Gamma \left( m+k\right) \left( -1\right) ^{k}}{%
	\Gamma \left( m\right) k!}\left[ L^{-2}\left( 1-L\right) ^{4}+\frac{1}{%
	\lambda }\right] ^{k}\left( 1-L\right) ^{4m}\left( \frac{t+2m}{n}\right) ^{j}%
\mathbf{1} \left\{ t+2m<\tau _{0}\right\}   \notag \\
&=&\sum_{k=0}^{\infty }\frac{\left( m\right) _{k}\left( -1\right) ^{k}}{k!}%
\left[ L^{-2}\left( 1-L\right) ^{4}+\frac{1}{\lambda }\right] ^{k}\left(
1-L\right) ^{4m}\left( \frac{t+2m}{n}\right) ^{j}\mathbf{1} \left\{
t+2m<\tau _{0}\right\} .  \label{tt56}
\end{eqnarray}%
Observe that if $4m>J+1$
\begin{eqnarray}
&&\left[ L^{-2}\left( 1-L\right) ^{4}+\frac{1}{\lambda }\right] ^{k}\left(
1-L\right) ^{4m}\left( \frac{t+2m}{n}\right) ^{j}\mathbf{1} \left\{
t+2m<\tau _{0}\right\}   \notag \\
&=&\sum_{i=0}^{k}\binom{k}{i}\frac{1}{\lambda ^{k-i}}L^{-2i}\left(
1-L\right) ^{4\left[ m+i\right] }\left( \frac{t+2m}{n}\right) ^{j}%
\mathbf{1} \left\{ t+2m<\tau _{0}\right\}   \notag \\
&=&\sum_{i=0}^{k}\binom{k}{i}\frac{1}{\lambda ^{k-i}}\left( 1-L\right) ^{4%
	\left[ m+i\right] }\left( \frac{t+2\left[ m+i\right] }{n}\right) ^{j}%
\mathbf{1} \left\{ t+2\left[ m+i\right] <\tau _{0}\right\}   \notag \\
&=&0  \label{tt52}
\end{eqnarray}%
for all $j\leq J$ and all $k\leq k_{n}=\lfloor n( r_{0}-r)/ 4  \rfloor $ because then for large enough $n$ and with $m=o\left(
n\right) $ we have 
\begin{equation}
t+2\left[ m+i\right] \leq \lfloor nr\rfloor +2\left[ m+k\right] \leq \lfloor
nr\rfloor +2m+\lfloor n ( r_{0}-r) / 2 ) \rfloor +1<\lfloor
nr_{0}\rfloor =\tau _{0},  \label{tt55}
\end{equation}%
in which case $\mathbf{1} \left\{ t+2\left[ m+i\right] <\tau _{0}\right\}
=1$ for all $i\leq k\leq k_{n},$ just as in (\ref{tt50}). In this event, it
follows that%
\begin{equation}
\left( 1-L\right) ^{4\left[ m+i\right] }\left( \frac{t+2\left[ m+i\right] }{n%
}\right) ^{j}\mathbf{1} \left\{ t+2\left[ m+i\right] <\tau _{0}\right\}
=\left( 1-L\right) ^{4\left[ m+i\right] }\left( \frac{t+2\left[ m+i\right] }{%
	n}\right) ^{j}=0,  \label{tt57}
\end{equation}%
and then 
\begin{equation}
\left[ L^{-2}\left( 1-L\right) ^{4}+\frac{1}{\lambda }\right] ^{k}\left(
1-L\right) ^{4m}\left( \frac{t+2m}{n}\right) ^{j}\mathbf{1} \left\{
t+2m<\tau _{0}\right\} =0,  \label{tt54}
\end{equation}%
for all $k\leq k_{n}.$ Hence, the first $k_{n}$ terms of the series in (\ref%
{tt56}) are zero for large enough $n$ and so 
\begin{eqnarray*}
	&&\sum_{k=0}^{\infty }\frac{\left( m\right) _{k}\left( -1\right) ^{k}}{k!}%
	\left[ L^{-2}\left( 1-L\right) ^{4}+\frac{1}{\lambda }\right] ^{k}\left(
	1-L\right) ^{4m}\left( \frac{t+2m}{n}\right) ^{j}\mathbf{1} \left\{
	t+2m<\tau _{0}\right\}  \\
	&=&\sum_{k>k_{n}= \lfloor n( r_{0}-r)/ 4  \rfloor }^{\infty }%
	\frac{\left( m\right) _{k}\left( -1\right) ^{k}}{k!}\left[ L^{-2}\left(
	1-L\right) ^{4}+\frac{1}{\lambda }\right] ^{k}\left( 1-L\right) ^{4m}\left( 
	\frac{t+2m}{n}\right) ^{j}\mathbf{1} \left\{ t+2m<\tau _{0}\right\} 
\end{eqnarray*}%
Next, for $k>k_{n}=\lfloor n ( r_{0}-r)/4   \rfloor $,  using (\ref{tt57}) we
have
\begin{eqnarray*} 
	&&\left[ L^{-2}\left( 1-L\right) ^{4}+\frac{1}{\lambda }\right] ^{k}\left(
	1-L\right) ^{4m}\left( \frac{t+2m}{n}\right) ^{j}\mathbf{1} \left\{
	t+2m<\tau _{0}\right\}  \\
	&=&\left( 1-L\right) ^{4m}\sum_{i=0}^{k}\binom{k}{i}\frac{1}{\lambda ^{k-i}}%
	L^{-2i}\left( 1-L\right) ^{4i}\left( \frac{t+2m}{n}\right) ^{j}\mathbf{1} %
	\left\{ t+2m<\tau _{0}\right\}  \\
	&=&\left( 1-L\right) ^{4m}\sum_{i=k_{n}+1}^{k}\binom{k}{i}\frac{1}{\lambda
		^{k-i}}\left( 1-L\right) ^{4i}\left( \frac{t+2\left[ m+i\right] }{n}\right)
	^{j}\mathbf{1} \left\{ t+2\left[ m+i\right] <\tau _{0}\right\}  \\
	&=&\left( 1-L\right) ^{4m}\sum_{i=k_{n}+1}^{k}\binom{k}{i}\frac{1}{\lambda
		^{k-i}}\left( 1-L\right) ^{4i}\left( \frac{t+2\left[ m+i\right] }{n}\right)
	^{j}\mathbf{1} \left\{ t+2\left[ m+i\right] <\tau _{0}\right\} 
\end{eqnarray*}%
for all $j\leq J.$ When $i\geq i_{n}:=\lfloor n ( r_{0}-r)/2 \rfloor $ we have 
\begin{equation*}
t+2\left[ m+i\right] \geq \lfloor nr\rfloor +2\left[ m+
 \lfloor n ( r_{0}-r )/2  \rfloor 
	\right] >\lfloor nr\rfloor +\lfloor n\left(
r_{0}-r\right) \rfloor +m>\lfloor nr_{0}\rfloor =\tau _{0},
\end{equation*}%
which implies $\mathbf{1} \left\{ t+2\left[ m+i\right] <\tau _{0}\right\}
=0.$ However, even in this case, application of powers of the differencing
operator $\left( 1-L\right) ^{4i}$ leads to adjacent integer values for
which $\left\{ t+2\left[ m+i\right] -\left( p-1\right) >\tau _{0}\right\} $
and $\left\{ t+2\left[ m+i\right] -p<\tau _{0}\right\} $ for some positive
integer $p.$ Hence, 
\begin{eqnarray*}
	&&\left( 1-L\right) ^{4i}\left[ \left( \frac{t+2\left[ m+i\right] }{n}%
	\right) ^{j}\mathbf{1} \left\{ t+2\left[ m+i\right] <\tau _{0}\right\} %
	\right]  \\
	&=&\sum_{p=0}^{4i}\binom{4i}{p}\left( -1\right) ^{p}L^{p}\left[ \left( \frac{%
		t+2\left[ m+i\right] }{n}\right) ^{j}\mathbf{1} \left\{ t+2\left[ m+i%
	\right] <\tau _{0}\right\} \right]  \\
	&=&\sum_{p=0}^{4i}\binom{4i}{p}\left( -1\right) ^{p}\left( \frac{t+2\left[
		m+i\right] -p}{n}\right) ^{j}\mathbf{1} \left\{ t+2\left[ m+i\right]
	-p<\tau _{0}\right\}  \\
	&=&\sum_{p=P_{\left( t\right) }}^{4i}\binom{4i}{p}\left( -1\right)
	^{k}\left( \frac{t+2\left[ m+i\right] -p}{n}\right) ^{j},
\end{eqnarray*}
where $P_{\left( t\right) }$ is the smallest $p$ for which $\left\{ 0\leq t+2%
\left[ m+i\right] -p<\tau _{0}\right\} ,$ in which case $\mathbf{1} %
\left\{ t+2\left[ m+i\right] -p<\tau _{0}\right\} =1$ for $p\geq P_{\left(
	t\right) }.$ We then deduce that%
\begin{eqnarray*}
	&&\sum_{k=0}^{\infty }\frac{\left( m\right) _{k}\left( -1\right) ^{k}}{k!}%
	\left[ L^{-2}\left( 1-L\right) ^{4}+\frac{1}{\lambda }\right] ^{k}\left(
	1-L\right) ^{4m}\left( \frac{t+2m}{n}\right) ^{j}\mathbf{1} \left\{
	t+2m<\tau _{0}\right\}  \\
	&=&\sum_{k>k_{n}= \lfloor n ( r_{0}-r )/4  \rfloor }^{\infty }%
	\frac{\left( m\right) _{k}\left( -1\right) ^{k}}{k!}\left[ L^{-2}\left(
	1-L\right) ^{4}+\frac{1}{\lambda }\right] ^{k}\left( 1-L\right) ^{4m}\left( 
	\frac{t+2m}{n}\right) ^{j}\mathbf{1} \left\{ t+2m<\tau _{0}\right\}  \\
	&=&\sum_{k>k_{n}=\lfloor n ( r_{0}-r )/4  \rfloor }^{\infty }%
	\frac{\left( m\right) _{k}\left( -1\right) ^{k}}{k!}\left( 1-L\right)
	^{4m}\sum_{i=k_{n}+1}^{k}\binom{k}{i}\frac{1}{\lambda ^{k-i}}\left(
	1-L\right) ^{4i}\left( \frac{t+2\left[ m+i\right] }{n}\right) ^{j}%
	\mathbf{1} \left\{ t+2\left[ m+i\right] <\tau _{0}\right\}  \\
	&=&\sum_{k>k_{n}=\lfloor n ( r_{0}-r )/4  \rfloor }^{\infty }%
	\frac{\left( m\right) _{k}\left( -1\right) ^{k}}{k!}\sum_{i=k_{n}+1}^{k}%
	\binom{k}{i}\frac{1}{\lambda ^{k-i}}\left( 1-L\right) ^{4m}\left[
	\sum_{p=P_{\left( t\right) }}^{4i}\binom{4i}{p}\left( -1\right) ^{k}\left( 
	\frac{t+2\left[ m+i\right] -p}{n}\right) ^{j}\right]  \\
	&=&0,
\end{eqnarray*}%
because $\left( 1-L\right) ^{4m}\left( t+2\left[ m+i\right] -p\right) ^{j}=0$
for all $j\leq J$ since $m\rightarrow \infty .$ This proves that on the
lower segment $t=\lfloor nr\rfloor <\tau _{0}=\lfloor nr_{0}\rfloor $ with $%
r<r_{0}$ we have 
\begin{equation*}
\left( 1-G_{\lambda }\right) ^{m}\left( \frac{t}{n}\right) ^{j}\mathbf{1} %
\left\{ t<\tau _{0}\right\} \rightarrow 0,\text{ \ as }m,n\rightarrow \infty
.
\end{equation*}%

\noindent \textbf{(ii) Upper Segment}

Treatment of the segment $t > \tau _{0}$ is similarly complicated because
application of powers of the differencing operator $\left( 1-L\right) $ to
the segmented polynomial $(t/n) ^{j}\mathbf{1} %
\left\{ t\geq \tau _{0}\right\} $ involves computations on either side of
the break point that occurs at $t=\tau _{0}$. \ Proceeding as in (\ref{tt56}%
) we have%
\begin{eqnarray}
&&\left( 1-G_{\lambda }\right) ^{m}\left( \frac{t}{n}\right) ^{j}=\left( 
\frac{L^{-2}\left( 1-L\right) ^{4}}{L^{-2}\left( 1-L\right) ^{4}+\frac{1}{%
		\lambda }}\right) ^{m}\left( \frac{t}{n}\right) ^{j}\mathbf{1} \left\{
t>\tau _{0}\right\}   \notag \\
&=&\frac{1}{\Gamma \left( m\right) }\int_{0}^{\infty }s^{m-1}e^{-s}e^{-s%
	\left[ L^{-2}\left( 1-L\right) ^{4}+\frac{1}{\lambda }\right] }ds\left(
1-L\right) ^{4m}\left( \frac{t+2m}{n}\right) ^{j}\mathbf{1} \left\{
t+2m>\tau _{0}\right\}   \notag \\
&=&\frac{1}{\Gamma \left( m\right) }\sum_{k=0}^{\infty }\frac{\left(
	-1\right) ^{k}}{k!}\int_{0}^{\infty }s^{m+k-1}e^{-s}ds\left[ L^{-2}\left(
1-L\right) ^{4}+\frac{1}{\lambda }\right] ^{k}\left( 1-L\right) ^{4m}\left( 
\frac{t+2m}{n}\right) ^{j}\mathbf{1} \left\{ t+2m>\tau _{0}\right\}  
\notag \\
&=&\sum_{k=0}^{\infty }\frac{\left( m\right) _{k}\left( -1\right) ^{k}}{k!}%
\left( 1-L\right) ^{4m}\left[ L^{-2}\left( 1-L\right) ^{4}+\frac{1}{\lambda }%
\right] ^{k}\left( \frac{t+2m}{n}\right) ^{j}\mathbf{1} \left\{ t+2m>\tau
_{0}\right\} .  \label{tt70}
\end{eqnarray}%
Next observe that if $4m>J+1$
\begin{eqnarray}
&&\left( 1-L\right) ^{4m}\left[ L^{-2}\left( 1-L\right) ^{4}+\frac{1}{%
	\lambda }\right] ^{k}\left( \frac{t+2m}{n}\right) ^{j}\mathbf{1} \left\{
t+2m>\tau _{0}\right\}   \notag \\
&=&\sum_{i=0}^{k}\binom{k}{i}\frac{1}{\lambda ^{k-i}}L^{-2i}\left(
1-L\right) ^{4\left[ m+i\right] }\left( \frac{t+2m}{n}\right) ^{j}%
\mathbf{1} \left\{ t+2m>\tau _{0}\right\}   \notag \\
&=&\sum_{i=0}^{k}\binom{k}{i}\frac{1}{\lambda ^{k-i}}\left( 1-L\right) ^{4%
	\left[ m+i\right] }\left( \frac{t+2\left[ m+i\right] }{n}\right) ^{j}%
\mathbf{1} \left\{ t+2\left[ m+i\right] >\tau _{0}\right\}   \notag \\
&=&0,  \label{tt71}
\end{eqnarray}%
for all $j\leq J$ and for all $k\leq k_{n}^{\varepsilon }=\lfloor 
n ( r-r_{0}-\varepsilon) /4   \rfloor $ for some small $\varepsilon
>0$ such that $r>r_{0}+\varepsilon .$ The final line (\ref{tt71}) follows
because for large enough $n$ and with $m=o\left( n\right) $ we have 
\begin{equation}
t+2\left[ m+i\right] -4\left[ m+i\right] \geq \lfloor nr\rfloor +2m-4\left[
m+k_{n}^{\varepsilon }\right] =\lfloor nr\rfloor -2m
- 4 \left \lfloor n\left( \frac{%
	r-r_{0}-\varepsilon }{4}\right) \right \rfloor 
	>\lfloor nr_{0}\rfloor =\tau _{0},
\label{tt72}
\end{equation}%
so that $\mathbf{1} \left\{ t+2\left[ m+i\right] -4\left[ m+i\right]
>\tau _{0}\right\} =1$ for all $i\leq k\leq k_{n}^{\varepsilon }$. It then follows that%
\begin{equation}
\left( 1-L\right) ^{4\left[ m+i\right] }\left( \frac{t+2\left[ m+i\right] }{n%
}\right) ^{j}\mathbf{1} \left\{ t+2\left[ m+i\right] >\tau _{0}\right\}
=\left( 1-L\right) ^{4\left[ m+i\right] }\left( \frac{t+2\left[ m+i\right] }{%
	n}\right) ^{j} =0 ,  \label{tt75}
\end{equation}%
when $4m\geq J+1.$ We deduce that 
\begin{eqnarray}
&&\sum_{k=0}^{\infty }\frac{\left( m\right) _{k}\left( -1\right) ^{k}}{k!}%
\left( 1-L\right) ^{4m}\left[ L^{-2}\left( 1-L\right) ^{4}+\frac{1}{\lambda }%
\right] ^{k}\left( \frac{t+2m}{n}\right) ^{j}\mathbf{1} \left\{ t+2m>\tau
_{0}\right\}   \notag \\
&=&\sum_{k=k_{n}^{\varepsilon }}^{\infty }\frac{\left( m\right) _{k}\left(
	-1\right) ^{k}}{k!}\left( 1-L\right) ^{4m}\left[ L^{-2}\left( 1-L\right)
^{4}+\frac{1}{\lambda }\right] ^{k}\left( \frac{t+2m}{n}\right) ^{j}%
\mathbf{1} \left\{ t+2m>\tau _{0}\right\} .  \label{tt74}
\end{eqnarray}%
Next, for $k>k_{n}^{\varepsilon }=\lfloor n ( r_{0}-r-\varepsilon ) / 4 \rfloor $ and in view of (\ref{tt75}), we have 
\begin{eqnarray}
&&\left[ L^{-2}\left( 1-L\right) ^{4}+\frac{1}{\lambda }\right] ^{k}\left(
1-L\right) ^{4m}\left( \frac{t+2m}{n}\right) ^{j}\mathbf{1} \left\{
t+2m>\tau _{0}\right\}   \notag \\
&=&\sum_{i=0}^{k}\binom{k}{i}\frac{1}{\lambda ^{k-i}}\left( 1-L\right) ^{4%
	\left[ m+i\right] }\left( \frac{t+2\left[ m+i\right] }{n}\right) ^{j}%
\mathbf{1} \left\{ t+2\left[ m+i\right] >\tau _{0}\right\}   \notag \\
&=&\sum_{i=k_{n}^{\varepsilon }+1}^{k}\binom{k}{i}\frac{1}{\lambda ^{k-i}}%
\left( 1-L\right) ^{4\left[ m+i\right] }\left( \frac{t+2\left[ m+i\right] }{n%
}\right) ^{j}\mathbf{1} \left\{ t+2\left[ m+i\right] >\tau _{0}\right\} .
\label{tt73}
\end{eqnarray}%
When $i\geq i_{n}^{\varepsilon }:=\lfloor n  ( r_{0}-r+\varepsilon )/2 \rfloor $  
\begin{equation*}
t+2\left[ m+i\right] \geq \lfloor nr\rfloor +
2\left[ m+ \left \lfloor n\left( \frac{%
	r_{0}-r+\varepsilon }{2}\right) \right \rfloor \right] >\lfloor nr\rfloor +\lfloor
n\left( r_{0}-r\right) \rfloor +m>\lfloor nr_{0}\rfloor =\tau _{0},
\end{equation*}%
which implies that $\mathbf{1} \left\{ t+2\left[ m+i\right] >\tau _{0}\right\}
=1.$ For large $k$ and $i$ in (\ref{tt73}) application of the operator $%
\left( 1-L\right) ^{4\left[ m+i\right] }$ involves powers of the
differencing operator $\left( 1-L\right) $ that lead to adjacent integer
values for which $\left\{ t+2\left[ m+i\right] -\left( p-1\right) >\tau
_{0}\right\} $ and $\left\{ t+2\left[ m+i\right] -p<\tau _{0}\right\} $ for
some positive integer $p.$ In this event, 
\begin{eqnarray*}
	&&\left( 1-L\right) ^{4i}\left[ \left( \frac{t+2\left[ m+i\right] }{n}%
	\right) ^{j}\mathbf{1} \left\{ t+2\left[ m+i\right] >\tau _{0}\right\} %
	\right]  \\
	&=&\sum_{p=0}^{4i}\binom{4i}{p}\left( -1\right) ^{p}L^{p}\left[ \left( \frac{%
		t+2\left[ m+i\right] }{n}\right) ^{j}\mathbf{1} \left\{ t+2\left[ m+i%
	\right] >\tau _{0}\right\} \right]  \\
	&=&\sum_{p=0}^{4i}\binom{4i}{p}\left( -1\right) ^{p}\left( \frac{t+2\left[
		m+i\right] -p}{n}\right) ^{j}\mathbf{1} \left\{ t+2\left[ m+i\right]
	-p>\tau _{0}\right\}  \\
	&=&\sum_{p=0}^{P_{\left( t\right) }}\binom{4i}{p}\left( -1\right) ^{k}\left( 
	\frac{t+2\left[ m+i\right] -p}{n}\right) ^{j},
\end{eqnarray*}%
where $P_{\left( t\right) }$ is the largest $p$ for which $\left\{ t+2\left[
m+i\right] -p>\tau _{0}\right\} ,$ in which case $\mathbf{1} \left\{ t+2%
\left[ m+i\right] -p<\tau _{0}\right\} =1$ for $p\leq P_{\left( t\right) }$
and $\mathbf{1} \left\{ t+2\left[ m+i\right] -p<\tau _{0}\right\} =0$ for
$p>P_{\left( t\right) }.$ We then deduce that for large enough $n$%
\begin{eqnarray*}
	&&\sum_{k=0}^{\infty }\frac{\left( m\right) _{k}\left( -1\right) ^{k}}{k!}%
	\left[ L^{-2}\left( 1-L\right) ^{4}+\frac{1}{\lambda }\right] ^{k}\left(
	1-L\right) ^{4m}\left( \frac{t+2m}{n}\right) ^{j}\mathbf{1} \left\{
	t+2m>\tau _{0}\right\}  \\
	&=&\sum_{k>k_{n}^{\varepsilon }=\lfloor n ( r_{0}-r )/4  \rfloor 
	}^{\infty }\frac{\left( m\right) _{k}\left( -1\right) ^{k}}{k!}\left[
	L^{-2}\left( 1-L\right) ^{4}+\frac{1}{\lambda }\right] ^{k}\left( 1-L\right)
	^{4m}\left( \frac{t+2m}{n}\right) ^{j}\mathbf{1} \left\{ t+2m>\tau
	_{0}\right\}  \\
	&=&\sum_{k>k_{n}^{\varepsilon }=\lfloor n ( r_{0}-r )/4  \rfloor
	 }^{\infty }\frac{\left( m\right) _{k}\left( -1\right) ^{k}}{k!}%
	\left( 1-L\right) ^{4m}\sum_{i=k_{n}^{\varepsilon }+1}^{k}\binom{k}{i}\frac{1%
	}{\lambda ^{k-i}}\left( 1-L\right) ^{4i}\left( \frac{t+2\left[ m+i\right] }{n%
	}\right) ^{j}\mathbf{1} \left\{ t+2\left[ m+i\right] >\tau _{0}\right\} 
	\\
	&=&\sum_{k>k_{n}^{\varepsilon }=\lfloor n ( r_{0}-r )/4  \rfloor
	 }^{\infty }\frac{\left( m\right) _{k}\left( -1\right) ^{k}}{k!}%
	\sum_{i=k_{n}^{\varepsilon }+1}^{k}\binom{k}{i}\frac{1}{\lambda ^{k-i}}%
	\left( 1-L\right) ^{4m}\left[ \sum_{p=0}^{P_{\left( t\right) }}\binom{4i}{p}%
	\left( -1\right) ^{k}\left( \frac{t+2\left[ m+i\right] -p}{n}\right) ^{j}%
	\right]  \\
	&=&0,
\end{eqnarray*}%
because $\left( 1-L\right) ^{4m}\left( t+2\left[ m+i\right] -p\right) ^{j}=0$
for all $j\leq J$ since $m\rightarrow \infty .$ This proves that on the
upper segment $t=\lfloor nr\rfloor >\tau _{0}=\lfloor nr_{0}\rfloor $ with $%
r>r_{0}$ we have 
\begin{equation*}
\left( 1-G_{\lambda }\right) ^{m}\left( t/n \right) ^{j}\mathbf{1} %
\left\{ t<\tau _{0}\right\} \rightarrow 0,\text{ \ as }m,n\rightarrow \infty.
\end{equation*}

By combining the results for both segments $t<\tau _{0}$ and $t > \tau _{0}$, it follows that the boosted filter
eventually reproduces accurately the polynomial component $g_{n}\left(
t\right) $ for all $t=\lfloor nr\rfloor $ with $r\ne r_{0}.$  The stochastic trend
component $x_{t}^{0}$ is treated in the same manner as the proof of Theorem
1, showing that as $m,n\rightarrow \infty $ and $\lambda =\mu n^{4}$ for any
fixed $\mu >0$, the boosted filter accurately captures the stochastic trend.
It then follows that 
\begin{equation*}
n^{-1/2} \widehat{f}_{\left\lfloor nr\right\rfloor }^{\left( m\right) }  
\rightsquigarrow g\left( r\right) +B\left( r\right) ,\text{\ for all }%
r\not=r_{0},
\end{equation*}%
when $ 1/m + m/n \rightarrow 0$ as $n\rightarrow \infty,$ giving the required result for all $r \ne r_0$.

\bigskip
	
\noindent \textbf{(iii) Limit theory at the Break Point}

From the above analysis, we know that as $m,n\rightarrow \infty $ with $%
m/n \rightarrow 0$  
\begin{equation}
n^{-1/2}\widehat{f}_{\left\lfloor nr\right\rfloor }^{\left( m\right)
}\rightsquigarrow B_{g}\left( r\right) =g\left( r\right) +B\left( r\right) ,%
\text{\ for all }r\not=r_{0}.  \label{tt79}
\end{equation}%
In particular, 
\begin{equation*}
\left[ 1-\left( 1-G_{\lambda }\right) ^{m}\right] g_{n}\left( t\right) 
\mathbf{1} \left\{ t=\lfloor nr\rfloor <\lfloor nr_{0}\rfloor \right\}
\rightarrow g\left( r\right) \text{ \ for\ all }r<r_{0},
\end{equation*}%
and 
\begin{equation*}
\left[ 1-\left( 1-G_{\lambda }\right) ^{m}\right] g_{n}\left( t\right) 
\mathbf{1} \left\{ t=\lfloor nr\rfloor >\lfloor nr_{0}\rfloor \right\}
\rightarrow g\left( r\right) \text{ \ for\ all }r>r_{0}.
\end{equation*}%
The limit function $g\left( r\right) $ of $g_{n}\left( t\right) $ as $%
n\rightarrow \infty $ has the form 
\begin{equation*}
g\left( r\right) =\left\{ 
\begin{array}{cc}
\alpha ^{0}+\beta _{1}^{0}r+...+\beta _{J}^{0}r^{J} & r<r_{0} \\ 
\alpha ^{1}+\beta _{1}^{1}r+...+\beta _{J}^{1}r^{J} & r\geq r_{0}%
\end{array}%
\right. 
\end{equation*}%
so that the limit function $B_{g}$ in (\ref{tt79}) satisfies $%
\lim_{r\nearrow r_{0}^-}B_{g}\left( r\right) =B_{g}\left( r_{0}^-\right)
=g\left( r_{0}^-\right) +B\left( r_{0}\right) $ because $B\left( r\right) $
is continuous and $g\left( r\right) $ is continuous for all $r<r_{0}$ and
has finite left limit $g\left( r_{0}^-\right) .$ Similarly, $\lim_{r\searrow
	r_{0}^+}B_{g}\left( r\right) =B_{g}\left( r_{0}^+\right) =g\left( r_{0}\right)
+B\left( r_{0}\right) $ because $g\left( r\right) $ is continuous for all $%
r>r_{0}$ and has finite right limit $g\left( r_{0}^+\right) =g\left(
r_{0}\right) .$ At the break point $\tau _{0}=\lfloor nr_{0}\rfloor $
itself, we can write the indicator $\mathbf{1} \left\{ t=\tau
_{0}\right\} $ as the simple symmetric average of the limits of the left and right
side indicators 
\begin{equation*}
\mathbf{1} \left\{ t=\lfloor nr\rfloor =\lfloor nr_{0}\rfloor \right\} =%
\frac{1}{2}\left[ \lim_{r_{\ast }\nearrow r_{0}}\mathbf{1} \left\{
t=\lfloor nr\rfloor \leq \lfloor nr_{\ast }\rfloor \right\} +\lim_{r^{\ast
	}\searrow r_{0}}\mathbf{1} \left\{ t=\lfloor nr\rfloor \geq \lfloor
nr^{\ast }\rfloor \right\} \right] .
\end{equation*}%
So by virtue of the continuity of the limit function $g\left( r\right) $ on the right and left sides of $r=r_{0}$, the existence of the right and left side limits of $g\left( r\right)$ at $r_0$,  and the asymptotic symmetry\footnote{%
	The asymptotic symmetry of the filter follows from the asymptotic symmetric
	Toeplitz representation of the filter given in (\ref{tt80}). Importantly, this
	asymptotic symmetry holds away from the end points and is valid therefore
	for an interior break point $\tau _{0}=\lfloor nr_{0}\rfloor $ when $%
	r_{0}\in \left( 0,1\right) .$} of the action of the filter about the interior point $%
\tau _{0}=\lfloor nr_{0}\rfloor $ when $r_{0}\in \left( 0,1\right) $ we
deduce that  
\begin{equation*}
n^{-1/2}\widehat{f}_{\left\lfloor nr_{0}\right\rfloor }^{\left( m\right)
}\rightsquigarrow B_{g}\left( r_{0}\right)
 =\frac{1}{2}\left[ g\left(
r_{0}^-\right) +g\left( r_{0}^+\right) \right] +B\left( r_{0}\right) ,
\end{equation*}%
giving the stated result.	
\end{proof}


\setcounter{table}{0} 
\setcounter{figure}{0} 

\renewcommand{\thefigure}{B\arabic{figure}} 
\renewcommand{\thetable}{B\arabic{table}} 

\section{Graphical Demonstrations}\label{sec:graphic}

This section displays additional graphs to illustrate the performance characteristics and empirical behavior of the trend determination methods
discussed in the main text, specifically the HP filter, the bHP-BIC filter, and the fitted AR(4) autoregression. The bHP-ADF filter generally lies between the conventional HP and bHP-BIC versions of the filter and is omitted to avoid overcrowding the graphics.

\subsection{Decomposition of Figure \ref{fig1}}\label{subsec:fig1-decomp}

\begin{figure}[htbp]
 \centering
  \subfloat[HP]{
  	\includegraphics[width=.7\linewidth]{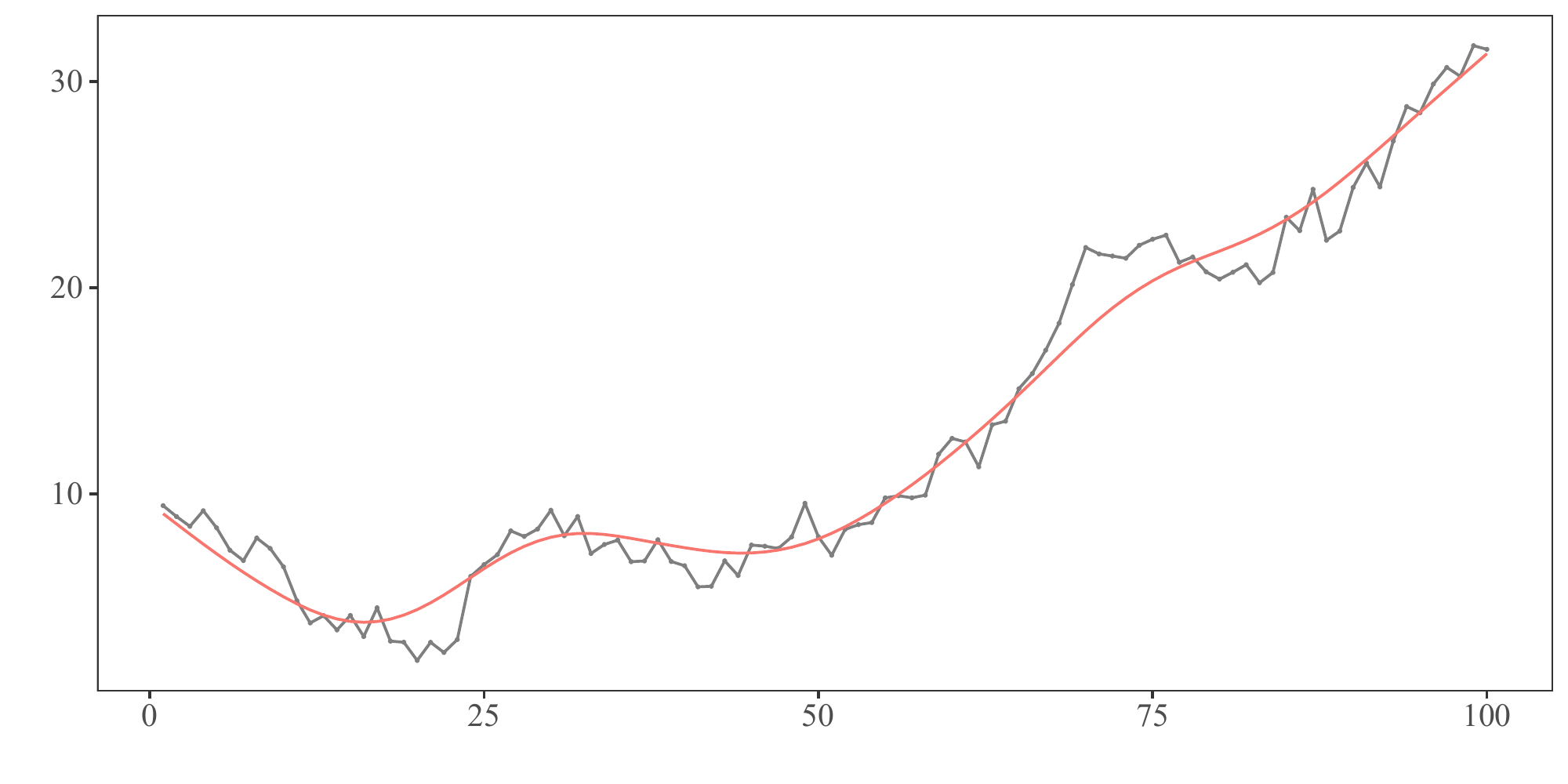}  
  }

  \subfloat[bHP-BIC (orange, data-determined $m=10$) and bHP with $m=128$ (dark blue)]{
  	\includegraphics[width=.7\linewidth]{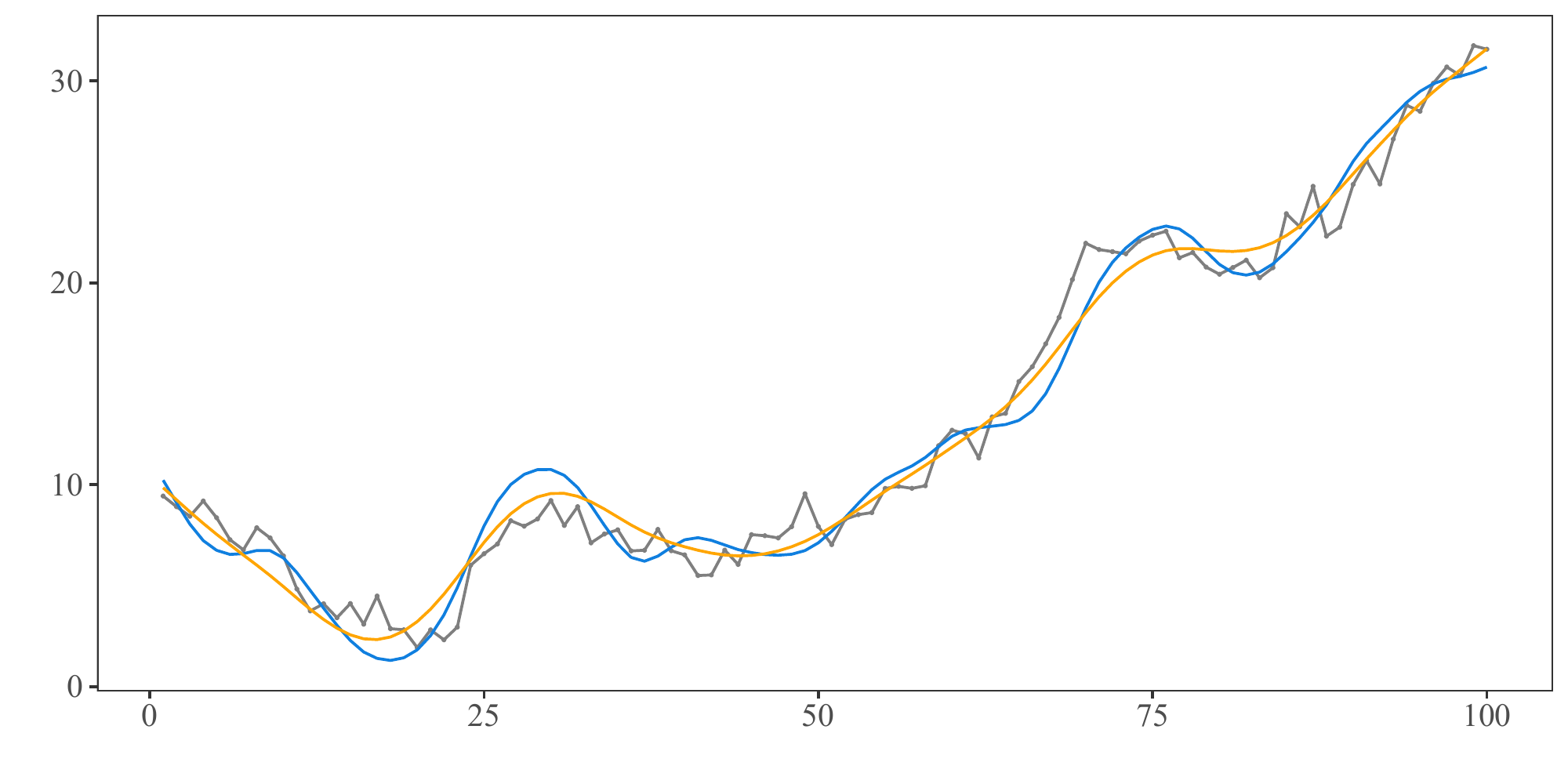}  
  }

  \subfloat[AR(4) fitted trend (violet)]{
    \includegraphics[width=.7\linewidth]{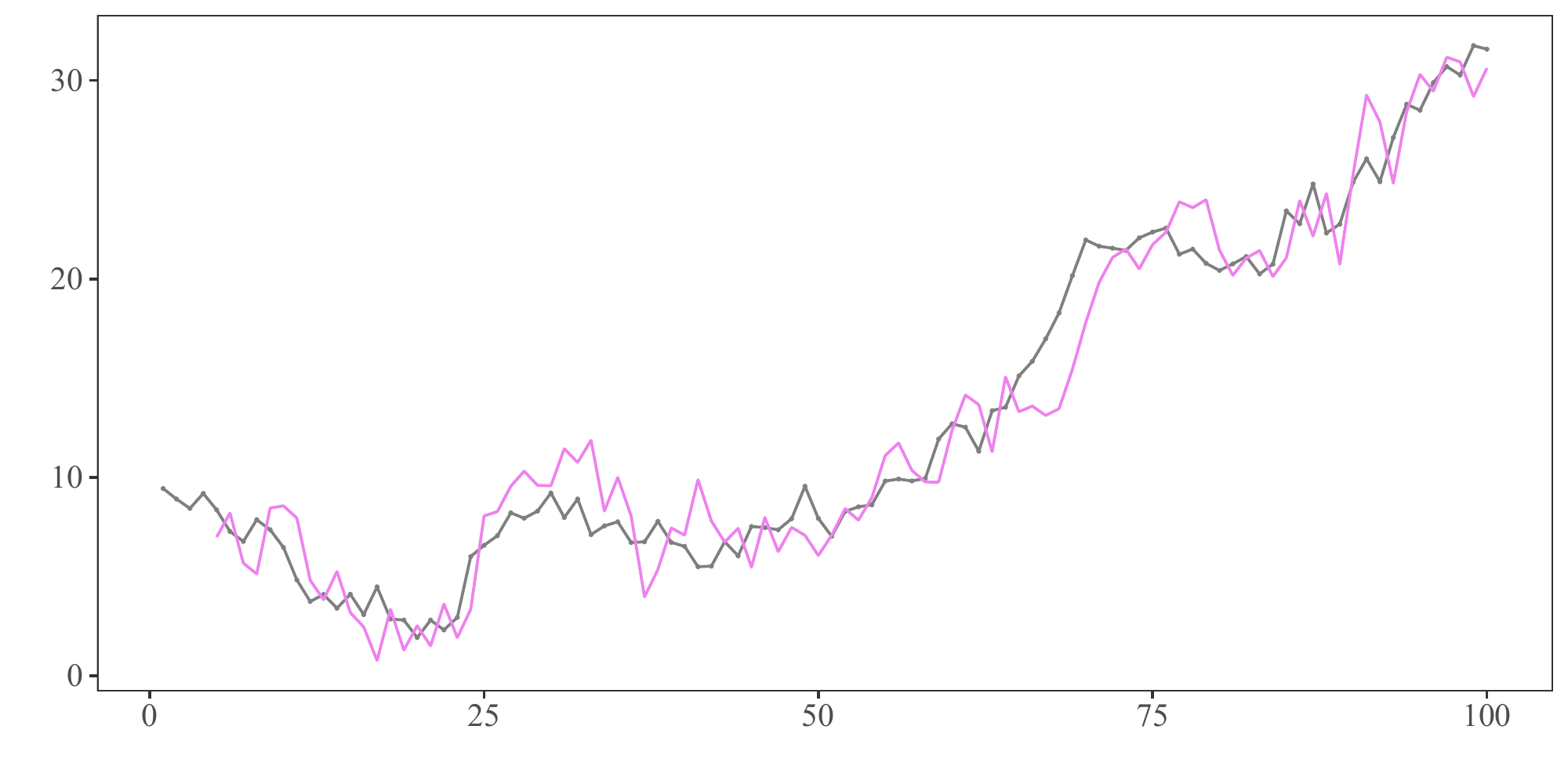} 
    }

\caption{ The same underlying trend (grey) in panels (a), (b), and (c) is shown against the HP trend in (a), bHP fitted trends in (b), and AR(4) fitted trend in (c). The true trend (grey) is the same as in the middle panel of Figure \ref{fig1}. }
\label{fig:fig1-decomp}
\end{figure}

To assist visualization of trend capturing performance by filters and autoregression, Figure \ref{fig:fig1-decomp} expands on Figure \ref{fig1} by displaying the trajectory (shown in grey) of a time series composed of (i) a stochastic trend, superposed with (ii) a deterministic fourth order time polynomial trend, and (iii) a stationary ARMA(1,1) disturbance --- see equation (\ref{eq:simulate_x0}). The trend function is shown in successive panels against the HP filter (shown in red in panel (a)), the bHP filter (shown in orange in panel (b)), and an AR(4) fitted trend (shown in violet in panel (c)). In the middle panel (b), the BIC selector determined the iteration number $m=10$ that led to the bHP-BIC fitted trend (shown in orange), which appears to track the underlying trend curve (shown in grey) somewhat more faithfully  than when many more iterations ($m=128$) of the filter are employed, as shown by the dark blue curve for $m=128$. The third (lower) panel (c) of Figure \ref{fig:fig1-decomp} displays the fitted trend (shown in violet) from an AR(4) autoregression for comparison. The AR fitted trend wanders around the true trend curve (in grey), showing greater susceptibility to noise and failure to capture the trend character of the deterministic fourth order time polynomial. This failure is manifested clearly in the systematic deviations of the AR(4) trend from the underlying trend that are apparent in the figure and are quantified in the MSE reported in Table \ref{tab:fig1_MSE} in the text.  
\vspace{2mm}

\subsection{Illustration of Robustness to Tuning Parameter Choices} \label{simu_lambda_robust}

\begin{figure}[htbp]
\begin{centering}
\subfloat[DGP 1 with $\lambda=6400$]{\includegraphics[scale=0.3]{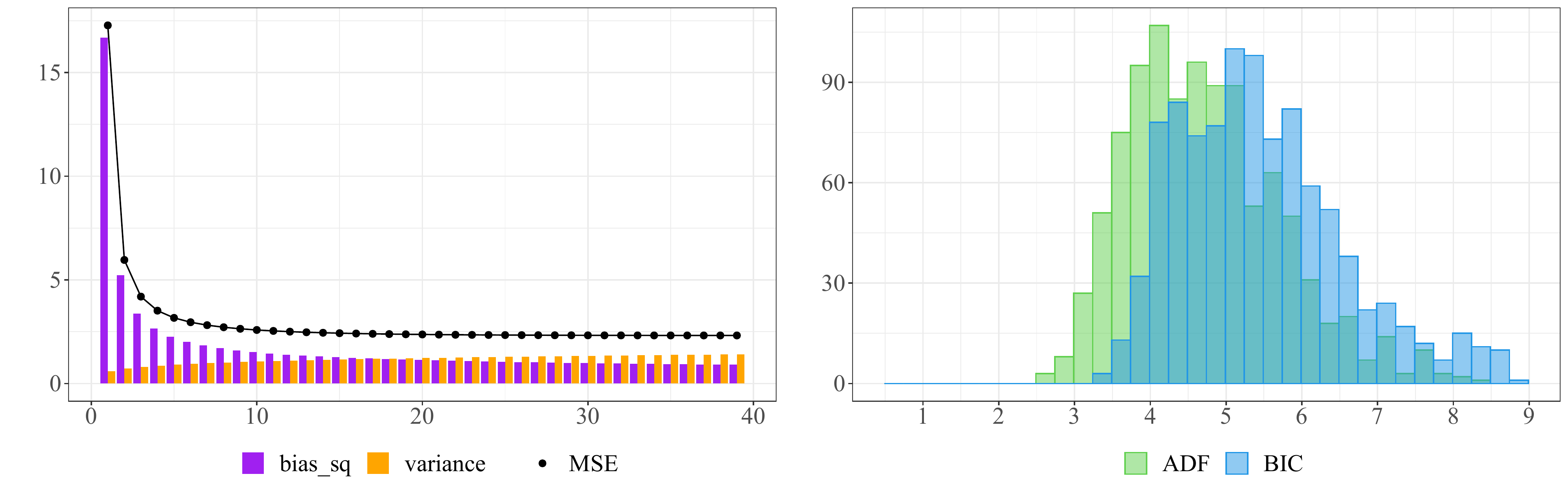}

}
\par\end{centering}
\begin{centering}
\subfloat[DGP 2 with $\lambda=6400$]{\includegraphics[scale=0.3]{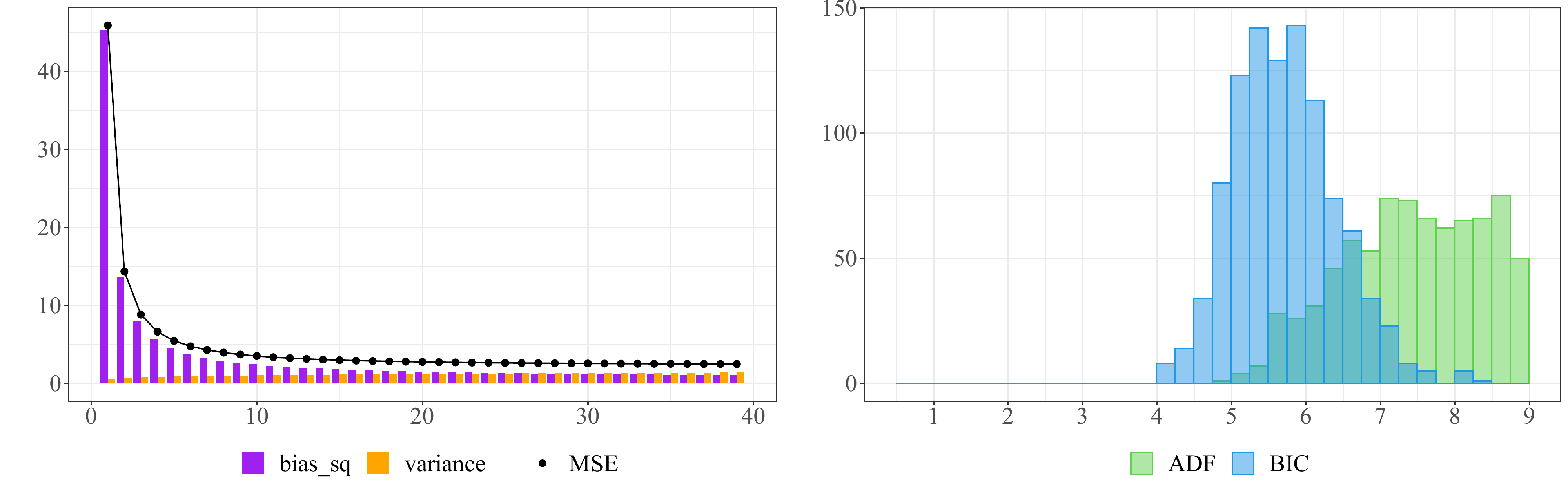}

}
\par\end{centering}
\begin{centering}
\subfloat[DGP 1 with $\lambda=400$]{\includegraphics[scale=0.3]{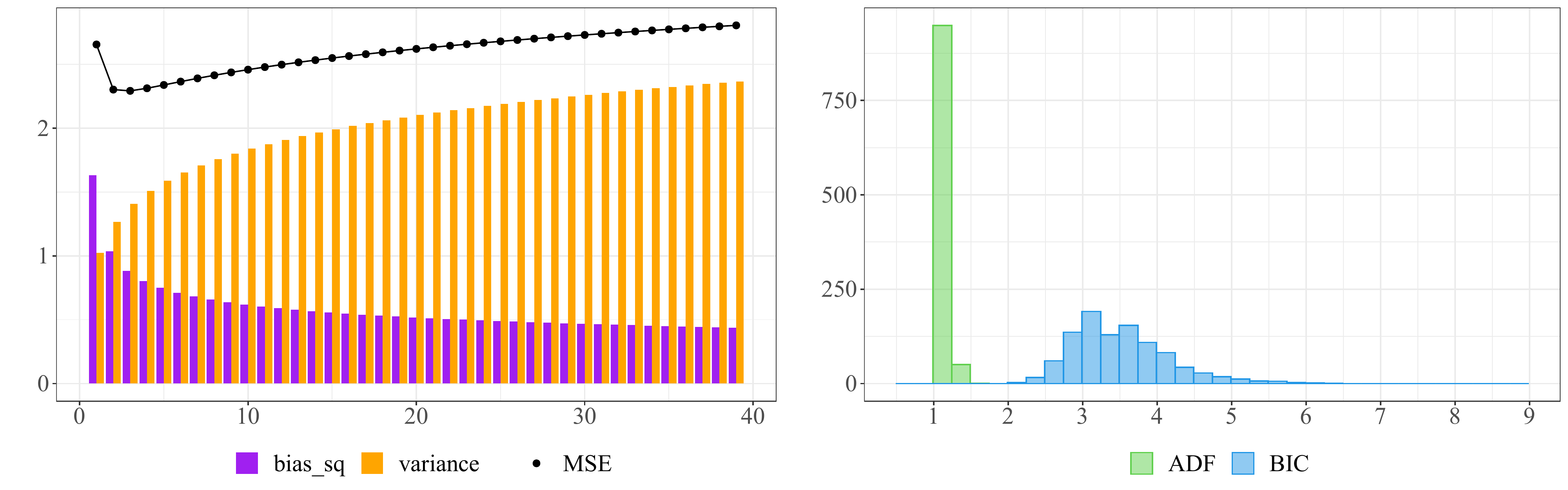}

}
\par\end{centering}
\begin{centering}
\subfloat[DGP 2 with $\lambda=400$]{\includegraphics[scale=0.3]{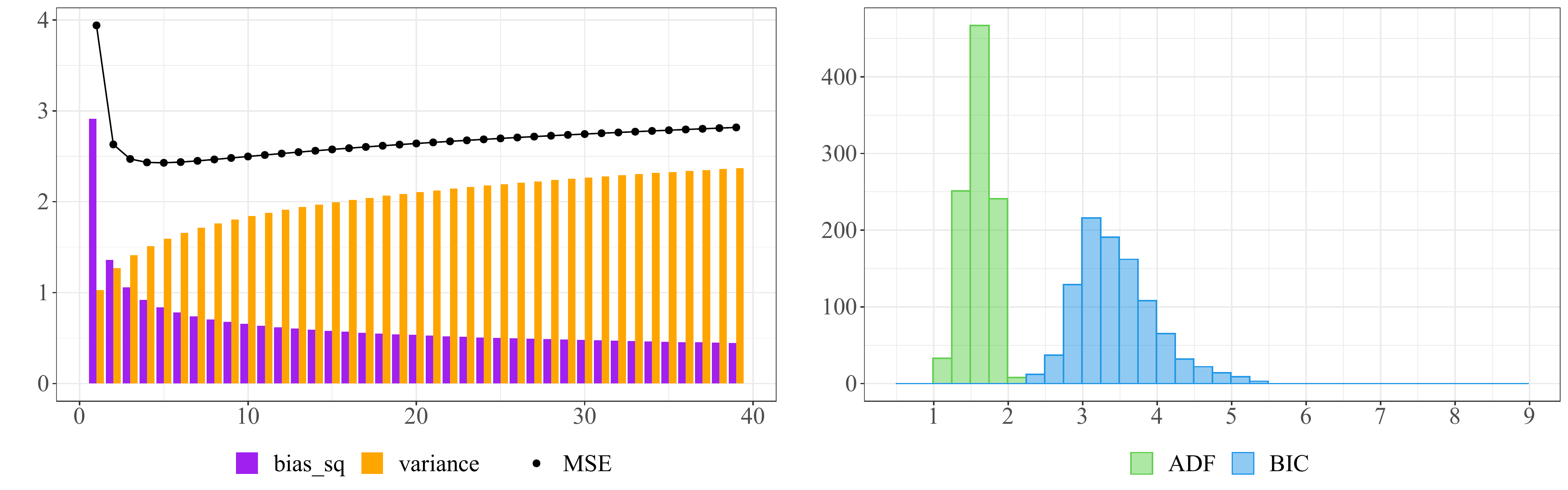}

}
\par\end{centering}
\caption{\label{fig:initial_lambda} 
Bias-variance tradeoffs for the bHP filter with tuning parameters $\lambda=6400$ and  $\lambda=400$ used as alternatives to the benchmark value $\lambda = 1600$. These graphs are the counterparts of those in Figure \ref{fig:bias-variance}.}
\end{figure}

This section analyzes the robustness of the bHP filter to the setting of the HP tuning parameter $\lambda$ that initiates the boosting iterations. A range of values for $\lambda$ are employed around the standard benchmark choice for the quarterly observation frequency. We use the same DGP 1 and DGP 2 as those in Section \ref{subsec:bias-variance} with settings $\lambda=6400$ (4 times the benchmark $1600$) and $\lambda=400$ (1/4 of 1600).
The results are reported in Figure \ref{fig:initial_lambda}, which serves as a counterpart to  Figure \ref{fig:bias-variance}.

Take DGP 1 for instance. 
When $\lambda=6400$, the original HP filter incurs a large bias (than HP with benchmark $\lambda=1600$) but this bias reduces quickly as iterations proceed. The minimal MSE is slightly
smaller than 2.5, which is very close to the value when $\lambda=1600$ and the MSE curve is flat in subsequent iterations.
Due to the enlarged initialization choice of $\lambda=6400$, more iterations in the bHP filter are needed until the ADF or BIC criteria terminate the algorithm. This phenomenon is consistent
with the concept of boosting based on a weak learner. 
When $\lambda=400$, the original HP is much less biased and it is already
close to the minimal MSE. The upturn of the MSE curve is witnessed
after 4 iterations. The minimal MSE is still slightly smaller than
2.5. ADF mostly indicates that 1 iteration is sufficient, whereas BIC mostly requires around 3--5 iterations. These results are consistent with the discussion of slower expansion rates for $\lambda$ in the HP filter given in 
PJ (Section 5). Similar patterns are observed in the case of DGP 2. 

In summary, over a reasonably wide range of initial values of $\lambda$ (delineated by a factor of 4) around the standard benchmark value of $\lambda=1600$,
we observe robust performance of the bHP filters in terms of
the variance-bias tradeoff and the minimal MSE fit to the true trend function. The number of iterations suggested by ADF or BIC respond to the choices of the initial values employed for $\lambda$ in recovering trend estimates that align with the benchmark $\lambda=1600$ setting.

\subsection{ Illustration of the DGPs in Section \ref{subsec:goodness} }

\begin{figure}[htbp]
	\centering
	\includegraphics[scale = 0.45 ]{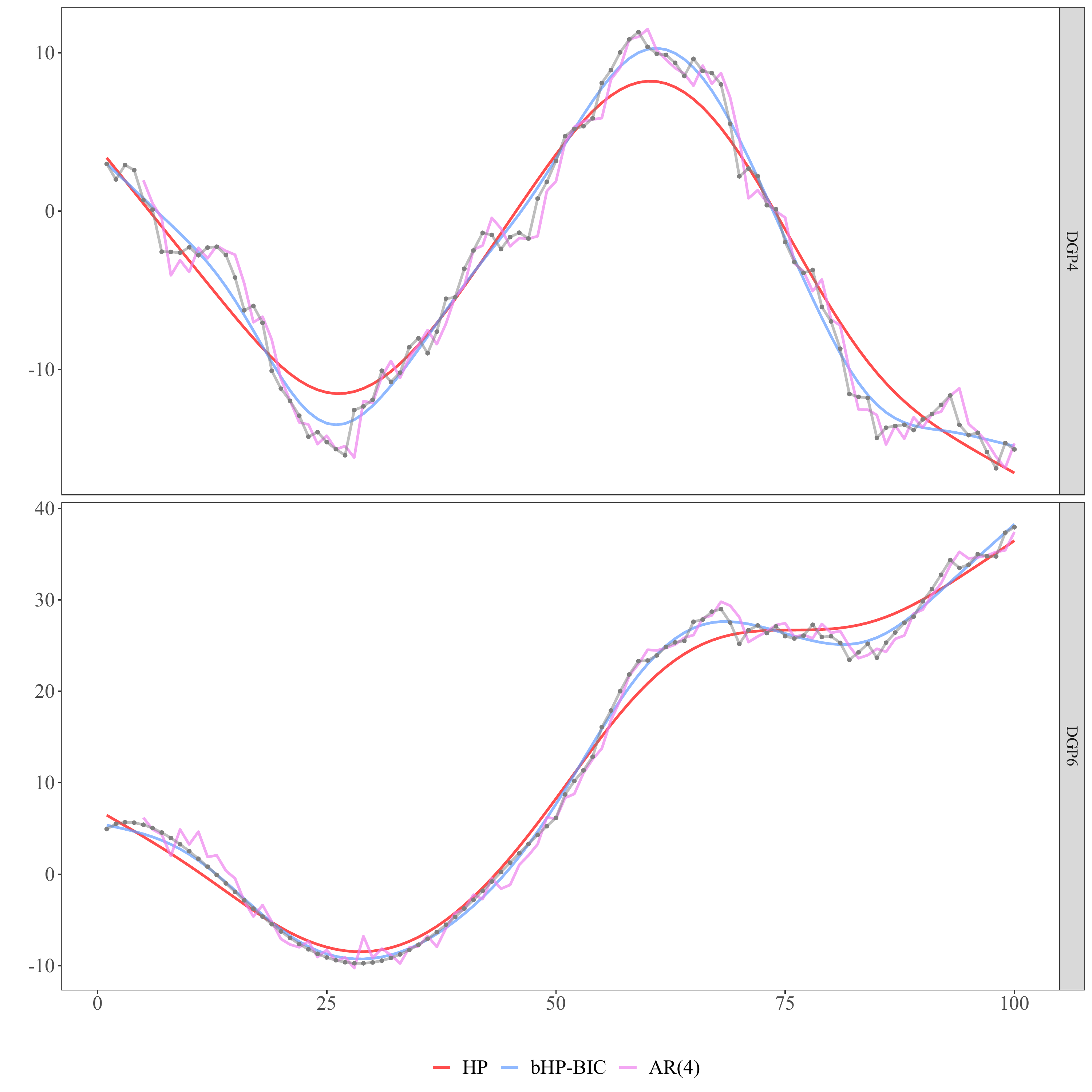} 
	\caption{\label{fig:AR4_simulation} Typical time series paths of the underlying trends in DGP 4 and DGP 6 (grey dotted curve) and the estimated trends (HP: red, bHP-BIC: blue, AR(4): violet). 
	}
\end{figure}

Figure \ref{fig:AR4_simulation} shows typical trend paths for DGPs 4 and 6
along with the corresponding estimated trends. 
In each case, the estimated bHP-BIC
trend component is significantly closer to the underlying trend 
than that of the HP filter. 
The AR(4) does not fit well in the stationary episode of DGP 6, most likely 
due to its nonsmooth nature and tendency to track the observations rather than the trend. 
Moreover, when the time series has no stationary episode as in DGP 4, the AR(4) fits also encounter difficulty in adapting to turning points in the time series. 
The latter is a feature
that echoes the empirical example in 
Section \ref{subsec:univariate} where the filter's behavior during economic crises is studied. 
Once the more complex deterministic mechanism is removed, the bHP filter and the AR(4) both fit well, as is evident in the MSE for DGP 3 in Table \ref{tab:MSE}.

\subsection{Additional Graphs For Section \ref{subsec:univariate}}
\label{subsec:additional_IP_figures}

\begin{figure}[htbp]
\centering
\includegraphics[width=.88\linewidth]{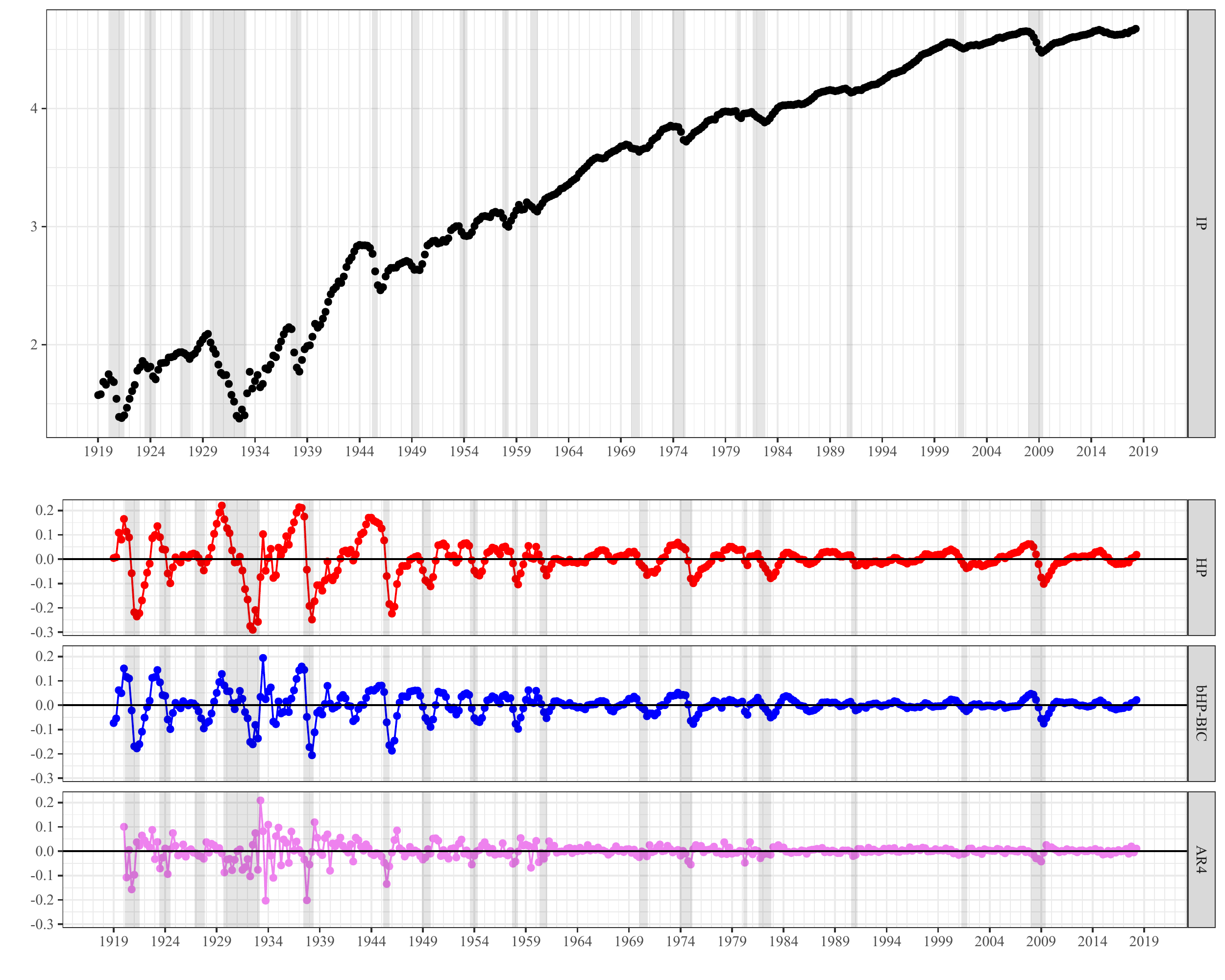}  

\caption{Industrial Production, fitted cycles, and NBER dated recessions starting from 1919. The 
black dots in the upper panel are the raw IP series data in logarithms. The lower panels show the fitted cycles. The shaded areas are the NBER dated recessions. }
\label{fig:IP_cycle1919}
\end{figure}

\begin{figure}[htbp]
\centering
\includegraphics[width=.88\linewidth]{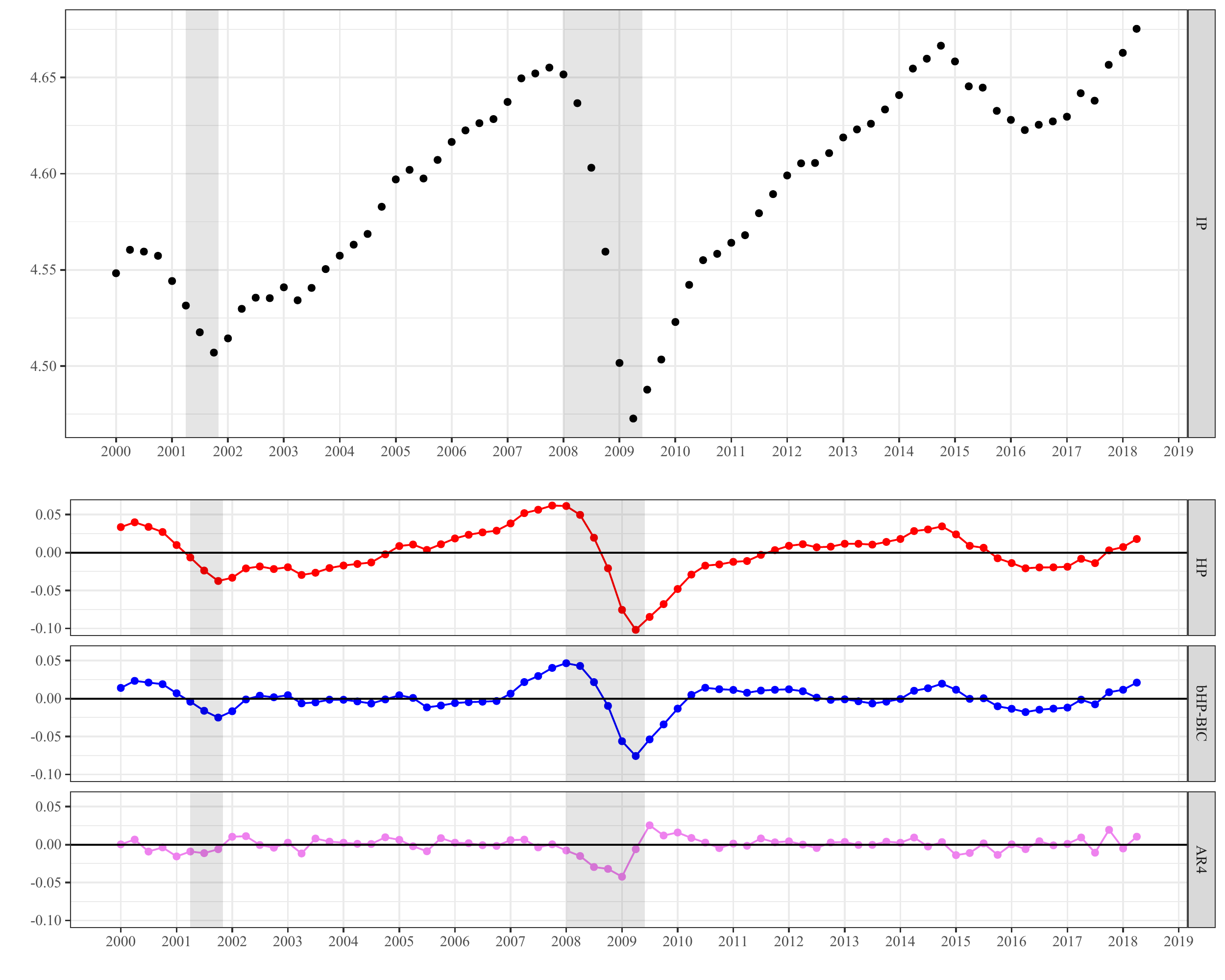}  

\caption{Industrial Production, fitted cycles, and NBER (shaded) recessions in the 21st century, zoomed from Figure \ref{fig:IP_cycle1919}.}
\label{fig:IP_cycle2000}
\end{figure}

Figures in this section detail the fitted 
cycles obtained in the filtering exercise for  
US industrial production index given in Section \ref{subsec:univariate}. 
Figure \ref{fig:IP_cycle1919} displays the entire series along
with the fitted cycles, and Figure \ref{fig:IP_cycle2000}
zooms in on the two decades of the 21st century.

Despite the large error magnitude and volatility in the early years, the HP filter cyclical component oscillates 
around zero and little evidence is observed of a tendency to drift away from the mean. 
Unsurprisingly, such a residual sequence leads to 
rejection of the null in an ADF test. The bHP-BIC cycle produces a similar test outcome as the standard HP filter, rejecting unit root nonstationarity. The resulting cyclical component of bHP-BIC has fewer large fluctuations than HP but shows cycles of irregular duration and intensity. In contrast, the AR(4)'s fitted trend closely tracks all observations in typical autoregressive fashion and there is little evidence of cycles in the residual.

\newpage

\bibliographystyle{ier}
\bibliography{hp_filter}

\end{document}